\documentclass[amsmath,onecolumn, floatfix,nofootinbib, prx, aps, usenames, dvipsnames,10pt,a4paper]{quantumarticle}
\pdfoutput=1

\usepackage[utf8]{inputenc}
\usepackage[english]{babel}
\usepackage[T1]{fontenc}
\usepackage{amsmath,amsthm,amsfonts,amssymb,amscd}
\usepackage{thmtools,thm-restate}
\usepackage{environ}
\usepackage{enumitem}
\usepackage{fancyhdr}
\usepackage{mathrsfs}
\usepackage[usenames,dvipsnames]{xcolor}
\usepackage{graphicx}
\usepackage{listings}
\usepackage[ colorlinks = true, 
             linkcolor = MidnightBlue,
             urlcolor  = MidnightBlue,
             citecolor = orange,
             anchorcolor = ForestGreen,
]{hyperref} 
\usepackage{pdfpages}
\usepackage{dsfont}
\usepackage[a4paper]{geometry}
\usepackage{scalerel,stackengine}
\usepackage[sort]{natbib}
\usepackage[page]{totalcount}
\usepackage{bm}
\usepackage{braket}
\usepackage{placeins} 
\usepackage{standalone}
\usepackage{lscape}
\usepackage{cleveref}
\usepackage{lipsum}
\usepackage{float}
\usepackage{etoolbox}
\usepackage{epigraph}
\usepackage{multicol}

\usepackage[many,breakable]{tcolorbox} 
\newtcolorbox[auto counter,number within=section]{protibox}[2][]{%
breakable,colback=orange!5,colframe=orange!75!yellow,enlarge top by=5pt,
title=Box~\thetcbcounter: #2,#1}
\definecolor{superlightgray}{gray}{0.999}

\tcbmaketheorem{figbox}{Fig}{float=ht!,colback=orange!5,colframe=orange!75!yellow}{fig}{fig}

\usepackage{tikz} 
\usetikzlibrary{shapes,arrows}

\usetikzlibrary{quantikz}

\usepackage{mathtools} 

\setlist[enumerate]{itemsep=0mm}

\newcommand{\constructionproofnoskip}[3]{\noindent\underline{\emph{#1}} #2 (Figure~#3).\ }
    
\newcommand{\constructionproof}[3]{\medskip\constructionproofnoskip{#1}{#2}{#3}}

\newcommand{\ifexportproofs}{\ifx\exportproofs\undefined\else}
\newcommand{\ifnexportproofs}{\ifx\exportproofs\undefined}
\newcommand{\ifcompleteproof}{\ifx\completeproofrequired\undefined\else}
\newcommand{\ifncompleteproof}{\ifx\completeproofrequired\undefined}

\NewEnviron{shortproof}{
    \ifncompleteproof
        \begin{proof}[Sketch of proof]
        \BODY
        \end{proof}
    \fi
}

\NewEnviron{completeproof}{
    \ifcompleteproof
        \begin{proof}
        \BODY
        \end{proof}
    \fi
}

\newcommand{\C}{\mathbb{C}}

\newcommand{\N}{\mathbb{N}}
\newcommand{\Q}{\mathbb{Q}}
\newcommand{\R}{\mathbb{R}}

\newcommand{\sL}[2]{{\mathfrak{s l}(2, \mathbb{C})}}
\newcommand{\bigO}{\mathcal{O}}
\newcommand{\Prob}[1]{\mathbb{P}\left( #1 \right)}
\newcommand{\given}{\,\middle|\,}
\newcommand{\E}[1]{\mathbb{E}\left[ #1 \right]}

\newcommand{\1}[1]{\mathds{1}}

\newcommand{\tr}{\mathrm{tr}}

\newcommand{\id}{\mathrm{id}}

\numberwithin{equation}{section}
\numberwithin{figure}{section}

\theoremstyle{remark}

\theoremstyle{plain}
\newtheorem{theorem}{Theorem}[section]
\newtheorem{corollary}[theorem]{Corollary}
\newtheorem{lemma}[theorem]{Lemma}

\newtheorem{definition}[theorem]{Definition}

\newcommand{\acframex}{0}
\newcommand{\acframey}{0}

\newcommand{\acreframe}[2]{
    \renewcommand{\acframex}{#1}
    \renewcommand{\acframey}{#2}
}

\newcommand{\acresetframe}{\acreframe{0}{0}}

\newcommand{\acrect}[4]{
    \draw[thick]
        ({\acframex+#1+0},  {\acframey+#2}) -- 
        ({\acframex+#1+#3}, {\acframey+#2}) -- 
        ({\acframex+#1+#3}, {\acframey+#2+#4}) -- 
        ({\acframex+#1+0},  {\acframey+#2+#4}) -- 
        ({\acframex+#1+0},  {\acframey+#2});
}

\newcommand{\acarrow}[4][->]{
    \draw[#1]
    (\acframex+#3,\acframey+#2) -- (\acframex+#4,\acframey+#2);
}

\newcommand{\acprotocol}[5]{
    \acrect{#2}{#3}{#4}{#5}
    \node[align=center] at ({\acframex+#2+#4/2}, {\acframey+#3+#5+0.2}) {#1};
}

\newcommand{\acresource}[5]{
    \acrect{#2}{#3}{#4}{#5}
    \node[align=center] at ({\acframex+#2+#4/2}, {\acframey+#3+#5/2}) {#1};
}

\newcommand{\acmessage}[5][->]{
    \acarrow[#1]{#3}{#4}{#5}
    \node[align=center] at ({\acframex+(#4+#5)/2}, {\acframey+#3+0.2}) {#2};
}

\newcommand{\acleftmessage}[5][->]{
    \acarrow[#1]{#3}{#4}{#5}
    \node[align=right] at ({\acframex+#4}, {\acframey+#3+0.2}) {#2};
}

\newcommand{\acrightmessage}[5][->]{
    \acarrow[#1]{#3}{#4}{#5}
    \node[align=left] at ({\acframex+#5}, {\acframey+#3+0.2}) {#2};
}

\newcommand{\indist}[3][]{
    \node at (\acframex+#2,\acframey+#3) {$\approx_{#1}$};
}

\newcommand{\accausality}[3]{
    \node[red] at (\acframex+#2,\acframey+#3) {#1};
}

\newcommand{\acnote}[3]{
    \node[blue,align=left] at (\acframex+#2,\acframey+#3) {#1};
}

\newcommand{\acpin}[3]{
    \node[align=center,font=\small] at (\acframex+#2,\acframey+#3) {#1};
}

\newcommand{\acellipsis}[2]{
    \node at (\acframex+#1, \acframey+#2) {$\vdots$};
}

\def\exportproofs{1}

\begin{document}

\title{Impossibility of composable Oblivious Transfer\newline in relativistic quantum cryptography}

\author{Lorenzo Laneve}
\affiliation{Department of Computer Science, ETH Zurich, 8092 Z\"urich, Switzerland}
\email{llaneve@ethz.ch}

\author{L\'idia del Rio}
\affiliation{Institute for Theoretical Physics, ETH Zurich, 8093 Z\"urich, Switzerland}
\email{lidia@phys.ethz.ch}

\maketitle

\begin{abstract}

We study the cryptographic primitive  \emph{Oblivious Transfer}; a composable construction of this resource would allow arbitrary multi-party computation to be carried out in a secure way, i.e.\ to compute functions in a distributed way while keeping inputs from different parties private~\cite{Kilian88, Ishai08, Unruh10}.
First we review a framework that allows us to analyze composability of classical and quantum cryptographic protocols in special relativity: Abstract Cryptography~\cite{MauRen11} instantiated by Causal Boxes~\cite{CausalBoxesFW, Vilasini19}. We then (1) explore and formalize different versions of oblivious transfer found in the literature, (2) prove that their equivalence holds also in  relativistic quantum settings, (3) show that it is impossible to composably construct any of these versions of oblivious transfer from only classical or quantum communication among distrusting agents in  relativistic settings, (4) prove that the impossibility also extends to multi-party computation, and (5) provide a mutual construction between oblivious transfer and bit commitment.
\end{abstract}

\setlength{\epigraphwidth}{4.5in}
\epigraph{Le immagini della memoria, una volta fissate con le parole, si cancellano. \vspace{5mm}\\
\emph{(Memory's images, once they are fixed in words, are erased.)}
}{Italo Calvino, \emph{Le città invisibili}}

\vspace{3cm}

\paragraph{Note:} We have structured this paper so that knowledge of quantum theory or special relativity is not required to follow the exposition of the main results, and is only necessary to understand some formal definitions and proofs in the appendix.  Throughout this work we use the word \emph{classical} as opposed to \emph{quantum}.

\vspace{0.5cm}

\paragraph{Acknowledgements.}
LdR acknowledges support from the Swiss National Science Foundation through SNSF project No.\ $200020\_165843$, the FQXi large grant \emph{Consciousness in the Physical World}, and from the Quantum Center of ETH Zurich.

\newpage
\tableofcontents


\section{Introduction}
\label{sec:introduction}

We address \emph{composable security} of cryptographic protocols: that is, to go beyond stand-alone security proofs and ensure that protocols are also secure when combined and composed with one another~\cite{Can01,MauRen11}. 
In \emph{multiparty cryptography}, we assume that any agent involved in a protocol may behave dishonestly, and try to build cryptographic resources robust against such behaviour. It is known to be impossible for mutually distrusting agents to build many desirable resources using only classical or quantum communication~\cite{CanFisch01, Mayers97, Lo1997, Lo1998, Lo97}.

\paragraph{Bit commitment and oblivious transfer.} There are two specially important two-party cryptographic primitives: \emph{bit commitment} and \emph{oblivious transfer}. Intuitively, in the former, the sender has to commit to a bit (commit phase), and then open the commitment at a later time (opening phase). The receiver cannot learn any information about the committed bit before the open phase, and the sender cannot change the value of the bit after the commit phase. Oblivious transfer consists of sending a number of messages to the receiver without knowing which of the messages are received. One version is the so-called \emph{1-out-of-2 oblivious transfer}, where the sender inputs two bits, and the receiver secretly chooses one of them, acquiring no information about the other one. We will formalize and discuss these primitives in later sections of this work. 

\paragraph{Known constructive results.}
These two primitives are complete (in the quantum setting) for arbitrary secure multi-party computation, i.e.\ a secure implementation of either of these two primitives can be used as a subroutine to implement complex primitives computing arbitrary functions. Kilian~\cite{Kilian88} first proved this for oblivious transfer in a classical setting,; his work was extended by Ishai~et~al.~\cite{Ishai08} who found a general construction for an efficient multi-party protocol, removing the assumption of honest majority. These results were shown to also hold in the quantum setting by Unruh~\cite{Unruh10}. In the same work, Unruh presented a quantum protocol achieving oblivious transfer using bit commitment, thus proving completeness of the latter in the quantum setting.

\paragraph{Known impossibility results.}
 Canetti and Fischlin~\cite{CanFisch01} showed that bit commitment is impossible in a classical, non-relativistic setting. Impossibility results for commitment protocols were extended to the quantum non-relativistic setting independently by Mayers~\cite{Mayers97} and Lo and Chau~\cite{Lo1997, Lo1998}. Finally, a recent work by Vilasini~et~al.~\cite{Vilasini19} showed that quantum bit commitment cannot be securely constructed even under relativistic constraints. Oblivious transfer was also proven to be impossible in non-relativistic settings: Lo~\cite{Lo97} showed that arbitrary one-sided two-party computation cannot achieve the desired security properties and, as a corollary, oblivious transfer cannot be securely constructed, even in the quantum setting.

\paragraph{Taking special relativity into account.}
Since quantum effects do not suffice to achieve security in multi-party computation, recent works try to also exploit constraints given by special relativity, like the maximum speed of light for propagation of messages: for example, Kent~\cite{Kent99, Kent12} proposed two relativistic protocols for bit commitment. However, the impossibility result proved by Vilasini~et~al.~\cite{Vilasini19} also implies their non-composability.
In the same work, Vilasini~et~al.\ showed that, under the assumption that a \emph{channel with delay} is possible (i.e.\ a channel that ensures a certain time interval between sending and receipt of a message), a (time-bounded) bit commitment protocol is possible, at least in principle. The possibility of Oblivious Transfer was left as an open question. 
A complete cryptographic framework that takes into account both quantum theory and special relativity is also relevant in the context of quantum communication in space~\cite{Ren2017, Yin1140, Liao2017}.

\paragraph{On composability of security.}
In order for these results to be relevant in practical applications, we need a theoretical notion of security that ensures \emph{composability}: a cryptographic primitive has to keep its security guarantees even when used in a broader context (e.g.\ many instances of the primitive are executed in parallel). For example, stand-alone security of a protocol is not enough to ensure that the protocol is not vulnerable to man-in-the-middle attacks, where dishonest agents could run two instances of the protocol in parallel and create undesirable correlations between the outputs. 
In order to model composability in a general way, Canetti~\cite{Can01} developed the \emph{Universally Composable framework}, where parties are represented as a network of machines exchanging messages, in the presence of a corrupting adversary and an environment that acts as a distinguisher. Protocols that are secure with respect to this framework can be proven to achieve strong security guarantees. Ben-Or and Mayers~\cite{BenMay04} and Unruh~\cite{Unruh10} proposed an extension of the Universally Composable framework to the quantum setting. More recently, Maurer and Renner~\cite{MauRen11} developed an alternative, the \emph{Abstract Cryptography} framework, which we use extensively throughout this work. The top-down approach used in this framework allows us to implement resources and protocols as Causal Boxes~\cite{CausalBoxesFW}, which gives us enough expressive power to model not only quantum protocols, but also relativistic ones, in a straightforward but general way. This approach was originally explored by Vilasini~et~al.~\cite{Vilasini19}, and our paper can be seen as an application to other primitives. We review the Abstract Cryptography framework in Section~\ref{sec:ac-framework}.

\paragraph{Contributions of this paper.}
In this work, we use the Abstract Cryptography framework to formalize various versions of oblivious transfer found in the literature (Section~\ref{sec:ot-definition}). We proceed by showing that the equivalence of all these versions (in the sense that they can be constructed from each other in a composable way) holds also in the relativistic quantum setting (Section~\ref{sec:ot-equivalence}). Then, our main result proves that all of these versions are impossible to construct without additional assumptions, by showing that a simple distinguisher is able to tell apart any construction with constant probability (Sections~\ref{sec:impossibility_OC}). We show that these proofs of impossibility extend to oblivious string transfer  and to multi-party computation (Section~\ref{sec:impossibility_string_MPC}).
We conclude by showing the equivalence between bit commitment and oblivious transfer in the relativistic quantum setting (Section~\ref{sec:BC}), allowing past and future work on minimal additional assumptions for bit commitment to be easily extended to oblivious transfer.
In the appendix, we present a more formal overview of  Abstract Cryptography  (Appendix~\ref{appendix:AC}) and of  Causal Boxes, which allow us to model relativistic quantum protocols (Appendix~\ref{appendix:CB}). All proofs can be found in Appendix~\ref{appendix:proofs}.

\section{Overview of the Abstract Cryptography framework}
\label{sec:ac-framework}
In this section we briefly review the Abstract Cryptography (AC) framework developed by Maurer and Renner~\cite{MauRen11}.  A more  formal review can be found in Appendix~\ref{appendix:AC}; here we present an  informal recap with the ingredients needed to follow the rest of the paper. Note that while the framework applies to general multipartite settings, in this summary we restrict ourselves to bipartite scenarios with two mutually distrusting agents, Alice and Bob.

\paragraph{Motivation.} Traditional theories of cryptography, built upon complexity and information theory with a bottom-up approach, formalize primitives and give notions of security directly from the underlying models of computation and communication. This makes it harder to generalize constructions and security to a different setting (for example, upgrading from classical to quantum communication channels).
In contrast, the AC framework follows a top-down approach: it defines primitives and protocols as abstract objects (resources) in a pseudo-metric space, where the chosen pseudo-metric depends on the notion of security we would like to use, and then lower levels of abstraction should define what these objects are. In this way, cryptography is formalized as a \textit{resource theory}: a secure primitive is seen as a resource, and a protocol implementing such primitive is said to \textit{construct} that resource.

\paragraph{Cryptographic resources.} The building blocks of the theory are `resources': we can think of them as trusted black boxes with interfaces that different agents can interact with and with a reliable input/output behaviour. For example we could think of the resource $R$ `addition' which takes in a bit $a$ from Alice and a bit $b$ from Bob and returns their sum $a \oplus b$ to both players at a later time. Resources can be composed (for example one could connect some of the interfaces of $R$ to another resource), and the resource theory is about what we can build from elementary building blocks, under cryptographic restrictions. Bit commitment and oblivious transfer are more complex types of resources, called \emph{cryptographic} primitives: in general, we want to know how the black box behaves when both players are honest and when either of them is dishonest, and this is formalized by a triple of resources, as follows.

\begin{definition}[Primitive]
    \label{def:primitive}
    A primitive is a triple of resources $\mathcal{R} = (R, R_A, R_B)$, where $R$ is a resource built by both honest Alice and Bob, and $R_A$ (resp.\ $R_B$) is a resource built under the assumption of dishonest Alice (resp.\ dishonest Bob).
\end{definition}

\noindent Splitting a primitive into three resources allows us to define primitives that behave differently upon dishonesty of one of the parties, which may be useful in some contexts where a primitive allowing some extra power to dishonest parties is still `secure enough' for us.

\begin{figure}
    \centering
    \begin{tikzpicture}[scale=1.5]
    \acmessage[<-]{}{0.25}{-0.5}{0}
    \acmessage[<-]{}{0.5}{-0.5}{0}
    \acmessage[->]{}{0.75}{-0.5}{0}
    \acresource{$R$}{0}{0}{1}{1}
    \acmessage[->]{}{0.35}{1}{1.5}
    \acmessage[<-]{}{0.65}{1}{1.5}
    
    \acprotocol{$\Pi_A$}{-1}{0}{0.5}{1}
    \acprotocol{$\Pi_B$}{1.5}{0}{0.5}{1}
    
    \acleftmessage[->]{}{0.65}{-1.5}{-1}
    \acleftmessage[<-]{}{0.35}{-1.5}{-1}
    
    \acrightmessage[<-]{}{0.5}{2}{2.5}
    
    \node at (3, 0.5) {$\approx^?$};
    
    \acleftmessage[->]{}{0.65}{3.5}{4}
    \acleftmessage[<-]{}{0.35}{3.5}{4}
    \acresource{$S$}{4}{0}{1}{1}
    \acmessage[<-]{}{0.5}{5}{5.5}
\end{tikzpicture} 
    \caption{\textbf{Bipartite resources and protocols}. Here $R$ is a shared cryptographic resource with interfaces towards Alice (left) and Bob (right). Resources like $R$ can be seen as trusted black boxes characterized by their input/output behaviour.
    The arrows represent input and output messages. $\Pi_A$ and $\Pi_B$ represent protocols implemented by Alice and Bob respectively. We may wonder whether the construction $\Pi_A \, R\,  \Pi_B$ can emulate another resource $S$ for all practical purposes (including when it is a subroutine of a larger protocol). This will be formalized ahead.}
    \label{fig:distinguishing-advantage}
\end{figure}
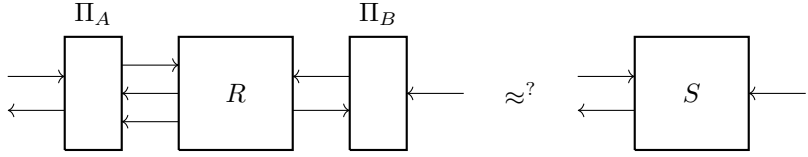

\paragraph{Protocols and constructions.} Generally, we are interested in constructing resources from one another. For example, suppose that Alice and Bob start from the above mentioned resource $R$ that returns $a\oplus b$ to both agents, and wanted to use it to build a two-way communication channel between the two, that is a resource $S$ that takes as inputs $a$ and $b$ and outputs Alice's bit $a$ to Bob, and Bob's bit $b$ to Alice. They can do this through a simple protocol: Alice's protocol $\Pi_A$ consists of sending $a$ to $R$, then receiving $R$'s output $x= a\oplus b$, and summing her bit $a$ again locally to $x$, obtaining $b$. Bob's protocol $\Pi_B$ is analogous, allowing him to recover $a$. We will represent this as $\Pi_A R \Pi_B \approx S$, for some equivalence relation $\approx$ we will define later: resource $S$ can be emulated by the construction of $R$ together with Alice's protocol $\Pi_A$ and Bob's protocol $\Pi_B$. We can think as the left side of $R$ as Alice's interface, where $\Pi_A$ is plugged in, and the right side as Bob's (Figure~\ref{fig:distinguishing-advantage}). We will see several examples in later sections. In practice constructions are often imperfect, and can only `approximate' the desired resource. In order to quantify these approximations and make statements we will denote with an equivalence relation $\approx_\varepsilon$, we need an operational measure of distinguishability between the two resources. We will define this now.  

\begin{figure}
        \centering
        \begin{tikzpicture}[scale=1.5]
    \acmessage[<-]{}{0.25}{-0.5}{0}
    \acmessage[<-]{}{0.5}{-0.5}{0}
    \acmessage[->]{}{0.75}{-0.5}{0}
    \acresource{$R$}{0}{0}{1}{1}
    \acmessage[->]{}{0.35}{1}{1.5}
    \acmessage[<-]{}{0.65}{1}{1.5}
    
    \draw[thick]
        (-0.5, 1) -- 
        (-0.5, -0.3) --
        (1.5, -0.3) --
        (1.5, 1) --
        (2, 1) --
        (2, -0.8) --
        (-1, -0.8) --
        (-1, 1) --
        (-0.5, 1);
        
    \node[align=center] at (0.5, -0.56) {$D$};
    
    \draw[->] (0.5,-0.8) -- (0.5,-1.1);
    \node[align=left] at (1.35, -1.1) {$D[R] \in \{ 0, 1 \}$};
    
     \node at (2.5, 0.5) {vs};
\end{tikzpicture}
\quad
\begin{tikzpicture}[scale=1.5]
    \acmessage[<-]{}{0.25}{-0.5}{0}
    \acmessage[<-]{}{0.5}{-0.5}{0}
    \acmessage[->]{}{0.75}{-0.5}{0}
    \acresource{$S$}{0}{0}{1}{1}
    \acmessage[->]{}{0.35}{1}{1.5}
    \acmessage[<-]{}{0.65}{1}{1.5}
    
    \draw[thick]
        (-0.5, 1) -- 
        (-0.5, -0.3) --
        (1.5, -0.3) --
        (1.5, 1) --
        (2, 1) --
        (2, -0.8) --
        (-1, -0.8) --
        (-1, 1) --
        (-0.5, 1);
        
    \node[align=center] at (0.5, -0.56) {$D$};
    
    \draw[->] (0.5,-0.8) -- (0.5,-1.1);
    \node[align=left] at (1.35, -1.1) {$D[S] \in \{ 0, 1 \}$};
\end{tikzpicture}
        \caption{\textbf{Distinguishers}. In abstract cryptography, attackers and the general environment of a resource are modelled by distinguishers. We use this tool to model how well a resource $R$ can emulate another resource $S$. For example, the distinguisher $D$ can represent the larger cryptographic setting in which $S$ would implement an ideal subroutine, and $R$ would be the resource constructed using a protocol.  \\
        A distinguisher $D$ interacts with the resource, either $R$ or $S$, and tries to identify which one it is. 
        It covers all the free interfaces of the resource and returns a bit on its outer interface (ideally $D[R]=0$ and $D[S]=1$). 
        The success of this task can be measured by the distinguishing advantage  $d^D(R, S) = \left|\Prob{D[R] = 1} - \Prob{D[S] = 1}\right|$, which is 1 if $D$ can perfectly distinguish the two resources, and 0 if $D$ cannot tell them apart at all. This creates a notion of proximity among resources that depends on the class of distinguishers used (e.g.\ classical, quantum, memory/time-bounded, etc.).}
        \label{fig:distinguisher-definition-main}
\end{figure}
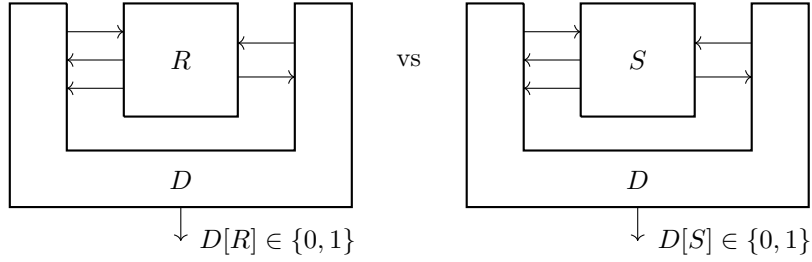

\paragraph{Distinguishing between resources.} Intuitively, we would like to relate how close two resources (say $R$ and $S$) are, to how easy it is for someone to distinguish them. If Alice and Bob think that their construction emulates perfect key distribution, then the quality of this emulation is measured by how secure their key is from adversaries. In abstract cryptography, we model the whole environment of a resource as a \emph{distinguisher}, a device with two interfaces: the inner interface attaches to all free interfaces of the observed resource, and the outer interface outputs a single bit. We call this bit $D[R]$ when the observed resource is $R$ (Figure~\ref{fig:distinguisher-definition-main}), and we can use it to define distinguishability between $R$ and $S$: intuitively, $D$ manages to distinguish $R$ from $S$ when $D[R]$ and $D[S]$ are different (e.g.\ $D[R] = 0$ and $D[S] = 1$ with high probability).

\begin{definition}[Distinguishing advantage~\cite{Portmann14}]
    Given a distinguisher $D$ and a resource set $\Omega$, the statistical advantage of $D$ in distinguishing between two resources $R, S \in \Omega$ is defined as
    \begin{align*}
        d^D(R, S) := \left|\Prob{D[R] = 1} - \Prob{D[S] = 1}\right|\ .
    \end{align*}
    Given a class of distinguishers $\mathbb{D}$, the distinguishing advantage relative to that class is defined as
    \begin{align*}
        d^{\mathbb{D}}(R, S) = \sup_{D \in \mathbb{D}} d^D(R, S).
    \end{align*}
    We use $R \approx_\varepsilon S$ to denote $d^{\mathbb{D}}(R, S) \le \varepsilon$. When $\varepsilon = 0$, we may also remove the subscript.
\end{definition}
\noindent This distinguishability notion is similar to those in classical game-based theories of cryptography (like in \cite{boneh2020}), but crucially here we did not fix a particular class of distinguishers, and this gives an extreme flexibility to the model. If we take $\mathbb{D}$ as the class of all possible distinguishers, we will obtain a definition of perfect (information-theoretic) security. On the other hand, we can model computational security by  restricting $\mathbb D$ to efficient distinguishers:  polynomial-time according to some complexity structure, classical, quantum or non-signalling distinguishers, or distinguishers with bounded memory, for example. It will be important for most proofs that $d^\mathbb D$ is well-behaved and in particular satisfies the triangle inequality.
 
\begin{lemma}[Distinguishers induce a pseudo-metric~\cite{MauRen11}]
    For any set of distinguishers $\mathbb D$, the distinguishing advantage  $d^{\mathbb D}$ is a pseudo-metric in the space of resources, that is: it is symmetric, satisfies the triangle inequality, and $d^{\mathbb D}(R,R)=0$ for all resources. 
\end{lemma}

\paragraph{Cryptographic security.} 
We have all the ingredients to quantify how well Alice and Bob can construct a primitive $\mathcal S$ starting from a primitive $\mathcal R$ and local protocols $\Pi_A$ and $\Pi_B$. For completeness, the emulation should work when both players are honest and when one of them is dishonest. There is a subtlety for the dishonest case: for example when Alice is dishonest we only consider $R_A \Pi_B$ without Alice's honest protocol (then we would be able to plug an arbitrary dishonest behaviour $\bar{\Pi}_A$ on the left), and the target resource is $S_A$. However, these two constructions may have different interfaces on Alice's side (in the example above, a dishonest Alice could simply not perform the sum $a \oplus x$). 
To account for this we allow the distinguisher to simulate Alice's behaviour through a protocol $\sigma_A$, called a \emph{simulator}.  

\begin{definition}[Cryptographic security~\cite{Portmann14}]
    \label{def:crypto-security}
    Let $\mathbb D$ be a class of distinguishers, and $\mathbb P$ be a class of possible protocols for Alice and Bob. 
    A protocol $\Pi = (\Pi_A, \Pi_B) \subset \mathbb P$ is said to $\varepsilon$-construct a primitive $\mathcal{S} = (S, S_A, S_B)$ from a primitive $\mathcal{R} = (R, R_A, R_B)$, with respect to  classes $\mathbb D$ and $\mathbb P$ if
    \begin{align*}
        \Pi_A\ R\ \ \Pi_B & \approx_\varepsilon S, \\
       \exists \ \sigma_A \in \mathbb P: \qquad R_A\ \Pi_B & \approx_\varepsilon \sigma_A\ S_A, \\
       \exists \ \sigma_B \in \mathbb P: \qquad
        \Pi_A\ R_B & \approx_\varepsilon S_B\ \sigma_B.
    \end{align*} The proximity is computed with respect to  the distinguishing advantage $d^{\mathbb D}$.
    If $\varepsilon = 0$, we also say that the construction is \textit{perfect}. Another way to denote this is $\mathcal{R} \stackrel{\Pi}{\longrightarrow} \mathcal{S}$.
\end{definition}

\noindent The strength of this notion depends on two customizations: the class $\mathbb{D}$ of distinguishers considered, and the class $\mathbb{P}$ of protocols from which we pick $\Pi_A, \Pi_B, \sigma_A$ and $\sigma_B$.  While it should be obvious it is desirable that honest protocols be efficient, it is worth noticing that the simulators of dishonest agents $\sigma_A$ and $\sigma_B$ should also be as simple as possible, since in a security argument a distinguisher will have to run these components internally, and thus the complexity requirements of such attack would also depend on the resources used by the protocols and simulators. This notion of security is further discussed in Appendix~\ref{appendix:AC}.

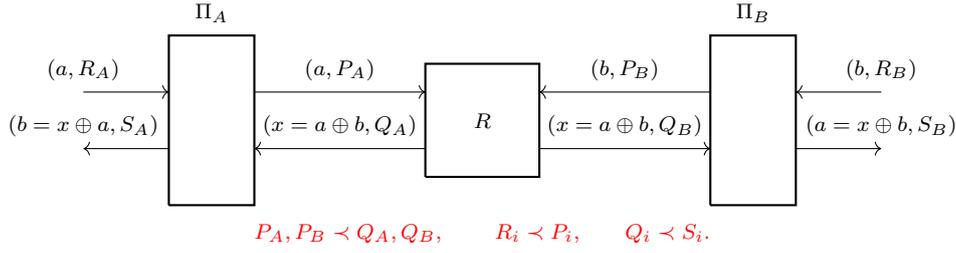
\begin{figure}
        \centering
        \begin{tikzpicture}[scale=1.5]
    \footnotesize
    
    \acleftmessage[->]{$(a, R_A)$}{0.75}{-3}{-2.25}
    \acleftmessage[<-]{$(b = x \oplus a, S_A)$}{0.25}{-3}{-2.25}

    \acprotocol{$\Pi_A$}{-2.25}{-0.25}{0.75}{1.5}

    \acresource{$R$}{0}{0}{1}{1}
    \acmessage[->]{$(a, P_A)$}{0.75}{-1.5}{0}
    \acmessage[<-]{$(x = a \oplus b, Q_A)$}{0.25}{-1.5}{0}
    
    \acmessage[<-]{$(b, P_B)$}{0.75}{1}{2.5}
    \acmessage[->]{$(x = a \oplus b, Q_B)$}{0.25}{1}{2.5}
    
    \acprotocol{$\Pi_B$}{2.5}{-0.25}{0.75}{1.5}
    
    \acrightmessage[<-]{$(b, R_B)$}{0.75}{3.25}{4}
    \acrightmessage[->]{$(a = x \oplus b, S_B)$}{0.25}{3.25}{4}
    
    \accausality{$P_A, P_B \prec Q_A, Q_B$,
    \qquad $R_i \prec P_i$,\qquad
    $Q_i \prec S_i$.}{0.5}{-0.5}
\end{tikzpicture}
        \caption{\textbf{Resources in spacetime}. To model quantum cryptographic protocols in relativistic spacetime, we implement resources as causal boxes~\cite{Portmann14,Vilasini19}, which are quantum information-processing devices whose inputs and outputs are of the form $(m, P)$ where $m$ is a message (for example a quantum state) and $P= (\vec x, t)$ its position in spacetime. (More generally messages can arrive at a quantum superposition of different positions, as described in  Appendix~\ref{appendix:CB}.)\\ 
        In this example, $R$ is a trusted resource that receives a bit $a$ from Alice at position $P_A$ and a bit $b$ from Bob at position $P_B$, and returns the sum $x= a \oplus b$ to Alice (at position $Q_A$) and to Bob (at $Q_B$). The resource must satisfy a simple causality condition: it can only return the outputs after it receives inputs, and not the other way around. This is formalized by the first condition in red, $P_A, P_B \prec Q_A, Q_B$, where the pre-order $\prec$ is given by special relativity: $Q_A$ and $Q_B$ must be in the future light cones of both $P_A$ and $P_B$.\\
        If Alice wants to recover Bob's bit $b$, she can apply a protocol $\Pi_A$ in which she sums $a$ to the output of $R$, as $x \oplus a = b$; Bob can act analogously on his side with a protocol $\Pi_B$ to recover $a$. This new construction $\Pi_A \, R \, \Pi_B$ is akin to a two-way communication channel between Alice and Bob; however we have not yet tested if it is composably secure. Note that now extra causality conditions apply ($R_i \prec P_i$ and $Q_i \prec S_i$, $i \in \{A, B\}$). 
        }
        \label{fig:causalbox_summ}
\end{figure}

\paragraph{Relativistic quantum resources.} Up to now the resources were unspecified, abstract objects. In order to treat quantum cryptographic tasks in relativistic settings, we need to instantiate resources as objects capable of processing quantum information in space-time. Causal boxes \cite{Portmann14} are suitable candidates: they are generalizations of quantum maps that also take into account the space-time position of input and output messages. The causal box framework is described in detail in Appendix~\ref{appendix:CB}; in order to follow the rest of the paper, we only need an intuition. We can think of each box as a closed physical experimental setup (like an optical table with mirrors and beam splitters that implements some quantum operation on incoming photons) together with input and output wires (like optical fiber cables) that connect boxes to one another.  Roughly speaking, wires transmit messages of the form $(m, P)$, where $m$ is the message and $P = (\vec x, t)$ is a point in Minkowski space-time marking where and when the message arrived --- that is, $P$ is composed of a 3D space position $\vec x$ and time $t$.  See Figure~\ref{fig:causalbox_summ} for an example. A point $P$ is in the causal past of another point $Q$ (also denoted as $P \prec Q$) if it is possible to reach $Q$ from $P$ by travelling at the speed of light (for physicists, if $Q$ is in the future light cone of $P$).  Both messages and their positions are quantum states, and in particular the framework can handle receiving or sending messages at a superposition of different times. Causal boxes must respect an internal causality condition, which allows them to be composed with each other arbitrarily, and as such we can model both basic resources and protocols as boxes.

\section{Results}
\label{sec:results}

\subsection{Definitions of Oblivious Transfer}
\label{sec:ot-definition}

In this section we formally introduce oblivious transfer as a primitive.  Different variants are found in the literature; here we implement as causal boxes in the Abstract Cryptography framework, and will later prove their equivalence in relativistic quantum settings. Unless otherwise noted, the dishonest versions of a resource $R_A, R_B$ are the same as the honest one, $R$. The causality conditions are given in red besides the causal box representation. 
The original definition of oblivious transfer was given by Rabin~\cite{rabin05}. 
\begin{definition}[Rabin Oblivious Transfer~\cite{rabin05}]
    A Rabin Oblivious Transfer is a primitive $\mathcal{OT}^R = (OT^R, OT_A^R, OT_B^R)$. Alice sends a bit $x$, and Bob receives it with probability $\frac{1}{2}$ (he is notified about the failure). Alice cannot infer whether the bit was received.
    
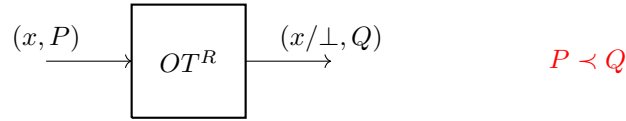
\begin{figure}[H]
        \centering
        \begin{tikzpicture}[scale=1.5]
    \acresource{$OT^R$}{0}{0}{1}{1}
    \acleftmessage[->]{$(x, P)$}{0.5}{-0.75}{0}
    \acrightmessage[->]{$(x / \bot, Q)$}{0.5}{1}{1.75}
    
    \accausality{$P \prec Q$}{4}{0.5}
\end{tikzpicture}
        \label{fig:rabin-ot-def}
    \end{figure}
\end{definition}

\noindent A more popular definition of oblivious transfer, called \textit{1-out-of-2}, or $\binom{2}{1}$-OT, was introduced by Kilian in its completeness proof~\cite{Kilian88}.
\begin{definition}[1-out-of-2 Oblivious Transfer~\cite{Kilian88}]
    A $\binom{2}{1}$-Oblivious Transfer is a primitive $\mathcal{OT} = (OT, OT_A, OT_B)$. Alice sends two bits $a_0, a_1$, and Bob chooses a bit $b$. Bob then receives $a_b$, Alice does not receive information about $b$, and Bob does not receive information about $a_{1-b}$.
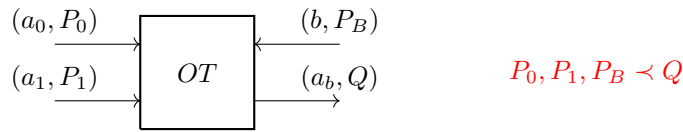
    \begin{figure}[H]
        \centering
        \begin{tikzpicture}[scale=1.5]
    \acresource{$OT$}{0}{0}{1}{1}
    \acleftmessage[->]{$(a_0, P_0)$}{0.75}{-0.75}{0}
    \acleftmessage[->]{$(a_1, P_1)$}{0.25}{-0.75}{0}
    \acrightmessage[<-]{$(b, P_B)$}{0.75}{1}{1.75}
    \acrightmessage[->]{$(a_b, Q)$}{0.25}{1}{1.75}
    
    \accausality{$P_0, P_1, P_B \prec Q$}{4}{0.5}
\end{tikzpicture}
        \label{fig:1-out-of-2-ot-def}
    \end{figure}
\end{definition}

\noindent Another version, called \textit{Randomized Oblivious Transfer}, was used by Unruh~\cite{Unruh10}.
\begin{definition}[Randomized Oblivious Transfer~\cite{Unruh10}]
    A Randomized Oblivious Transfer can be defined as a primitive $\mathcal{ROT} = (ROT, ROT_A, ROT_B)$.
    Alice receives two bits $s_0, s_1$, chosen uniformly at random by the primitive. Bob sends a bit $b$, and receives only $s_b$. Alice does not receive information about $b$, and Bob does not receive information about $s_{1-b}$. A dishonest Alice may be allowed to choose $s_0, s_1$.
    \begin{figure}[H]
        \centering
        \begin{tikzpicture}[scale=1.5]
    \acreframe{-2}{0}
    \acresource{$ROT$}{0}{0}{1}{1}
    \acleftmessage[<-]{$(s_0, P_0)$}{0.75}{-0.75}{0}
    \acleftmessage[<-]{$(s_1, P_1)$}{0.25}{-0.75}{0}
    \acrightmessage[<-]{$(b, P_B)$}{0.75}{1}{1.75}
    \acrightmessage[->]{$(s_b, Q)$}{0.25}{1}{1.75}
    
    \accausality{$P_B \prec Q$}{0.5}{-0.25}

    \acreframe{2}{0}
    \acresource{$ROT_A$}{0}{0}{1}{1}
    \acleftmessage[->]{$(s_0, P_0)$}{0.75}{-0.75}{0}
    \acleftmessage[->]{$(s_1, P_1)$}{0.25}{-0.75}{0}
    \acrightmessage[<-]{$(b, P_B)$}{0.75}{1}{1.75}
    \acrightmessage[->]{$(s_b, Q)$}{0.25}{1}{1.75}
    
    \accausality{$P_0, P_1, P_B \prec Q$}{0.5}{-0.25}
    \acresetframe
    
\end{tikzpicture}
        \label{fig:randomized-ot-def}
    \end{figure}
\end{definition}
\noindent It is worth noticing that allowing a dishonest Alice to choose $s_0, s_1$ significantly weakens the definition: if $s_0, s_1$ were chosen at random regardless of the honesty of the two parties, then one can see that a coin flip (a notoriously impossible~\cite{Vilasini19} primitive returning the same uniformly random bit to both parties) can be easily constructed (say, using $s_0$ as outcome). On the other hand, such weakening is fundamental to achieve equivalence with the other versions of oblivious transfer, as we will see in the next section.

\subsection{Equivalence of Oblivious Transfer primitives}
\label{sec:ot-equivalence}
Now we show that the three primitives are equivalent even in a relativistic quantum setting, in the sense that each of them can be constructed using secure (and composable) instances of the other two, either perfectly or with an exponentially decaying distinguishing probability. Using the results of this subsection, it will be sufficient to prove the impossibility of one of them, and we will get the impossibility of the other versions almost for free.
\ifexportproofs
    Here we present the statements and protocols for the different constructions, while the full security proofs (including the spacetime stamps of messages) can be found in Appendix~\ref{apx:equivalence-proofs}.
    Firstly, the statements of the equivalences. 
\fi

\begin{protibox}[label={box:OT_equivalences}]{Equivalence of Oblivious Transfer primitives}

\begin{restatable}[Construction $\mathcal{ROT} \rightarrow \mathcal{OT}$]{lemma}{rototconstruction}
    \label{thm:rot-ot-construction}
    The $\binom{2}{1}$-oblivious transfer $\mathcal{OT}$ can be perfectly constructed from the randomized oblivious transfer $\mathcal{ROT}$. 
    The constructing protocol $ \Pi^1$ (Definition~\ref{def:protocol-1}) and the simulators $\sigma^1_A, \sigma^1_B$ are classical and use only $\bigO(1)$ elementary, local operations and classical communication.
\end{restatable}
\begin{completeproof}
    \begin{figure}[ht]
        \centering
        \begin{tikzpicture}[scale=1]
    \scriptsize
    
    \acleftmessage[->]{$(a_0, P'_0)$}{0.25}{-2.75}{-2.25}
    \acleftmessage[->]{$(a_1, P'_1)$}{-0.25}{-2.75}{-2.25}
    \acprotocol{$\Pi^1_A$}{-2.25}{-1.25}{0.75}{2.5}
    \acmessage[<-]{$(s_0, P_0)$}{0.75}{-1.5}{-0.5}
    \acmessage[<-]{$(s_1, P_1)$}{0.25}{-1.5}{-0.5}
    \acresource{$ROT$}{-0.5}{0}{1}{1}
    \acmessage[<-]{$(b, P_B)$}{0.75}{0.5}{1.5}
    \acmessage[->]{$(s_b, Q)$}{0.25}{0.5}{1.5}
    \acmessage[->]{$(c_0 = a_0 \oplus s_0, Q_0)$}{-0.5}{-1.5}{1.5}
    \acmessage[->]{$(c_1 = a_1 \oplus s_1, Q_1)$}{-1}{-1.5}{1.5}
    \acprotocol{$\Pi^1_B$}{1.5}{-1.25}{0.75}{2.5}
    \acrightmessage[<-]{$(b, P'_B)$}{0.25}{2.25}{3}
    \acrightmessage[->]{$(c_b \oplus s_b, Q')$}{-0.25}{2.25}{3}

    \indist{4.25}{0}
    \acpin{(a)}{4.25}{-1}

    \acleftmessage[->]{$(a_0, P'_0)$}{0.25}{5.25}{6}
    \acleftmessage[->]{$(a_1, P'_1)$}{-0.25}{5.25}{6}
    \acresource{$OT$}{6}{-0.5}{1}{1}
    \acrightmessage[<-]{$(b, P'_B)$}{0.25}{7}{7.75}
    \acrightmessage[->]{$(a_b, Q')$}{-0.25}{7}{7.75}
    
    \accausality{$P_0, P_1, P'_0, P'_1 \prec Q_0, Q_1 \prec Q'$}{0}{-1.7}

    \acreframe{0}{-4}
    
    \acmessage[->]{$(s_0, P_0)$}{0.75}{-1.5}{-0.5}
    \acmessage[->]{$(s_1, P_1)$}{0.25}{-1.5}{-0.5}
    \acresource{$ROT_A$}{-0.5}{0}{1}{1}
    \acmessage[<-]{$(b, P_B)$}{0.75}{0.5}{1.5}
    \acmessage[->]{$(s_b, Q)$}{0.25}{0.5}{1.5}
    \acmessage[->]{$(c_0, Q_0)$}{-0.5}{-1.5}{1.5}
    \acmessage[->]{$(c_1, Q_1)$}{-1}{-1.5}{1.5}
    \acprotocol{$\Pi^1_B$}{1.5}{-1.25}{0.75}{2.5}
    \acrightmessage[<-]{$(b, P'_B)$}{0.25}{2.25}{3}
    \acrightmessage[->]{$(c_b \oplus s_b, Q')$}{-0.25}{2.25}{3}

    \indist{4.25}{0};
    \acpin{(b)}{4.25}{-1}

    \acleftmessage[->]{$(s_0, P_0)$}{0.75}{5.25}{6}
    \acleftmessage[->]{$(s_1, P_1)$}{0.25}{5.25}{6}
    \acleftmessage[->]{$(c_0, Q_0)$}{-0.25}{5.25}{6}
    \acleftmessage[->]{$(c_1, Q_1)$}{-0.75}{5.25}{6}
    \acprotocol{$\sigma^1_A$}{6}{-1}{1}{2}
    \acmessage[->]{$(a_0, P''_0)$}{0.25}{7}{8}
    \acmessage[->]{$(a_1, P''_1)$}{-0.25}{7}{8}
    \acresource{$OT_A$}{8}{-0.5}{1}{1}
    \acrightmessage[<-]{$(b, P'_B)$}{0.25}{9}{9.75}
    \acrightmessage[->]{$(a_b, Q')$}{-0.25}{9}{9.75}
    
    \acnote{$a_i \leftarrow s_i \oplus c_i$}{6.5}{-1.25}
    
    \accausality{$P_0, P_1, Q_0, Q_1 \prec P''_0, P''_1 \prec Q'$}{7}{-1.9}

    \acreframe{0}{-8}
    
    \acleftmessage[->]{$(a_0, P'_0)$}{0.25}{-2.75}{-2.25}
    \acleftmessage[->]{$(a_1, P'_1)$}{-0.25}{-2.75}{-2.25}
    \acprotocol{$\Pi^1_A$}{-2.25}{-1.25}{0.75}{2.5}
    \acmessage[<-]{$(s_0, P_0)$}{0.75}{-1.5}{-0.5}
    \acmessage[<-]{$(s_1, P_1)$}{0.25}{-1.5}{-0.5}
    \acresource{$ROT_B$}{-0.5}{0}{1}{1}
    \acmessage[<-]{$(b, P_B)$}{0.75}{0.5}{1.5}
    \acmessage[->]{$(s_b, Q)$}{0.25}{0.5}{1.5}
    \acmessage[->]{$(c_0 = a_0 \oplus s_0, Q_0)$}{-0.5}{-1.5}{1.5}
    \acmessage[->]{$(c_1 = a_1 \oplus s_1, Q_1)$}{-1}{-1.5}{1.5}

    \indist{4.25}{0}
    \acpin{(c)}{4.25}{-1}

    \acleftmessage[->]{$(a_0, P'_0)$}{0.25}{5.25}{6}
    \acleftmessage[->]{$(a_1, P'_1)$}{-0.25}{5.25}{6}
    \acresource{$OT_B$}{6}{-0.5}{1}{1}
    \acmessage[<-]{$(b, P''_B)$}{0.25}{7}{8}
    \acmessage[->]{$(a_b, Q'')$}{-0.25}{7}{8}
    \acprotocol{$\sigma^1_B$}{8}{-1}{1}{2}
    \acrightmessage[<-]{$(b, P_B)$}{0.75}{9}{9.75}
    \acrightmessage[->]{$(s_b, Q)$}{0.25}{9}{9.75}
    \acrightmessage[->]{$(c_0, Q_0)$}{-0.25}{9}{9.75}
    \acrightmessage[->]{$(c_1, Q_1)$}{-0.75}{9}{9.75}

    \acnote{$c_b \leftarrow a_b \oplus s_b$}{8.5}{-1.25}
    \acnote{$c_{1-b} \leftarrow Be(1/2)$}{8.42}{-1.6}
    
    \accausality{$P'_0, P'_1, P''_B \prec Q'' \prec Q_0, Q_1$ and $P_B \prec P''_B$}{7}{-2.1}
    
    \acresetframe
\end{tikzpicture}
        \caption{Construction of a 1-out-of-2 Oblivious Transfer from a Randomized Oblivious Transfer (Lemma~\ref{thm:rot-ot-construction}). Alice encrypts $a_0, a_1$ with a One-Time Pad, and sends them to Bob. Bob  only receives  one of the two keys.}
        \label{fig:rot-ot-construction}
    \end{figure}
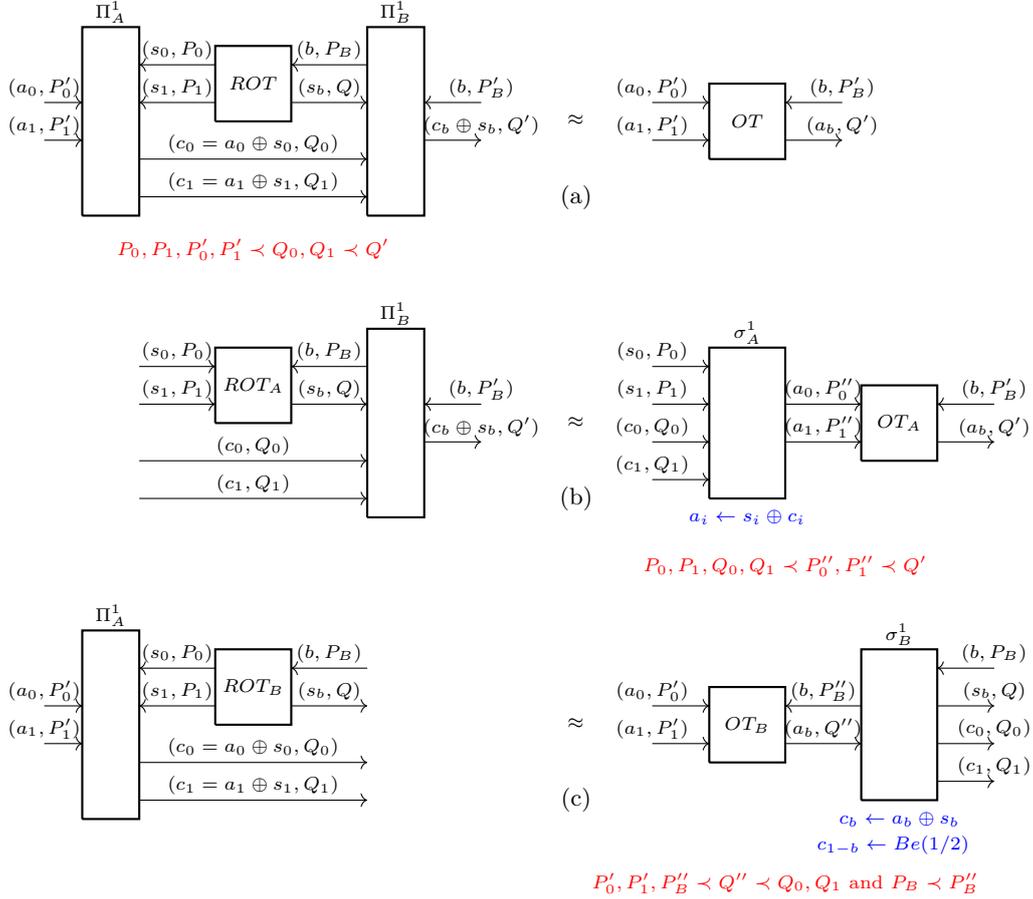
    When we provide a construction we need to argue that the three conditions imposed by Definition~\ref{def:crypto-security} hold (in this case, with $\varepsilon = 0$ since we claim perfect construction).
    
    \constructionproof{Honest protocol}{$\Pi^1_A ROT\ \Pi^1_B \approx OT$}{\ref{fig:rot-ot-construction}(a)} One can see that, under honest assumption, the protocol implements a $\binom{2}{1}$-oblivious transfer via a simple one-time pad protocol by Alice. Thus, the honest construction is indistinguishable from $OT$.
    
    \constructionproof{Simulation against dishonest Alice}{$ROT_A\Pi^1_B \approx \sigma^1_A OT_A$}{\ref{fig:rot-ot-construction}(b)}
    The simulator $\sigma^1_A$ receives $s_0, s_1, c_0, c_1$ on its left interface (recall that $ROT_A$ allows dishonest Alice to choose $s_0, s_1$), and has to output $a_0, a_1$ in such a way that $a_b = c_b \oplus s_b$. To achieve this, $\sigma^1_A$ simply sets $a_i = s_i \oplus c_i$.
    
    \constructionproof{Simulation against dishonest Bob}{$\Pi^1_A ROT_B \approx OT_B \sigma^1_B$}{\ref{fig:rot-ot-construction}(c)}
    The simulator $\sigma^1_B$ receives $b$ on its right interface and it will have to output $s_b, c_0, c_1$. In order to have indistinguishability in this case we need to see $c_b = a_b \oplus s_b$ and $s_b$ uniformly random. The other ciphertext $c_{1-b}$, on the other hand, can be simply extracted at random, since $s_{1-b}$ does not appear anywhere in the outer interfaces.
\end{completeproof}

\begin{restatable}[Construction $\mathcal{OT} \rightarrow \mathcal{ROT}$]{lemma}{otrotconstruction}
    \label{thm:ot-rot-construction}
   The randomized oblivious transfer $\mathcal{ROT}$ can be perfectly constructed from the $\binom{2}{1}$-oblivious transfer $\mathcal{OT}$. The constructing protocol $\Pi^2$ (Definition~\ref{def:protocol-2}) and the simulators $\sigma^2_A, \sigma^2_B$ are classical and use only  elementary, local operations and classical communication.
\end{restatable}
\begin{completeproof}
    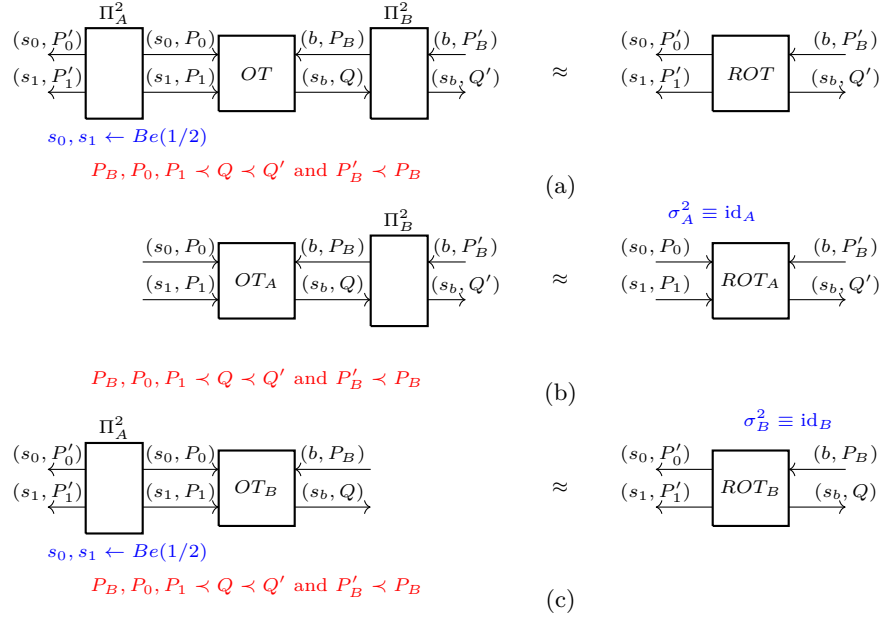
\begin{figure}[ht]
        \centering
        \begin{tikzpicture}[scale=1]
    \scriptsize
    
    \acleftmessage[<-]{$(s_0, P'_0)$}{0.75}{-2.75}{-2.25}
    \acleftmessage[<-]{$(s_1, P'_1)$}{0.25}{-2.75}{-2.25}
    \acprotocol{$\Pi^2_A$}{-2.25}{-0.1}{0.75}{1.2}
    \acmessage[->]{$(s_0, P_0)$}{0.75}{-1.5}{-0.5}
    \acmessage[->]{$(s_1, P_1)$}{0.25}{-1.5}{-0.5}
    \acresource{$OT$}{-0.5}{0}{1}{1}
    \acmessage[<-]{$(b, P_B)$}{0.75}{0.5}{1.5}
    \acmessage[->]{$(s_b, Q)$}{0.25}{0.5}{1.5}
    \acprotocol{$\Pi^2_B$}{1.5}{-0.1}{0.75}{1.2}
    \acrightmessage[<-]{$(b, P'_B)$}{0.75}{2.25}{2.75}
    \acrightmessage[->]{$(s_b, Q')$}{0.25}{2.25}{2.75}
    
    \acnote{$s_0, s_1 \leftarrow Be(1/2)$}{-1.7}{-0.35}
    
    \accausality{$P_B, P_0, P_1 \prec Q \prec Q'$ and $P'_B \prec P_B$}{0}{-0.8}

    \indist{4}{0.5}
    \acpin{(a)}{4}{-1}

    \acleftmessage[<-]{$(s_0, P'_0)$}{0.75}{5.25}{6}
    \acleftmessage[<-]{$(s_1, P'_1)$}{0.25}{5.25}{6}
    \acresource{$ROT$}{6}{0}{1}{1}
    \acrightmessage[<-]{$(b, P'_B)$}{0.75}{7}{7.75}
    \acrightmessage[->]{$(s_b, Q')$}{0.25}{7}{7.75}

    \acreframe{0}{-2.75}
    
    \acmessage[->]{$(s_0, P_0)$}{0.75}{-1.5}{-0.5}
    \acmessage[->]{$(s_1, P_1)$}{0.25}{-1.5}{-0.5}
    \acresource{$OT_A$}{-0.5}{0}{1}{1}
    \acmessage[<-]{$(b, P_B)$}{0.75}{0.5}{1.5}
    \acmessage[->]{$(s_b, Q)$}{0.25}{0.5}{1.5}
    \acprotocol{$\Pi^2_B$}{1.5}{-0.1}{0.75}{1.2}
    \acrightmessage[<-]{$(b, P'_B)$}{0.75}{2.25}{2.75}
    \acrightmessage[->]{$(s_b, Q')$}{0.25}{2.25}{2.75}

    \indist{4}{0.5}
    \acpin{(b)}{4}{-1}

    \acleftmessage[->]{$(s_0, P_0)$}{0.75}{5.25}{6}
    \acleftmessage[->]{$(s_1, P_1)$}{0.25}{5.25}{6}
    \acresource{$ROT_A$}{6}{0}{1}{1}
    \acrightmessage[<-]{$(b, P'_B)$}{0.75}{7}{7.75}
    \acrightmessage[->]{$(s_b, Q')$}{0.25}{7}{7.75}
    
    \acnote{$\sigma^2_A \equiv \id_A$}{6}{1.4}
    
    \accausality{$P_B, P_0, P_1 \prec Q \prec Q'$ and $P'_B \prec P_B$}{0}{-0.8}

    \acreframe{0}{-5.5}
    
    \acleftmessage[<-]{$(s_0, P'_0)$}{0.75}{-2.75}{-2.25}
    \acleftmessage[<-]{$(s_1, P'_1)$}{0.25}{-2.75}{-2.25}
    \acprotocol{$\Pi^2_A$}{-2.25}{-0.1}{0.75}{1.2}
    \acmessage[->]{$(s_0, P_0)$}{0.75}{-1.5}{-0.5}
    \acmessage[->]{$(s_1, P_1)$}{0.25}{-1.5}{-0.5}
    \acresource{$OT_B$}{-0.5}{0}{1}{1}
    \acmessage[<-]{$(b, P_B)$}{0.75}{0.5}{1.5}
    \acmessage[->]{$(s_b, Q)$}{0.25}{0.5}{1.5}
    
    \acnote{$s_0, s_1 \leftarrow Be(1/2)$}{-1.7}{-0.35}
    
    \accausality{$P_B, P_0, P_1 \prec Q \prec Q'$ and $P'_B \prec P_B$}{0}{-0.8}

    \indist{4}{0.5}
    \acpin{(c)}{4}{-1}
    
    \acleftmessage[<-]{$(s_0, P'_0)$}{0.75}{5.25}{6}
    \acleftmessage[<-]{$(s_1, P'_1)$}{0.25}{5.25}{6}
    \acresource{$ROT_B$}{6}{0}{1}{1}
    \acrightmessage[<-]{$(b, P_B)$}{0.75}{7}{7.75}
    \acrightmessage[->]{$(s_b, Q)$}{0.25}{7}{7.75}
    
    \acnote{$\sigma^2_B \equiv \id_B$}{7}{1.4}
\end{tikzpicture}
        \caption{Construction of a Randomized Oblivious Transfer from a 1-out-of-2 Oblivious Transfer (Lemma~\ref{thm:ot-rot-construction}). Alice chooses $s_0, s_1$ uniformly at random, and uses an instance of the 1-out-of-2 OT to send them to Bob.}
        \label{fig:ot-rot-construction}
    \end{figure}
    In the protocol $\Pi^2$, Alice and Bob run an instance of the $\binom{2}{1}$-Oblivious Transfer, when (honest) Alice picks $a_0, a_1$ uniformly at random. $\Pi^2_B$, as well as $\sigma^2_A, \sigma^2_B$, simply act as identity (forwarding messages). The construction is summarized in Figure~\ref{fig:ot-rot-construction}. It is worth noticing that the definition of the Randomized OT under dishonest Alice (which is nothing more than a $\binom{2}{1}$-OT itself) is important for this construction to work.
\end{completeproof}

\begin{restatable}[Construction $\mathcal{OT} \rightarrow \mathcal{OT}^R$]{lemma}{otrabinconstruction}
    \label{thm:ot-rabin-construction}
    Rabin's oblivious transfer $\mathcal{OT}^R$ can be perfectly constructed from $\binom{2}{1}$-oblivious transfer $\mathcal{OT}$. The constructing protocol $\Pi^3$ (Definition~\ref{def:protocol-3}) and the simulators $\sigma^3_A, \sigma^3_B$ are classical and use only  elementary, local operations and classical communication.
\end{restatable}
\begin{completeproof}
    \begin{figure}[ht]
        \centering
        \begin{tikzpicture}[scale=1]
    \scriptsize
    
    \acleftmessage[->]{$(x, P')$}{0}{-3}{-2.25}
    \acprotocol{$\Pi^3_A$}{-2.25}{-0.75}{0.75}{2}
    \acmessage[->]{$(a_0, P_0)$}{0.75}{-1.5}{-0.5}
    \acmessage[->]{$(a_1, P_1)$}{0.25}{-1.5}{-0.5}
    \acresource{$OT$}{-0.5}{0}{1}{1}
    \acmessage[<-]{$(b, P_B)$}{0.75}{0.5}{1.5}
    \acmessage[->]{$(a_b, Q)$}{0.25}{0.5}{1.5}
    \acmessage[->]{$(b^*, S)$}{-0.5}{-1.5}{1.5}
    \acprotocol{$\Pi^3_B$}{1.5}{-0.75}{0.75}{2}
    \acrightmessage[->]{$(a_b/\bot, Q')$}{0}{2.25}{3}
    
    \acnote{$b^* \leftarrow Be(1/2)$}{-1.75}{-1}
    \acnote{$a_{b^*} \leftarrow x$}{-1.75}{-1.4}
    \acnote{$a_{1 - b^*} \leftarrow Be(1/2)$}{-1.75}{-1.75}
    
    \acnote{$b \leftarrow Be(1/2)$}{2}{-1}
    \acnote{$b \stackrel{?}{=} b^*$}{2}{-1.3}
    
    \accausality{$P' \prec P_0, P_1 \prec Q \prec S \prec Q'$}{2.3}{-1.9}

    \indist{4.25}{0}
    \acpin{(a)}{4.25}{-1}

    \acleftmessage[->]{$(x, P')$}{0}{5.25}{6}
    \acresource{$OT^R$}{6}{-0.5}{1}{1}
    \acrightmessage[->]{$(x/\bot, Q')$}{0}{7}{7.75}

    \acreframe{0}{-4}

    \acmessage[->]{$(a_0, P_0)$}{0.75}{-1.5}{-0.5}
    \acmessage[->]{$(a_1, P_1)$}{0.25}{-1.5}{-0.5}
    \acresource{$OT_A$}{-0.5}{0}{1}{1}
    \acmessage[<-]{$(b, P_B)$}{0.75}{0.5}{1.5}
    \acmessage[->]{$(a_b, Q)$}{0.25}{0.5}{1.5}
    
    \acmessage[->]{$(b^*, S)$}{-0.5}{-1.5}{1.5}
    \acprotocol{$\Pi^3_B$}{1.5}{-0.75}{0.75}{2}
    \acrightmessage[->]{$(a_b/\bot, Q')$}{0.25}{2.25}{3}

    \indist{4.25}{0.25};
    \acpin{(b)}{4.25}{-1}

    \acleftmessage[->]{$(a_0, P_0)$}{0.75}{5.25}{6}
    \acleftmessage[->]{$(a_1, P_1)$}{0.25}{5.25}{6}
    \acleftmessage[->]{$(b^*, S)$}{-0.5}{5.25}{6}
    \acprotocol{$\sigma^3_A$}{6}{-0.75}{1}{1.75}
    \acmessage[->]{$(a_{b^*}, P)$}{0.25}{7}{8}
    \acresource{$OT^R_A$}{8}{-0.25}{1}{1}
    \acrightmessage[->]{$(a_{b^*}/\bot, Q')$}{0.25}{9}{9.75}
    
    \acnote{$b \leftarrow Be(1/2)$}{2}{-1}
    \acnote{$b \stackrel{?}{=} b^*$}{2}{-1.3}
    
    \accausality{$P_0, P_1, S \prec P \prec Q'$}{6}{-1.5}

    \acreframe{0}{-8}

    \acleftmessage[->]{$(x, P')$}{0.25}{-3}{-2.25}
    \acprotocol{$\Pi^3_A$}{-2.25}{-0.75}{0.75}{2}
    \acmessage[->]{$(a_0, P_0)$}{0.75}{-1.5}{-0.5}
    \acmessage[->]{$(a_1, P_1)$}{0.25}{-1.5}{-0.5}
    \acresource{$OT_B$}{-0.5}{0}{1}{1}
    \acmessage[<-]{$(b, P_B)$}{0.75}{0.5}{1.5}
    \acmessage[->]{$(a_b, Q)$}{0.25}{0.5}{1.5}
    \acmessage[->]{$(b^*, S)$}{-0.5}{-1.5}{1.5}

    \indist{4.25}{0.25};
    \acpin{(c)}{4.25}{-1}

    \acleftmessage[->]{$(x, P')$}{0.25}{5}{5.75}
    \acresource{$OT^R_B$}{5.75}{-0.25}{1}{1}
    \acmessage[->]{$(x/\bot, Q'')$}{0.25}{6.75}{8.25}
    \acprotocol{$\sigma^3_B$}{8.25}{-0.75}{1}{1.75}
    \acrightmessage[<-]{$(b, P_B)$}{0.75}{9.25}{10}
    \acrightmessage[->]{$(a_b, Q)$}{0.25}{9.25}{10}
    \acrightmessage[->]{$(b^*, S)$}{-0.5}{9.25}{10}

    \acnote{$b^* \leftarrow Be(1/2)$}{-1.75}{-1}
    \acnote{$a_{b^*} \leftarrow x$}{-1.75}{-1.4}
    \acnote{$a_{1 - b^*} \leftarrow Be(1/2)$}{-1.75}{-1.75}
    
    \accausality{$Q'', P_B \prec Q \prec S$}{6}{-1.5}
\end{tikzpicture}
        \caption{Construction of a Rabin OT from a $\binom{2}{1}$-OT (Lemma~\ref{thm:ot-rabin-construction}). Honest Alice chooses at random in which interface to insert $x$, and honest Bob chooses at random which of the two bits to receive.}
        \label{fig:ot-rabin-construction}
    \end{figure}
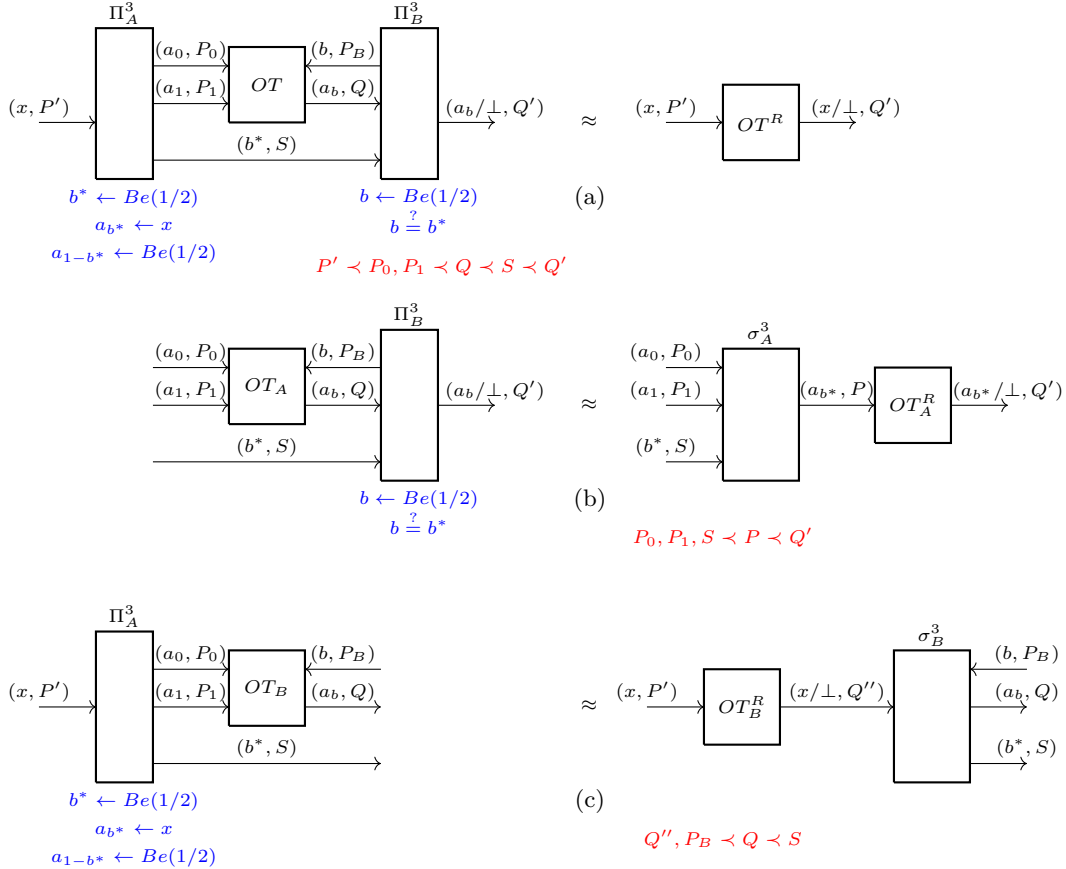
    \constructionproofnoskip{Honest protocol}{$\Pi^3_A OT \Pi^3_B \approx OT^R$}{\ref{fig:ot-rabin-construction}(a)} Note that $\bot$ will be returned if and only if $b^* \neq b$, and this happens with probability exactly $\frac{1}{2}$. Thus, Bob will output $x$ on its outer interface with this probability, exactly like the ideal Rabin OT.
    
    \constructionproof{Simulation against dishonest Alice}{$OT_A \Pi^3_B \approx \sigma^3_A OT^R_A$}{\ref{fig:ot-rabin-construction}(b)} The simulator
    $\sigma^3_A$ takes the bits $a_0, a_1, b^*$ as inputs on its left interface and, in order to achieve perfect construction, it can output $x = a_{b^*}$ to $OT^R_A$.
    
    \constructionproof{Simulation against dishonest Bob}{$\Pi^3_A OT_B \approx OT^R_B \sigma^3_B$}{\ref{fig:ot-rabin-construction}(c)} Bob's simulator
    $\sigma^3_B$ takes as input either $x$ or $\bot$ on the left interface: if $x$ is received, we output $b^* = b$ and $a_b = x$, otherwise, $\bot$ is received, and we set $b^* = 1 - b$, choosing $a_b$ uniformly at random.
\end{completeproof}

\begin{restatable}[Construction $\mathcal{OT}^R \rightarrow \mathcal{OT}$ (adapted from~\cite{crepeau87})]{lemma}{rabinotconstruction}
    \label{thm:rabin-ot-construction}
    The $\binom{2}{1}$ oblivious transfer $\mathcal{OT}$ can be $e^{-\Omega(k)}$-constructed from $3k$ instances of Rabin's oblivious transfer $\mathcal{OT}^R$. The constructing protocol $ \Pi^4$  (Definition~\ref{def:protocol-4}) and the simulators $\sigma^4_A, \sigma^4_B$ are classical, run in $\bigO(k)$ time,  use $\bigO(k)$ space and $\bigO(k)$ bits of classical communication.
\end{restatable}
\begin{completeproof}
    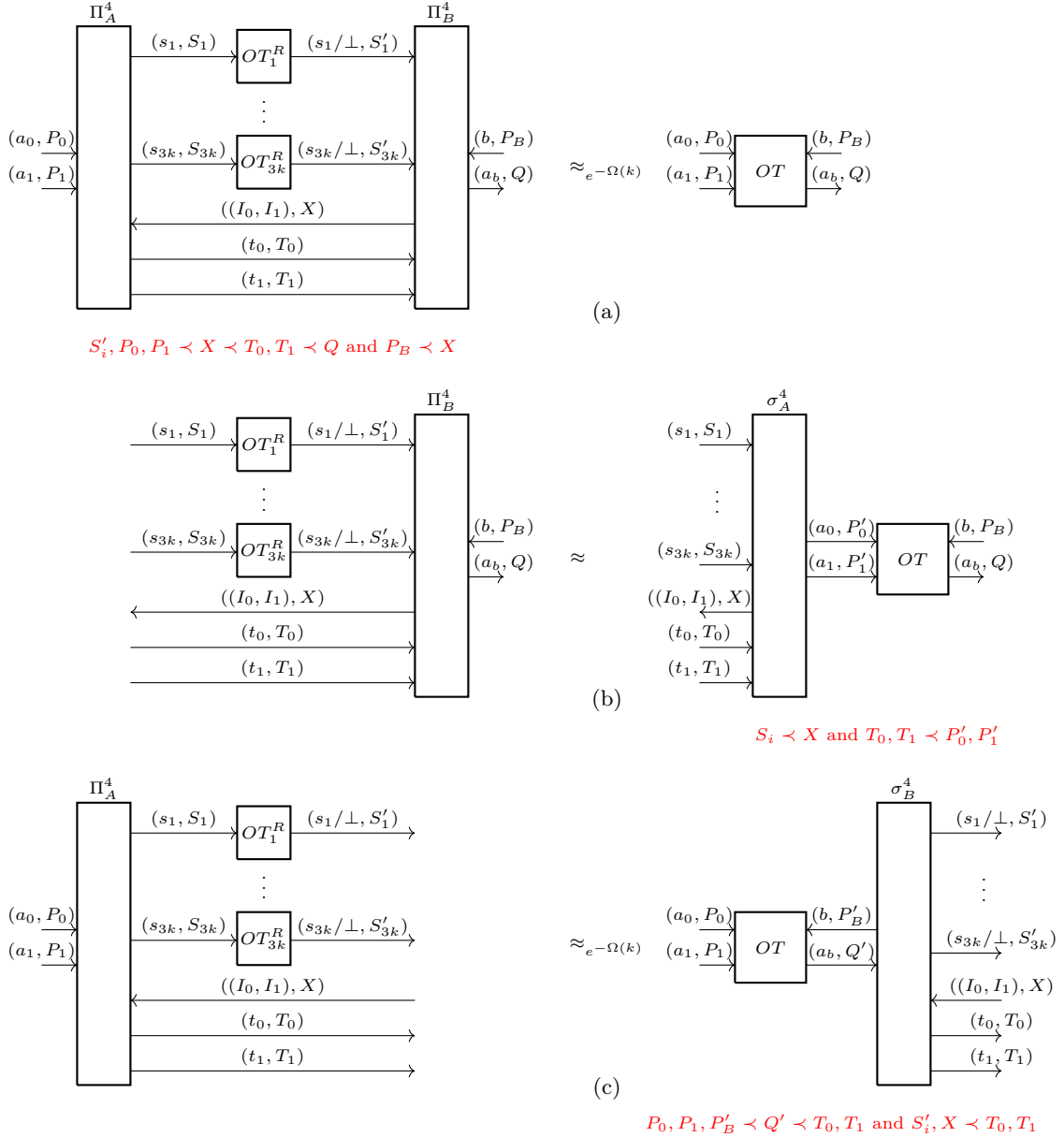
\begin{figure}[!ht]
        \centering
        \begin{tikzpicture}[scale=1]
    \scriptsize
    
    \acleftmessage[->]{$(a_0, P_0)$}{1.75}{-7.75}{-7.25}
    \acleftmessage[->]{$(a_1, P_1)$}{1.25}{-7.75}{-7.25}
    \acprotocol{$\Pi^4_A$}{-7.25}{-0.45}{0.75}{4}
    \acmessage[->]{$(s_1, S_1)$}{3.12}{-6.5}{-5}
    \acmessage[->]{$(s_{3k}, S_{3k})$}{1.6}{-6.5}{-5}
    
    \acresource{$OT^R_1$}{-5}{2.75}{0.75}{0.75}
    \acellipsis{-4.62}{2.45}
    \acresource{$OT^R_{3k}$}{-5}{1.25}{0.75}{0.75}
    
    \acmessage[->]{$(s_1/\bot, S'_1)$}{3.12}{-4.25}{-2.5}
    \acmessage[->]{$(s_{3k}/\bot, S'_{3k})$}{1.6}{-4.25}{-2.5}

    \acmessage[<-]{$((I_0, I_1), X)$}{0.75}{-6.5}{-2.5}
    \acmessage[->]{$(t_0, T_0)$}{0.25}{-6.5}{-2.5}
    \acmessage[->]{$(t_1, T_1)$}{-0.25}{-6.5}{-2.5}
    
    \acprotocol{$\Pi^4_B$}{-2.5}{-0.45}{0.75}{4}
    \acrightmessage[<-]{$(b, P_B)$}{1.75}{-1.75}{-1.25}
    \acrightmessage[->]{$(a_b, Q)$}{1.25}{-1.75}{-1.25}
    
    \accausality{$S'_i, P_0, P_1 \prec X \prec T_0, T_1 \prec Q$ and $P_B \prec X$}{-4.5}{-1}

    \indist[e^{-\Omega(k)}]{0.2}{1.5}
    \acpin{(a)}{0.2}{-0.5}

    \acleftmessage[->]{$(a_0, P_0)$}{1.75}{1.5}{2}
    \acleftmessage[->]{$(a_1, P_1)$}{1.25}{1.5}{2}
    \acresource{$OT$}{2}{1}{1}{1}
    \acrightmessage[<-]{$(b, P_B)$}{1.75}{3}{3.5}
    \acrightmessage[->]{$(a_b, Q)$}{1.25}{3}{3.5}

    \acreframe{0}{-5.5}
    
    \acmessage[->]{$(s_1, S_1)$}{3.12}{-6.5}{-5}
    \acmessage[->]{$(s_{3k}, S_{3k})$}{1.6}{-6.5}{-5}
    
    \acresource{$OT^R_1$}{-5}{2.75}{0.75}{0.75}
    \acellipsis{-4.62}{2.45}
    \acresource{$OT^R_{3k}$}{-5}{1.25}{0.75}{0.75}
    
    \acmessage[->]{$(s_1/\bot, S'_1)$}{3.12}{-4.25}{-2.5}
    \acmessage[->]{$(s_{3k}/\bot, S'_{3k})$}{1.6}{-4.25}{-2.5}

    \acmessage[<-]{$((I_0, I_1), X)$}{0.75}{-6.5}{-2.5}
    \acmessage[->]{$(t_0, T_0)$}{0.25}{-6.5}{-2.5}
    \acmessage[->]{$(t_1, T_1)$}{-0.25}{-6.5}{-2.5}
    
    \acprotocol{$\Pi^4_B$}{-2.5}{-0.45}{0.75}{4}
    \acrightmessage[<-]{$(b, P_B)$}{1.75}{-1.75}{-1.25}
    \acrightmessage[->]{$(a_b, Q)$}{1.25}{-1.75}{-1.25}
    
    \accausality{$S_i \prec X$ and $T_0, T_1 \prec P'_0, P'_1$}{4}{-1}

    \indist{-0.2}{1.5}
    \acpin{(b)}{0.2}{-0.5}

    \acleftmessage[->]{$(s_1, S_1)$}{3.12}{1.5}{2.25}
    \acellipsis{1.75}{2.42}
    \acleftmessage[->]{$(s_{3k}, S_{3k})$}{1.42}{1.5}{2.25}
    
    \acleftmessage[<-]{$((I_0, I_1), X)$}{0.75}{1.5}{2.25}
    \acleftmessage[->]{$(t_0, T_0)$}{0.25}{1.5}{2.25}
    \acleftmessage[->]{$(t_1, T_1)$}{-0.25}{1.5}{2.25}
    
    \acprotocol{$\sigma^4_A$}{2.25}{-0.45}{0.75}{4}
    \acmessage[->]{$(a_0, P'_0)$}{1.75}{3}{4}
    \acmessage[->]{$(a_1, P'_1)$}{1.25}{3}{4}
    \acresource{$OT$}{4}{1}{1}{1}
    \acrightmessage[<-]{$(b, P_B)$}{1.75}{5}{5.5}
    \acrightmessage[->]{$(a_b, Q)$}{1.25}{5}{5.5}

    \acreframe{0}{-11}
    
    \acleftmessage[->]{$(a_0, P_0)$}{1.75}{-7.75}{-7.25}
    \acleftmessage[->]{$(a_1, P_1)$}{1.25}{-7.75}{-7.25}
    \acprotocol{$\Pi^4_A$}{-7.25}{-0.45}{0.75}{4}
    \acmessage[->]{$(s_1, S_1)$}{3.12}{-6.5}{-5}
    \acmessage[->]{$(s_{3k}, S_{3k})$}{1.6}{-6.5}{-5}
    
    \acresource{$OT^R_1$}{-5}{2.75}{0.75}{0.75}
    \acellipsis{-4.62}{2.45}
    \acresource{$OT^R_{3k}$}{-5}{1.25}{0.75}{0.75}
    
    \acmessage[->]{$(s_1/\bot, S'_1)$}{3.12}{-4.25}{-2.5}
    \acmessage[->]{$(s_{3k}/\bot, S'_{3k})$}{1.6}{-4.25}{-2.5}

    \acmessage[<-]{$((I_0, I_1), X)$}{0.75}{-6.5}{-2.5}
    \acmessage[->]{$(t_0, T_0)$}{0.25}{-6.5}{-2.5}
    \acmessage[->]{$(t_1, T_1)$}{-0.25}{-6.5}{-2.5}

    \indist[e^{-\Omega(k)}]{0.2}{1.5}
    \acpin{(c)}{0.2}{-0.5}
    
    \acleftmessage[->]{$(a_0, P_0)$}{1.75}{1.5}{2}
    \acleftmessage[->]{$(a_1, P_1)$}{1.25}{1.5}{2}
    \acresource{$OT$}{2}{1}{1}{1}
    \acmessage[<-]{$(b, P'_B)$}{1.75}{3}{4}
    \acmessage[->]{$(a_b, Q')$}{1.25}{3}{4}
    
    \acprotocol{$\sigma^4_B$}{4}{-0.45}{0.75}{4}
    \acrightmessage[->]{$(s_1/\bot, S'_1)$}{3.12}{4.75}{5.75}
    \acellipsis{5.5}{2.42}
    \acrightmessage[->]{$(s_{3k}/\bot, S'_{3k})$}{1.42}{4.75}{5.75}
    
    \acrightmessage[<-]{$((I_0, I_1), X)$}{0.75}{4.75}{5.75}
    \acrightmessage[->]{$(t_0, T_0)$}{0.25}{4.75}{5.75}
    \acrightmessage[->]{$(t_1, T_1)$}{-0.25}{4.75}{5.75}
    
    \accausality{$P_0, P_1, P'_B \prec Q' \prec T_0, T_1$ and $S'_i, X \prec T_0, T_1$}{3.5}{-1}
\end{tikzpicture}
        \caption{Construction of a $\binom{2}{1}$-Oblivious Transfer from $3k$ instances of Rabin Oblivious Transfer (Lemma~\ref{thm:rabin-ot-construction}).}
        \label{fig:rabin-ot-construction}
    \end{figure}
    For the rest of the proof, let $OT^R_* = OT^R_1 || \cdots || OT^R_{3k}$ be the parallel composition of the $3k$ instances of the Rabin OT (notice that it is also the same under dishonest Alice and Bob).
    
    \constructionproof{Honest protocol}{$\Pi^4_A OT^R_* \Pi^4_B \approx_{e^{-\Omega(k)}} OT$}{\ref{fig:rabin-ot-construction}(a)}
    The only problem with the honest protocol arises when Bob receives too few bits from the Rabin OTs: if the number $X$ of received bits is less than $k$, Bob will abort\footnote{It is important that this check is done \textit{before} Alice sends $(t_0, t_1)$. Otherwise, in a context where the protocol is repeated upon abortion, Bob can abort after learning $a_0$, and then ask for $b = 1$ in the next iteration.}, as it will not be possible for him to construct a completely known subset. If $X$ exceeds $2k$, then two disjoint subsets can be constructed, and Bob would be able to know both $a_0, a_1$ (this is not a problem for the honest protocol). However, $X \sim Binom(3k, \frac{1}{2})$, and by a multiplicative Chernoff bound (Theorem~\ref{thm:chernoff-bound}) we obtain:
    \begin{align*}
        \Prob{X < k} = \Prob{X < \frac{3k}{2} \left( 1 - \frac{1}{3} \right)} \le e^{-\Omega(k)}
    \end{align*}
    Since the protocol and the ideal OT are indistinguishable unless $X < k$, the difference lemma (Lemma~\ref{thm:diff-lemma-1}) gives us that the above probability is an upper bound for the distinguishing advantage of any distinguisher.
    
    \constructionproof{Simulation against dishonest Alice}{$OT^R_* \Pi^4_B \approx \sigma^4_A OT_A$}{\ref{fig:rabin-ot-construction}(b)}
    $\sigma^4_A$ takes the bits $s_1, \ldots, s_{3k}$, and it should output $(I_0, I_1)$. In order to do this:
    \begin{enumerate}[label=(\arabic*)]
        \item It activates each incoming $s_i$ with probability $1/2$ (simulating the failure/success of the Rabin OT), then it selects $I_0, I_1$ accordingly (unless $X < k$, in which case $\sigma^4_A$ aborts exactly like $\Pi^4_B$). This is to emulate the choice of the subsets made by $\Pi^4_B$.
        
        \item It computes $a_i = t_i \oplus (\bigoplus_{j \in I_i} s_j)$ (remember that all $s_i$ are known to $\sigma^4_A$) and sends them to the $\binom{2}{1}$-OT on its right interface.
    \end{enumerate}
    This simulation is perfectly indistinguishable from $OT^R_* \Pi_B$.

    \constructionproof{Simulation against dishonest Bob}{$\Pi^4_A OT^R_* \approx_{e^{-\Omega(k)}} OT_B \sigma^4_B$}{\ref{fig:rabin-ot-construction}(c)} $\sigma^4_B$ gets to choose $b$ from the ideal $\binom{2}{1}$-OT on its left interface.
    \begin{enumerate}[label=(\arabic*)]
        \item The simulator activates each $s_i$ with probability $1/2$ (again, in order to simulate the failure/success of the Rabin OT), outputting either $\bot$ or a uniformly random bit.
        
        \item When $(I_0, I_1)$ arrives, the simulator checks for disjointness and then checks which one of them is completely known (the simulator knows it, as it chose which bits failed), and asks for the according bit to the OT on the left.
        
        \item At this point, $t_b$ can be computed using the bit $a_b$ returned by the OT and the bits of the interval $I_b$, while $t_{1-b}$ is chosen uniformly at random (since $I_{1-b}$ is not completely known, no information about $a_{1-b}$ could be extracted anyway).
    \end{enumerate}
    The construction is indistinguishable unless $(I_0, I_1)$ are both completely known (in this case $\sigma^4_B$ lacks information to construct $t_{1-b}$ in step 3), but this can happen only if $X \ge 2k$. By a Chernoff bound:
    \begin{align*}
        \Prob{X \ge 2k} = \Prob{X \ge \frac{3k}{2} \left( 1 + \frac{1}{3} \right)} \le e^{-\Omega(k)}
    \end{align*}
    and using the difference lemma we bound the distinguishing advantage of any distinguisher with this probability.
\end{completeproof}

\end{protibox}

The explicit protocols for the constructions are all very simple, except for the last one, which requires linear resources.

\begin{definition}[Protocol $\Pi^1$ for $\mathcal{ROT} \rightarrow \mathcal{OT}$]
    \label{def:protocol-1}
    The following  protocol $\Pi^1 = (\Pi^1_A, \Pi^1_B)$ uses one instance of randomized oblivious transfer to construct a  $\binom{2}{1}$-OT:
    \begin{enumerate}[label=(\arabic*)]
        \item Alice is given $a_0, a_1$, and Bob is given $b$.
        \item Bob inputs $b$ to the randomized OT, obtaining $s_b$.
        \item Alice encrypts $a_i$ using a one-time pad with key $s_i$, and sends both encrypted bits $c_0, c_1$ to Bob.
        \item Bob will be able to decrypt only $c_b$ as he received only $s_b$, but not $s_{1-b}$. Thus, he can decrypt and output $a_b$.
    \end{enumerate}
\end{definition}

\begin{definition}[Protocol  $\Pi^2$ for $\mathcal{OT} \rightarrow \mathcal{ROT}$]
    \label{def:protocol-2}
    The protocol $\Pi^2 = (\Pi^2_A, \Pi^2_B)$ uses an instance of $\binom{2}{1}$-oblivious transfer to construct a randomized oblivious transfer. In the protocol, honest Alice simply inputs two bits chosen uniformly at random to the OT as $a_0, a_1$, and outputs them also to her outer interface.
\end{definition}

\begin{definition}[Protocol $\Pi^3$ for $\mathcal{OT} \rightarrow \mathcal{OT}^R$]
    \label{def:protocol-3}
    The protocol $\Pi^3 = (\Pi^3_A, \Pi^3_B)$ constructs a Rabin oblivious transfer using an instance of $\binom{2}{1}$-oblivious transfer and works as follows:
    \begin{enumerate}[label=(\arabic*)]
        \item Both Alice and Bob choose a uniformly random bit ($b^*$ and $b$, respectively);
        \item An instance of the $\binom{2}{1}$-OT is executed, where Alice sets $a_{b^*} = x$ ($x$ is the input of the Rabin OT), and chooses $a_{1 - b^*}$ at random. Bob picks $a_b$.
        \item Alice reveals $b^*$ to Bob. 
        \item Bob checks whether $b = b^*$, and determines  whether he received the bit $x$ or a uniformly random bit. In the latter case, $\bot$ will be returned to the outer interface.
    \end{enumerate}
\end{definition}

\begin{definition}[Protocol $\Pi^4$ for $\mathcal{OT}^R \rightarrow \mathcal{OT}$ ~\cite{crepeau87}]
    \label{def:protocol-4}
    Fixed a security parameter $k$, the protocol $\Pi^4 = (\Pi^4_A, \Pi^4_B)$ constructs a $\binom{2}{1}$-oblivious transfer using $3k$ instances of Rabin oblivious transfer and works as follows:
    \begin{enumerate}[label=(\arabic*)]
        \item Alice chooses $3k$ bits $s_1, \ldots, s_{3k}$ uniformly and independently at random, and sends them to Bob using the $3k$ instances of Rabin OT;
        
        \item We say that a subset $I \subseteq [3k]$ with $|I| = k$ is \emph{completely known} if Bob knows $s_i$ for every $i \in I$. Bob chooses a completely known subset as $I_b$, and chooses another subset of $k$ bits as $I_{1-b}$, disjoint from $I_b$ at random. At this point, $(I_0, I_1)$ is sent to Alice.
        
        \item Alice checks that $I_0 \cap I_1 = \emptyset$, then she sends $(t_0, t_1)$ to Bob such that:
        \begin{align*}
            t_i = \left( \bigoplus_{j \in I_i} s_j \right) \oplus a_i
        \end{align*}
        Note that Bob can compute $\bigoplus_{j \in I} s_j$ if and only if $I$ is completely known (otherwise, he learns no information).
    \end{enumerate}
\end{definition}

\noindent This last protocol is adapted from Cr\'epeau's   construction from Rabin OT to $\binom{2}{1}$-OT~\cite{crepeau87}. The main idea there is that, with very high probability, Bob will have enough bits to create a completely known subset, but not enough to create two of them. Therefore, he will be able to retrieve $a_b$ but not $a_{1-b}$.
Note that the construction of Lemma~\ref{thm:rabin-ot-construction} is different from the other two constructions we presented in this section: first of all, the honest protocol may fail completely, in the sense that there is a (small) probability that Bob cannot retrieve either of $a_0, a_1$, and this is captured in the proof by the non-zero distinguishing advantage of the honest construction. Secondly, there is a small chance that Bob can cheat, namely when the number of bits received from the Rabin OTs exceeds $2k$ (in this case he can retrieve both $a_0, a_1$ by constructing two disjoint completely known subsets), and this is a consequence of the imperfection of the simulation of $\sigma^4_B$ against dishonest Bob. However, in the security proof we bound both these `imperfections' with a term $e^{-{\Omega(k)}}$, and one can obtain an exponentially  small  cheating/failure probability by linearly increasing the security parameter $k$.



\subsection{Impossibility of composable Oblivious Transfer}
\label{sec:impossibility_OC}

In this section we present our proof of impossibility for oblivious transfer. The idea of the proof here is similar to the one given by Vilasini~et~al.~\cite{Vilasini19} for the impossibility of coin flip: we use this to prove that constructing $\mathcal{ROT}$ is impossible, then Lemma~\ref{thm:ot-rot-construction} will extend the result to $\mathcal{OT}$.
Then, we will turn our attention to Rabin's oblivious transfer: in principle, one can use Lemma~\ref{thm:rabin-ot-construction} to entail its impossibility from the impossibility of the other two versions. However, the distinguishing advantages would be poor, and decaying with respect to the security parameter $k$ of the construction. Instead, we will first present a generalization of the Rabin's oblivious transfer, and then prove its impossibility directly.
\ifexportproofs
    The complete proofs can be found in Appendix~\ref{apx:impossibility-proofs}.
\fi

\begin{protibox}[label={box:OT_impossibility}]{Impossibility of randomized and 1-out-of-2 OT}

\begin{restatable}[Impossibility of $\mathcal{ROT}$]{theorem}{rotimpossibility}
    \label{thm:rot-impossibility}
    For any $\varepsilon < \frac{1}{12}$, it is impossible to $\varepsilon$-construct $\mathcal{ROT}$ between two mutually distrusting parties with a mere exchange of messages, be it classical, quantum, non-signalling or relativistic. A distinguisher achieving this advantage has the same computational requirements as the protocol or the simulators.
\end{restatable}
\begin{shortproof}
    Suppose we have a two-party protocol $\Pi = (\Pi_A, \Pi_B)$ ran by Alice and Bob such that, for some $\varepsilon < \frac{1}{12}$
    \begin{align}
        \Pi_A \Pi_B & \approx_\varepsilon ROT \label{eq:rot-imp-sketch-1} \\
        \Pi_B & \approx_\varepsilon \sigma_A ROT_A \label{eq:rot-imp-sketch-2} \\
        \Pi_A & \approx_\varepsilon ROT_B \sigma_B \label{eq:rot-imp-sketch-3}
    \end{align}
    By applying the composability properties of the resources guaranteed by the Abstract Cryptography framework, we obtain
    \begin{align*}
        ROT_B \sigma_B \sigma_A ROT_A \approx_{3\varepsilon} ROT
    \end{align*}
    We reach a contradiction by exhibiting a simple distinguisher $D$ attacking this last construction with advantage at least $\frac{1}{4}$, implying $3\varepsilon \ge \frac{1}{4}$.
\end{shortproof}

\begin{completeproof}
    \begin{figure}[ht]
        \centering
        \begin{tikzpicture}[scale=1]
    \scriptsize
    
    \acleftmessage[<-]{$(s_0, P_0)$}{0.25}{-1.5}{-1}
    \acleftmessage[<-]{$(s_1, P_1)$}{-0.25}{-1.5}{-1}
    \acprotocol{$\Pi_A$}{-1}{-1}{0.75}{2}
    \acmessage[->]{}{0.75}{-0.25}{1}
    \acmessage[<-]{}{0.5}{-0.25}{1}
    \acmessage[->]{}{0.25}{-0.25}{1}
    \acellipsis{0.35}{-0.2}
    \acmessage[->]{}{-0.75}{-0.25}{1}
    \acprotocol{$\Pi_B$}{1}{-1}{0.75}{2}
    \acrightmessage[<-]{$(b, P_B)$}{0.25}{1.75}{2.25}
    \acrightmessage[->]{$(s_b, Q)$}{-0.25}{1.75}{2.25}

    \indist[\varepsilon]{3.5}{0}
    \acpin{(a)}{3.5}{-1}

    \acleftmessage[<-]{$(s_0, P_0)$}{0.25}{4.5}{5}
    \acleftmessage[<-]{$(s_1, P_1)$}{-0.25}{4.5}{5}
    \acresource{$ROT$}{5}{-0.5}{1}{1}
    \acrightmessage[<-]{$(b, P_B)$}{0.25}{6}{6.5}
    \acrightmessage[->]{$(s_b, Q)$}{-0.25}{6}{6.5}

    \acreframe{0}{-3}
    
    \acmessage[->]{}{0.75}{-0.25}{1}
    \acmessage[<-]{}{0.5}{-0.25}{1}
    \acmessage[->]{}{0.25}{-0.25}{1}
    \acellipsis{0.35}{-0.2}
    \acmessage[->]{}{-0.75}{-0.25}{1}
    \acprotocol{$\Pi_B$}{1}{-1}{0.75}{2}
    \acrightmessage[<-]{$(b, P_B)$}{0.25}{1.75}{2.25}
    \acrightmessage[->]{$(s_b, Q)$}{-0.25}{1.75}{2.25}

    \indist[\varepsilon]{3.5}{0}
    \acpin{(b)}{3.5}{-1}
    
    \acmessage[->]{}{0.75}{4.75}{5.25}
    \acmessage[<-]{}{0.5}{4.75}{5.25}
    \acmessage[->]{}{0.25}{4.75}{5.25}
    \acellipsis{5}{-0.25}
    \acmessage[->]{}{-0.75}{4.75}{5.25}
    \acprotocol{$\sigma_A$}{5.25}{-1}{0.75}{2}
    \acmessage[->]{$(s_0, P_0)$}{0.25}{6}{7}
    \acmessage[->]{$(s_1, P_1)$}{-0.25}{6}{7}
    \acresource{$ROT_A$}{7}{-0.5}{1}{1}
    \acrightmessage[<-]{$(b, P_B)$}{0.25}{8}{8.5}
    \acrightmessage[->]{$(s_b, Q)$}{-0.25}{8}{8.5}

    \acreframe{0}{-6}
    
    \acleftmessage[<-]{$(s_0, P_0)$}{0.25}{-1.5}{-1}
    \acleftmessage[<-]{$(s_1, P_1)$}{-0.25}{-1.5}{-1}
    \acprotocol{$\Pi_A$}{-1}{-1}{0.75}{2}
    \acmessage[->]{}{0.75}{-0.25}{1}
    \acmessage[<-]{}{0.5}{-0.25}{1}
    \acmessage[->]{}{0.25}{-0.25}{1}
    \acellipsis{0.35}{-0.2}
    \acmessage[->]{}{-0.75}{-0.25}{1}

    \indist[\varepsilon]{3.5}{0}
    \acpin{(c)}{3.5}{-1}
    
    \acleftmessage[<-]{$(s_0, P_0)$}{0.25}{4.75}{5.25}
    \acleftmessage[<-]{$(s_1, P_1)$}{-0.25}{4.75}{5.25}
    \acresource{$ROT_B$}{5.25}{-0.5}{1}{1}
    \acmessage[<-]{$(b, P_B)$}{0.25}{6.25}{7.25}
    \acmessage[->]{$(s_b, Q)$}{-0.25}{6.25}{7.25}
    \acprotocol{$\sigma_B$}{7.25}{-1}{0.75}{2}
    \acmessage[->]{}{0.75}{8}{8.5}
    \acmessage[<-]{}{0.5}{8}{8.5}
    \acmessage[->]{}{0.25}{8}{8.5}
    \acellipsis{8.3}{-0.2}
    \acmessage[->]{}{-0.75}{8}{8.5}

    \acreframe{0}{-9}
    \acleftmessage[<-]{$(s_0, P_0)$}{0.25}{-3.5}{-3}
    \acleftmessage[<-]{$(s_1, P_1)$}{-0.25}{-3.5}{-3}
    \acresource{$ROT_B$}{-3}{-0.5}{1}{1}
    
    \acmessage[<-]{$(b, P_B)$}{0.25}{-2}{-1}
    \acmessage[->]{$(s_b, Q)$}{-0.25}{-2}{-1}
    \acprotocol{$\sigma_{BA}$}{-1}{-0.5}{1}{1}
    \acmessage[->]{$(s'_0, P'_0)$}{0.25}{0}{1}
    \acmessage[->]{$(s'_1, P'_1)$}{-0.25}{0}{1}
    
    \acresource{$ROT_A$}{1}{-0.5}{1}{1}
    \acrightmessage[<-]{$(b', P'_B)$}{0.25}{2}{2.5}
    \acrightmessage[->]{$(s'_{b'}, Q')$}{-0.25}{2}{2.5}
    
    \indist[3\varepsilon]{3.5}{0}
    \acpin{(d)}{3.5}{-1}
    
    \acresource{$ROT$}{5}{-0.5}{1}{1}
    
    \acleftmessage[<-]{$(s_0, P_0)$}{0.25}{4.5}{5}
    \acleftmessage[<-]{$(s_1, P_1)$}{-0.25}{4.5}{5}
    
    \acrightmessage[<-]{$(b, P_B)$}{0.25}{6}{6.5}
    \acrightmessage[->]{$(s_b, Q)$}{-0.25}{6}{6.5}
    
    \acresetframe
\end{tikzpicture}
        \caption{Graphical representation of the proof of Theorem~\ref{thm:rot-impossibility}. The last construction is obtained by plugging (b) and (c), and then applying (a) using the triangle inequality. The two simulators $\sigma_B \sigma_A$ are merged into a single simulator $\sigma_{BA}$.}
        \label{fig:rot-no-go-construction}
    \end{figure}
    Suppose for a contradiction there is a two-party protocol $\Pi = (\Pi_A, \Pi_B)$, ran by Alice and Bob respectively, such that
    \begin{align}
        \Pi_A \Pi_B & \approx_\varepsilon ROT \label{eq:rot-imp-1} \\
        \Pi_B & \approx_\varepsilon \sigma_A ROT_A \label{eq:rot-imp-2} \\
        \Pi_A & \approx_\varepsilon ROT_B \sigma_B \label{eq:rot-imp-3}
    \end{align}
    for $\varepsilon < \frac{1}{12}$ and some simulators $\sigma_A, \sigma_B$. By triangle inequality we can infer that:
    \begin{align}
        \Pi_A \Pi_B & \approx_{2\varepsilon} ROT_B \sigma_B \sigma_A ROT_A & \text{by \eqref{eq:rot-imp-2} + \eqref{eq:rot-imp-3}} \label{eq:rot-imp-4} \\
        ROT & \approx_{3\varepsilon} ROT_B \sigma_B \sigma_A ROT_A & \text{by \eqref{eq:rot-imp-1} + \eqref{eq:rot-imp-4}} \nonumber
    \end{align}
    For simplicity, we now consider $\sigma_B \sigma_A$ as a single simulator $\sigma_{BA}$. This does not hinder the correctness of the proof as we are simply quantifying over a broader set of simulators ($\sigma_{BA}$ internally simulates the exchange of messages between $\sigma_B, \sigma_A$). Hence, we found the following inequality:
    \begin{align*}
        3\varepsilon \ge d^{\mathbb{D}}(ROT, ROT_B \sigma_{BA} ROT_A)
    \end{align*}
    The construction, along with the names of the variables we are going to use for the rest of this proof is given in Figure~\ref{fig:rot-no-go-construction}.
    Now consider a distinguisher $D$ which inputs a uniformly random bit $b'$ on the right interface of $ROT_B \sigma_{BA} ROT_A$ and guesses the non-ideal resource if and only if it observes $s_{b'} \neq s'_{b'}$. On the other side, $ROT_B$ will output uniformly random $s_0, s_1$. $D$ will certainly be able to distinguish the two systems if $s_{b'} \neq s'_{b'}$, which tells us, by the statistical separation lemma (Lemma~\ref{thm:diff-lemma-2}):
    \begin{align*}
        3\varepsilon & \ge d^D(ROT, ROT_B \sigma_{BA} ROT_A) \ge \Prob{s_{b'} \neq s'_{b'}}
    \end{align*}
    We conclude the argument by finding a lower bound for this probability: assume without loss of generality that $Q \prec P'_0, P'_1$, i.e.\ $s_b$ is used by $\sigma_{BA}$ for the choice of $s'_0, s'_1$ (otherwise, $s_{b'}$ and $s'_{b'}$ would be unconditionally independent and the claim would follow).
    \begin{align*}
        \Prob{s_{b'} \neq s'_{b'}} & = \Prob{s_{b'} \neq s'_{b'} \given b = b'} \Prob{b = b'} + \Prob{s_{b'} \neq s'_{b'} \given b \neq b'} \Prob{b \neq b'} \\
        & = \frac{1}{2} \Prob{s_{b'} \neq s'_{b'} \given b = b'} +  \frac{1}{2} \Prob{s_{b'} \neq s'_{b'} \given b \neq b'} \\
        & \ge \frac{1}{2} \Prob{s_{b'} \neq s'_{b'} \given b \neq b'}
    \end{align*}
    When $b \neq b'$, $ROT_B$ will deliver $s_b$ to $\sigma_{BA}$, which is independent from $s_{1-b} = s_{b'}$. Therefore, $\sigma_{BA}$ has to return to $ROT_A$ a bit that needs to match $s_{b'}$, of which it has no information and is uniformly random. Therefore,
    \begin{align*}
        \Prob{s_{b'} \neq s'_{b'} \given b \neq b'} = \frac{1}{2}
    \end{align*}
    concluding that $\varepsilon \ge \frac{1}{12}$ for any possible causal order chosen by $\sigma_{BA}$ and this leads to a contradiction.
\end{completeproof}

\begin{corollary}[Impossibility of $\mathcal{OT}$]
    \label{thm:ot-impossibility}
    For any $\varepsilon < \frac{1}{12}$, it is impossible to $\varepsilon$-construct $\mathcal{OT}$ between two mutually distrusting parties with a mere exchange of messages, be it classical, quantum, non-signalling or relativistic. A distinguisher achieving this advantage has the same computational requirements as the protocol or the simulators.
\end{corollary}
\begin{proof}
    Follows directly from the impossibility of $\mathcal{ROT}$ (Theorem~\ref{thm:rot-impossibility}) along with the perfect construction $\mathcal{OT} \stackrel{\Pi^2}{\longrightarrow} \mathcal{ROT}$ (Lemma~\ref{thm:ot-rot-construction}).
\end{proof}

\end{protibox}

\noindent We would like to point out that these results (and the subsequent ones) do not only include information-theoretic secure constructions of OT, but also computationally secure ones: as mentioned in Section~\ref{sec:ac-framework}, it all depends on the class of distinguishers $\mathbb{D}$ we consider when we state Equations~\eqref{eq:rot-imp-sketch-1}--\eqref{eq:rot-imp-sketch-3}. Since we then exhibit a distinguisher $D$ that is classical and efficient (it only needs three bits of memory and one comparison), the impossibility holds as long as we consider a class $\mathbb{D}$ containing $D$. Vilasini~et~al.~\cite{Vilasini19} show how to derive three explicit distinguishers from $D$, each attacking one of the three constructions required by Definition~\ref{def:crypto-security}, with the same complexity requirements as the protocol or the simulators. We omit this detail here for conciseness. Now, as explained above, we need to provide a direct result for the impossibility of Rabin OT and, in the meanwhile, we take the chance to make a more general statement. First we derive a generalization of Rabin's OT (which reduces to it for $p=\frac12$, $\mathcal{OT}^{1/2} \equiv \mathcal{OT}^R$).
\begin{definition}[Probabilistic transfer]
    A $p$-Rabin Oblivious Transfer (or probabilistic transfer) is a primitive $\mathcal{OT}^p = (OT^p, OT^p_A, OT^p_B)$. Alice sends a bit $x$, and Bob receives $x$ with probability $p$ (and $\bot$ otherwise).
\end{definition}

\begin{protibox}[label={box:pOT_impossibility}]{Impossibility of probabilistic OT}
\begin{restatable}[Impossibility of $\mathcal{OT}^p$]{theorem}{rabinimpossibility}
    \label{thm:rabin-impossibility}
    For any $\varepsilon < \frac{1}{6} p (1 - p)$, it is impossible to $\varepsilon$-construct $\mathcal{OT}^p$ between two mutually distrusting parties with a mere exchange of messages, be it classical, quantum, non-signalling or relativistic. A distinguisher achieving this advantage has the same computational requirements as the protocol or the simulators.
\end{restatable}
\begin{completeproof}
    \begin{figure}[!ht]
        \centering
        \begin{tikzpicture}[scale=1]
    \scriptsize
    
    \acleftmessage[->]{$(x, P)$}{0}{-1.5}{-1}
    \acprotocol{$\Pi_A$}{-1}{-1}{0.75}{2}
    \acmessage[->]{}{0.75}{-0.25}{1}
    \acmessage[<-]{}{0.5}{-0.25}{1}
    \acmessage[->]{}{0.25}{-0.25}{1}
    \acellipsis{0.35}{-0.2}
    \acmessage[->]{}{-0.75}{-0.25}{1}
    \acprotocol{$\Pi_B$}{1}{-1}{0.75}{2}
    \acrightmessage[->]{$(x/\bot, Q)$}{0}{1.75}{2.5}

    \indist[\varepsilon]{4}{0}
    \acpin{(a)}{4}{-1}
    
    \acleftmessage[->]{$(x, P)$}{0}{4.75}{5.25}
    \acresource{$OT^p$}{5.25}{-0.5}{1}{1}
    \acrightmessage[->]{$(x/\bot, Q)$}{0}{6.25}{7}

    \acreframe{0}{-3}
    
    \acmessage[->]{}{0.75}{-0.25}{1}
    \acmessage[<-]{}{0.5}{-0.25}{1}
    \acmessage[->]{}{0.25}{-0.25}{1}
    \acellipsis{0.35}{-0.2}
    \acmessage[->]{}{-0.75}{-0.25}{1}
    \acprotocol{$\Pi_B$}{1}{-1}{0.75}{2}
    \acrightmessage[->]{$(x/\bot, Q)$}{0}{1.75}{2.5}

    \indist[\varepsilon]{4}{0}
    \acpin{(b)}{4}{-1}
    
    \acmessage[->]{}{0.75}{4.75}{5.25}
    \acmessage[<-]{}{0.5}{4.75}{5.25}
    \acmessage[->]{}{0.25}{4.75}{5.25}
    \acellipsis{5}{-0.25}
    \acmessage[->]{}{-0.75}{4.75}{5.25}
    \acprotocol{$\sigma_A$}{5.25}{-1}{0.75}{2}
    \acmessage[->]{$(x, P)$}{0}{6}{7}
    \acresource{$OT^p_A$}{7}{-0.5}{1}{1}
    \acrightmessage[->]{$(x / \bot, Q)$}{0}{8}{8.75}

    \acreframe{0}{-6}
    
    \acleftmessage[->]{$(x, P)$}{0}{-1.5}{-1}
    \acprotocol{$\Pi_A$}{-1}{-1}{0.75}{2}
    \acmessage[->]{}{0.75}{-0.25}{1}
    \acmessage[<-]{}{0.5}{-0.25}{1}
    \acmessage[->]{}{0.25}{-0.25}{1}
    \acellipsis{0.35}{-0.2}
    \acmessage[->]{}{-0.75}{-0.25}{1}

    \indist[\varepsilon]{4}{0}
    \acpin{(c)}{4}{-1}
    
    \acleftmessage[->]{$(x, P)$}{0}{4.75}{5.25}
    \acresource{$OT^p_B$}{5.25}{-0.5}{1}{1}
    \acmessage[->]{$(x/\bot, Q)$}{0}{6.25}{7.5}
    \acprotocol{$\sigma_B$}{7.5}{-1}{0.75}{2}
    \acmessage[->]{}{0.75}{8.25}{8.75}
    \acmessage[<-]{}{0.5}{8.25}{8.75}
    \acmessage[->]{}{0.25}{8.25}{8.75}
    \acellipsis{8.55}{-0.2}
    \acmessage[->]{}{-0.75}{8.25}{8.75}

    \acreframe{0}{-8.5}
    
    \acleftmessage[->]{$(x_A, P)$}{0}{-3.5}{-3}
    \acresource{$OT^p_B$}{-3}{-0.5}{1}{1}
    
    \acmessage[->]{$(x_A / \bot, Q)$}{0}{-2}{-0.6}
    \acprotocol{$\sigma_{BA}$}{-0.6}{-0.5}{0.5}{1}
    \acmessage[->]{$(x_B, P')$}{0}{-0.1}{1}
    
    \acresource{$OT^p_A$}{1}{-0.5}{1}{1}
    \acrightmessage[->]{$(x_B/\bot, Q')$}{0}{2}{2.75}
    
    \indist[3\varepsilon]{4}{0}
    \acpin{(d)}{4}{-1}
    
    \acleftmessage[->]{$(x, P)$}{0}{4.75}{5.25}
    \acresource{$OT^p_B$}{5.25}{-0.5}{1}{1}
    \acrightmessage[->]{$(x/\bot, Q)$}{0}{6.25}{7}
    
    \acresetframe
\end{tikzpicture}
        \caption{Graphical representation of the proof of Theorem~\ref{thm:rabin-impossibility}. The last construction is obtained by plugging (b) to (c), and then applying (a) using the triangle inequality. The two simulators $\sigma_B \sigma_A$ are merged into a single simulator $\sigma_{BA}$.}
        \label{fig:rabin-no-go-construction}
    \end{figure}
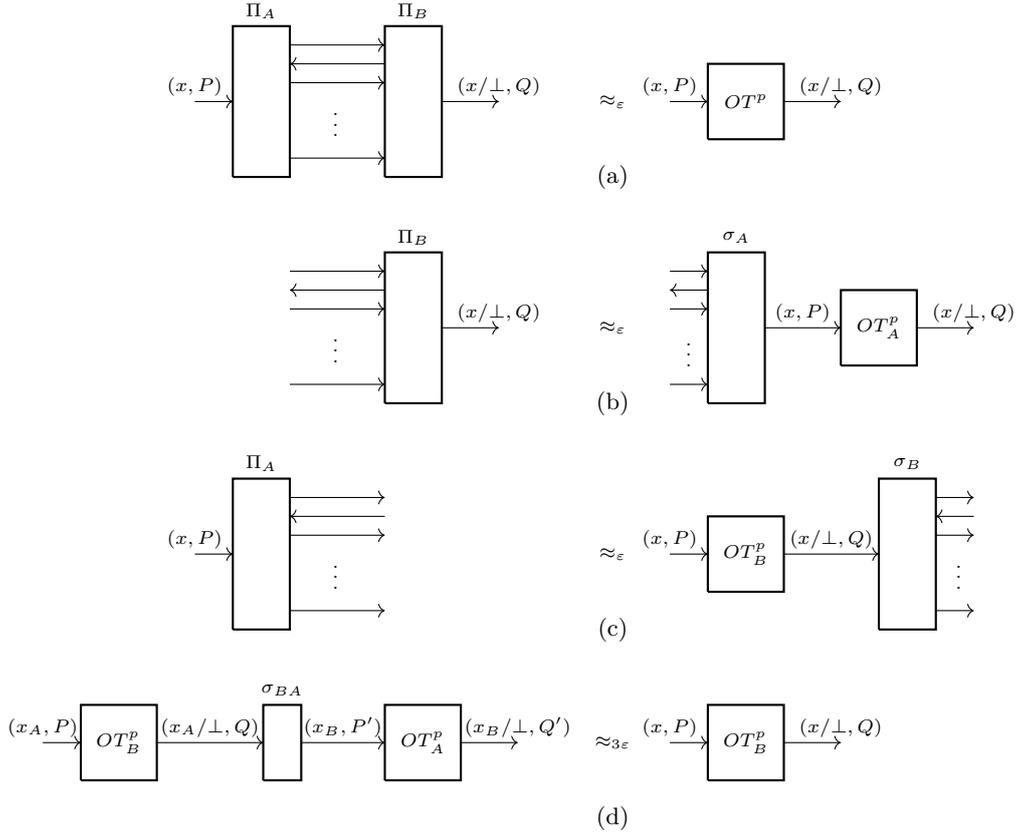
    Suppose for a contradiction there is a two-party protocol $\Pi = (\Pi_A, \Pi_B)$, ran by Alice and Bob respectively, such that
    \begin{align*}
        \Pi_A \Pi_B & \approx_\varepsilon OT^p \\
        \Pi_B & \approx_\varepsilon \sigma_A OT^p_A \\
        \Pi_A & \approx_\varepsilon OT^p_B \sigma_B
    \end{align*}
    for $\varepsilon < \frac{1}{6} p (1 - p)$ and some simulators $\sigma_A, \sigma_B$. Following the exact same reasoning as in Theorem~\ref{thm:rot-impossibility}, we obtain:
    \begin{align*}
        OT^p & \approx_{3\varepsilon} OT^p_B \sigma_{BA} OT^p_A
    \end{align*}
    Now, we take a distinguisher $D$ which inputs $x_A$ chosen uniformly at random on the left interface and compares it with the output $x_B$ received on the right interface, guessing the constructed resource if and only if $x_A \neq x_B$. By the statistical separation lemma we have:
    \begin{align*}
        3\varepsilon & \ge d^D(OT^p, OT^p_B \sigma_{BA} OT^p_A) \ge \Prob{x_A \neq x_B}
    \end{align*}
    We conclude the argument by finding a lower bound for this probability. Let $E$ be the event where $OT^p_B$ fails (i.e.\ returns $\bot$) and $OT^p_A$ does not (i.e.\ copies the input to its output). Notice that, whenever $E$ occurs, $x_B$ will be equal to what $\sigma_{BA}$ gave on its right interface. However, since $OT^p_B$ failed, $\sigma_{BA}$ received no information about $x_A$, therefore:
    \begin{align*}
        \Prob{x_A \neq x_B} & = \Prob{x_A \neq x_B \given E} \Prob{E} + \Prob{x_A \neq x_B \given \neg E} \Prob{\neg E} \\
        & \ge \Prob{x_A \neq x_B \given E} \Prob{E} \\
        & = p(1 - p) \cdot \Prob{x_A \neq x_B \given E}  = \frac{1}{2} p (1 - p)
    \end{align*}
    which means that $\varepsilon \ge \frac{1}{6} p (1 - p)$. Contradiction.
\end{completeproof}

\end{protibox}

\noindent We would like to highlight an intuition formalized by this result: when $p = 0$ or $p = 1$, the theorem becomes meaningless, as $\mathcal{OT}^1$ is the identity resource, and $\mathcal{OT}^0$ is a resource blocking any incoming message, which can be both constructed. As the information about the delivery of the bit becomes more and more hidden from Alice (i.e.\ $p \rightarrow \frac{1}{2}$), the impossibility becomes more and more evident.

The reader may also wonder why we did not simply use Theorem~\ref{thm:rabin-impossibility} along with the perfect constructions of Lemmas~\ref{thm:ot-rabin-construction} and \ref{thm:rot-ot-construction} in order to prove the impossibility of $\mathcal{OT}$ and $\mathcal{ROT}$. The reason is that we would obtain a slightly weaker result: Theorem~\ref{thm:rot-impossibility} proves impossibility up to $\varepsilon = \frac{1}{12}$, while here we could only show the one for $\mathcal{OT}^R$ up to $\varepsilon = \frac{1}{24}$.



\subsection{Impossibility of Oblivious String Transfer and Multi-Party Computation}
\label{sec:impossibility_string_MPC}

In the literature, oblivious transfer primitives are also intended for multiple bit strings~\cite{Kilian88}: let $\mathcal{OT}^s, \mathcal{ROT}^s, \mathcal{OT}^{p, s}$ be an extension of $\mathcal{OT}, \mathcal{ROT}, \mathcal{OT}^p$ where, instead of bits, strings of $s$ bits are transferred\footnote{Formally, Alice sends elements of a set $\mathcal{X}$, with $|\mathcal{X}| = 2^{s}$: for this purpose, $s$ does not need to be an integer, so $|\mathcal{X}|$ does not need to be a power of two.}. For these primitives, stronger impossibility results can be proven. The equivalence given by the constructions of Lemmas~\ref{thm:rot-ot-construction}--\ref{thm:rabin-ot-construction} are naturally extended to the $s$-bit case.
\ifexportproofs
    The proofs of this section can be found in Appendix~\ref{apx:impossibility-string-proofs}.
\fi
\begin{protibox}[label={box:string_impossibility}]{Impossibility of oblivious string transfer}
\begin{restatable}[Impossibility of $\mathcal{ROT}^s$]{theorem}{rotstringimpossibility}
    \label{thm:mult-rot-impossibility}
    For any $\varepsilon < \frac{1}{6} \left( 1 - \frac{1}{2^s} \right)$, it is impossible to $\varepsilon$-construct $\mathcal{ROT}^s$ between two mutually distrusting parties with a mere exchange of messages, be it classical, quantum, non-signalling or relativistic. A distinguisher achieving this advantage has the same computational requirements as the protocol or the simulators.
\end{restatable}
\begin{completeproof}
    The proof is identical to the one given for Theorem~\ref{thm:rot-impossibility}, with the only difference that $s_{b'}, s'_{b'}$ are strings of $s$ bits. Therefore, if $\sigma_{BA}$ does not receive information about $s_{b'}$, the probability that the output on the right is different is:
    \begin{align*}
        \Prob{s_{b'} \neq s'_{b'} \given b \neq b'} = 1 - \frac{1}{2^s}
    \end{align*}
\end{completeproof}

\noindent The perfect construction of Lemma~\ref{thm:ot-rot-construction} extends the bound also to $\mathcal{OT}^s$. Also the impossibility for Rabin OT is extended:
\begin{restatable}[Impossibility of $\mathcal{OT}^{p, s}$]{theorem}{rabinstringimpossibility}
    \label{thm:mult-rabin-impossibility}
    For any $\varepsilon < \frac{1}{3} \left( 1 - \frac{1}{2^s} \right) p (1 - p)$, it is impossible to $\varepsilon$-construct $\mathcal{OT}^{p, s}$ between two mutually distrusting parties with a mere exchange of messages, be it classical, quantum, non-signalling or relativistic. A distinguisher achieving this advantage has the same computational requirements as the protocol or the simulators.
\end{restatable}
\begin{completeproof}
    We follow the same reasoning as for Theorem~\ref{thm:rabin-impossibility}, except for the fact that, when the event $E$ occurs, $\sigma_{BA}$ needs to match a string of $s$ bits instead of a single bit, which translates to:
    \begin{align*}
        \Prob{x_A \neq x_B \given E} = 1 - \frac{1}{2^s}
    \end{align*}
\end{completeproof}
\end{protibox}

\noindent These two results tell us that the impossibility bound grows exponentially fast for an oblivious transfer of elements carrying $s$ bits of information. In particular, if we let $s \rightarrow \infty$, allowing arbitrarily large strings, the bound becomes as high as $\frac{1}{6}$ (for $\mathcal{ROT}^s$ and $\mathcal{OT}^s$) and $\frac{1}{3} p (1 - p)$ (for $\mathcal{OT}^{p,s}$).



Since we proved that oblivious transfer is impossible, the only remaining question to address is whether arbitrary multi-party computation is also impossible, or there may be another complete primitive that is constructible in the relativistic quantum setting. In this section we show, with the same technique as above, that even composably secure two-party computation of a simple boolean function such as the \emph{and} of two bits is impossible to construct.
\ifexportproofs
    Appendix~\ref{apx:impossibility-mpc-proofs} contains the proof for the claim below.
\fi

\begin{definition}
    Let $f : \{ 0, 1 \}^{m + n} \rightarrow \{ 0, 1 \}^r$ be a boolean function. A two-party computation of $f$ is a primitive $\mathcal{C}^f = (C^f, C^f_A, C^f_B)$. Alice inputs $x \in \{ 0, 1 \}^m$ and Bob inputs $y \in \{ 0, 1 \}^n$. Both receive the value $f(x, y)$, but none of them can retrieve information about the input given by the other party (aside from what can be inferred by the final value).
    
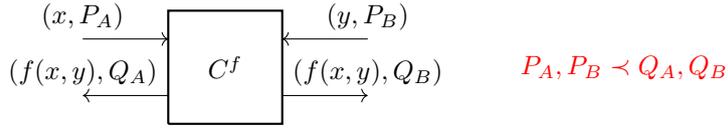
\begin{figure}[H]
        \centering
        \begin{tikzpicture}[scale=1.5]
    \acresource{$C^f$}{0}{0}{1}{1}
    \acleftmessage[->]{$(x, P_A)$}{0.75}{-0.75}{0}
    \acleftmessage[<-]{$(f(x, y), Q_A)$}{0.25}{-0.75}{0}
    \acrightmessage[<-]{$(y, P_B)$}{0.75}{1}{1.75}
    \acrightmessage[->]{$(f(x, y), Q_B)$}{0.25}{1}{1.75}
    
    \accausality{$P_A, P_B \prec Q_A, Q_B$}{4}{0.5}
\end{tikzpicture}
        \label{fig:multi-party-comp-def}
    \end{figure}
\end{definition}
\noindent We define the function $and: \{ 0, 1 \}^2 \rightarrow \{ 0, 1 \}$ where $and(x, y) = x \cdot y$.

\begin{protibox}[label={box:MPC_impossibility}]{Impossibility of multi-party computation}

\begin{restatable}[Impossibility of multi-party \emph{and} computation]{theorem}{mpcimpossibility}
    \label{thm:mpc-impossibility}
    For any $\varepsilon < \frac{1}{12}$, it is impossible to $\varepsilon$-construct $\mathcal{C}^{and}$ between two mutually distrusting parties with a mere exchange of messages, be it classical, quantum, non-signalling or relativistic. A distinguisher achieving this advantage has the same computational requirements as the protocol or the simulators.
\end{restatable}
\noindent With a similar argument, one can prove that also a multi-party computation of the \emph{or} function is impossible.

\end{protibox}



\subsection{Mutual constructions between Oblivious Transfer and Bit Commitment}
\label{sec:BC}

We conclude this work by reviewing the formal definition of bit commitment, and proving that composably secure commitment and oblivious transfer primitives can be constructed from each other in the relativistic quantum setting. The error probability decays exponentially on the resources used. This sort of equivalence allows us to extend results on minimal assumptions for composable bit commitment (such as the one by Prokop~\cite{prokop20}) also to oblivious transfer. Moreover, the equivalence lemmas proved in Section~\ref{sec:ot-equivalence} extend the argument to every version of the oblivious transfer primitive. \ifexportproofs

\fi


\begin{definition}
    A bit commitment is a primitive $\mathcal{BC} = (BC, BC_A, BC_B)$. Alice commits to a bit $x$ at position $P$, Bob is notified about the commitment but does not receive any information about $x$ until Alice decides to open the commit at some position $Q \succ P$. 
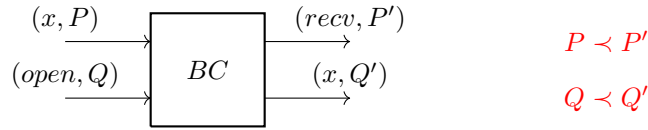
\begin{figure}[H]
        \centering
        \begin{tikzpicture}[scale=1.5]
    \acresource{$BC$}{0}{0}{1}{1}
    \acleftmessage[->]{$(x, P)$}{0.75}{-0.75}{0}
    \acleftmessage[->]{$(open, Q)$}{0.25}{-0.75}{0}
    \acrightmessage[->]{$(recv, P')$}{0.75}{1}{1.75}
    \acrightmessage[->]{$(x, Q')$}{0.25}{1}{1.75}
    
    \accausality{$P \prec P'$}{4}{0.75}
    \accausality{$Q \prec Q'$}{4}{0.25}
\end{tikzpicture}
        \label{fig:bit-commitment-def}
    \end{figure}
\end{definition}
\noindent We include in this definition the possibility for Alice to never open the commitment (or, equivalently, to \textit{abort} it).
Here we present the Unruh's construction $\mathcal{BC} \to \mathcal{OT}$~\cite{Unruh10}, translated to the AC framework. The security proof gives a probabilistic analysis of  cheating behaviour (Appendix~\ref{apx:impossibility-string-proofs}).

\begin{protibox}[label={box:BC_constructions}]{Constructions between bit commitment and oblivious transfer}

\begin{restatable}[Construction $\mathcal{BC} \rightarrow \mathcal{OT}$]{lemma}{unruhconstruction}
    \label{thm:bc-ot-construction}
    $\mathcal{OT}$ can be $e^{-\Omega(n)}$-constructed from $2n$ instances of $\mathcal{BC}$. The constructing protocol $\Pi^5$  (Definition~\ref{def:protocol-5}) and  the simulators are quantum, run in time $\bigO(n)$ and use $\bigO(n)$ space.
\end{restatable}
\begin{completeproof}
    In the following analysis, let $BC_* = BC^1 || \cdots || BC^{2n}$.
    
    \constructionproof{Honest protocol}{$\Pi^5_A BC_* \Pi^5_B \approx_{e^{-\Omega(n)}} OT$}{\ref{fig:bc-ot-construction}(a)}
    The test carried out by Alice to check that Bob measured the states gives us no problems, since it will always succeed under assumption of honest Bob. The only case where the honest construction diverges from the ideal OT is when $X < \frac{k}{3}$, whose probability is:
    \begin{align*}
        \Prob{X < \frac{k}{3}} = \Prob{X < \frac{k}{2} \left( 1 - \frac{1}{3} \right)} \le e^{-\Omega(k)} = e^{-\Omega(n)}
    \end{align*}
    using a Chernoff bound. This, along with an application of the difference lemma, completes the honest construction.
    
    \constructionproof{Simulation against dishonest Alice}{$BC_* \Pi^5_B \approx \sigma^5_A OT_A$}{\ref{fig:bc-ot-construction}(b)}
    Notice that $\sigma^5_A$ internally fakes the commitments, and thus will have complete control over the transmissions of $\bar{x}_i$ and $\bar\theta_i$:
    \begin{enumerate}[label=(\arabic*)]
        \item $\sigma^5_A$ can avoid measuring $\{ \ket{\psi_i} \}_i$ until it receives $T$. At this point only those in $T$ are measured with randomly chosen bases and their fake commitments are opened.
        
        \item When $\{ \theta_i \}_{i \in R}$ are received, they are used to correctly measure all the bits, allowing $\sigma^5_A$ to choose $I_0, I_1$ that are both completely known.
        
        \item $\sigma^5_A$ received both $a_0, a_1$, and it can give them as input to the ideal OT.
    \end{enumerate}
    This gives a perfect construction.
    
    \constructionproof{Simulation against dishonest Bob}{$\Pi^5_A BC_* \approx_{e^{-\Omega(n)}} OT_B \sigma^5_B$}{\ref{fig:bc-ot-construction}(c)}
    Also $\sigma^5_B$ fakes the commitments. Hence, it receives $\{ \bar{x}_i, \bar\theta_i \}_i$ immediately in the (fake) commit phase.
    \begin{enumerate}[label=(\arabic*)]
        \item The test is carried out normally, ignoring the additional information (if it fails, $\sigma^5_B$ aborts exactly like $\Pi^5_A$).
        
        \item When $I_0, I_1$ arrive, $\sigma^5_B$ can deduce which of the two is completely known: if $I_b$ is the completely known interval, $\sigma^5_B$ can ask for $a_b$ to the ideal OT, compute $t_b \leftarrow a_b \oplus \left( \bigoplus_{i \in I_b} x_i \right)$, and choose $t_{1-b}$ uniformly at random.
    \end{enumerate}
    The construction is indistinguishable unless $I_0, I_1$ are both completely known (in which case we let $\sigma^5_B$ abort). Suppose dishonest Bob avoids measuring $x$ of the $n$ states. Denote with $X'$ the number of correctly measured bits among the $k' \in [k - x, k]$ ones that were measured honestly (excluding the ones used for the test). Thus, we have $X \le X' + x$, with $X' \sim Binom(k', \frac{1}{2})$ (the upper bound is because also some of the $x$ states might be used for the test). We now split into two cases:
    \begin{itemize}
        \item if $x \le \frac{k}{12}$, then we certainly need $X' \ge \frac{2k}{3} - \frac{k}{12}$ in order to have $X \ge \frac{2k}{3}$. By a Chernoff bound, the probability that this happens is:
        \begin{align*}
            \Prob{X' \ge \frac{2k}{3} - \frac{k}{12}} & = \Prob{X' \ge \frac{k}{2} \left(1 + \frac{1}{6} \right)} \\
            & \le \Prob{X' \ge \frac{k'}{2} \left(1 + \frac{1}{6} \right)} \\
            & \le e^{-\Omega(k')} = e^{-\Omega(k)} = e^{-\Omega(n)} & \text{since $k' \ge \frac{11}{12} k$}
        \end{align*}
            
        \item if $x > \frac{k}{12}$, we want to argue that the event $A$ in which Bob passes the test has negligible probability. We analyze the number $Z$ of bits among the $x$ that were not measured which are chosen to be in $T$ by Alice. Then, each of these bits has $\frac{1}{4}$ chance of being detected by Alice (correct basis but wrong bit in the commitment). Therefore, the probability of $A$ conditioned on $Z$ is
        \begin{align*}
            \Prob{A \given Z = z} = \left(\frac{3}{4}\right)^z
        \end{align*}
        We conclude this analysis by bounding the probability that $Z$ is low: notice that $Z$ is an hypergeometric random variable, taking $h$ out of $n$ elements without replacement, and $x$ of them are marked. By using Hoeffding's inequality (Theorem~\ref{thm:hoeffding-hyperg}), we obtain:
        \begin{align*}
            \Prob{Z \le \frac{xh}{2n}} = \Prob{Z \le \frac{xh}{n} - \frac{xh}{2n}} \le e^{-2\left(\frac{x}{2n}\right)^2 h} = e^{-\Omega(n)}
        \end{align*}
        Using the law of total probability we combine our results:
        \begin{align*}
            \Prob{A} & = \Prob{A \given Z \le \frac{xh}{2n}} \Prob{Z \le \frac{xh}{2n}} + \Prob{A \given Z > \frac{xh}{2n}} \Prob{Z > \frac{xh}{2n}} \\
                & \le \Prob{Z \le \frac{xh}{2n}} + \Prob{A \given Z > \frac{xh}{2n}} \\
                & \le e^{-\Omega(n)} + \left(\frac{3}{4}\right)^{\frac{xh}{2n}} = e^{-\Omega(n)} + e^{-\Omega(n)} = e^{-\Omega(n)}
        \end{align*}
    \end{itemize}
    The above argument tells us that dishonest Bob manages to cheat with probability at most $e^{-\Omega(n)}$ regardless of the number $x$ of states he avoids measuring (even if $x$ is randomized, we can use the law of total probability conditioning on its value). Applying the difference lemma we conclude the third construction.
    \begin{figure}
        \centering
        \begin{tikzpicture}[scale=1]
    \scriptsize
    
    \acleftmessage[->]{$(a_0, P_0)$}{0.75}{-7.75}{-7.25}
    \acleftmessage[->]{$(a_1, P_1)$}{0.25}{-7.75}{-7.25}
    \acprotocol{$\Pi^5_A$}{-7.25}{-2.45}{0.75}{6}
    \acmessage[->]{$(\{ \ket{\psi_i} \}_i, A)$}{3.25}{-6.5}{-2.5}
    \acmessage[<-]{$(recv, B'_1)$}{2.75}{-6.5}{-5.1}
    \acmessage[<-]{$(\bar{x}_1, D'_1)$}{2.25}{-6.5}{-5.1}
    \acresource{$BC^1$}{-5.1}{2}{1}{1}
    \acmessage[<-]{$(\bar{x}_1, B_1)$}{2.75}{-4.1}{-2.5}
    \acmessage[<-]{$(open, D_1)$}{2.25}{-4.1}{-2.5}
    
    \acmessage[<-]{$(recv, B'_2)$}{1.5}{-6.5}{-5.1}
    \acmessage[<-]{$(\bar\theta_1, D'_2)$}{1}{-6.5}{-5.1}
    \acresource{$BC^2$}{-5.1}{0.75}{1}{1}
    \acmessage[<-]{$(\bar\theta_1, B_2)$}{1.5}{-4.1}{-2.5}
    \acmessage[<-]{$(open, D_2)$}{1}{-4.1}{-2.5}
    
    \acellipsis{-4.6}{0.5}
    
    \acmessage[->]{$(T, C)$}{-0.25}{-6.5}{-2.5}
    \acmessage[->]{$(\{ \theta_i \}_i, E)$}{-0.75}{-6.5}{-2.5}
    \acmessage[<-]{$((I_0, I_1), X)$}{-1.25}{-6.5}{-2.5}
    \acmessage[->]{$(t_0, T_0)$}{-1.75}{-6.5}{-2.5}
    \acmessage[->]{$(t_1, T_1)$}{-2.25}{-6.5}{-2.5}
    \acprotocol{$\Pi^5_B$}{-2.5}{-2.45}{0.75}{6}
    \acrightmessage[<-]{$(b, P_B)$}{0.75}{-1.75}{-1.25}
    \acrightmessage[->]{$(a_b, Q)$}{0.25}{-1.75}{-1.25}
    
    \accausality{$A \prec B_i \prec B'_i \prec C \prec D_i \prec D'_i \prec E \prec X \prec T_0, T_1 \prec Q$ and $P_B \prec X$}{-3.5}{-3}

    \indist[e^{-\Omega(k)}]{0.2}{0.5}
    \acpin{(a)}{0.2}{-1.5}
    
    \acleftmessage[->]{$(a_0, P_0)$}{0.75}{1.5}{2}
    \acleftmessage[->]{$(a_1, P_1)$}{0.25}{1.5}{2}
    \acresource{$OT$}{2}{0}{1}{1}
    \acrightmessage[<-]{$(b, P_B)$}{0.75}{3}{3.5}
    \acrightmessage[->]{$(a_b, Q)$}{0.25}{3}{3.5}

    \acreframe{0}{-7.5}

    \acmessage[->]{$(\{ \ket{\psi_i} \}_i, A)$}{3.25}{-6.5}{-2.5}
    \acmessage[<-]{$(recv, B'_1)$}{2.75}{-6.5}{-5.1}
    \acmessage[<-]{$(\bar{x}_1, D'_1)$}{2.25}{-6.5}{-5.1}
    \acresource{$BC^1_A$}{-5.1}{2}{1}{1}
    \acmessage[<-]{$(\bar{x}_1, B_1)$}{2.75}{-4.1}{-2.5}
    \acmessage[<-]{$(open, D_1)$}{2.25}{-4.1}{-2.5}
    
    \acmessage[<-]{$(recv, B'_2)$}{1.5}{-6.5}{-5.1}
    \acmessage[<-]{$(\bar\theta_1, D'_2)$}{1}{-6.5}{-5.1}
    \acresource{$BC^2_A$}{-5.1}{0.75}{1}{1}
    \acmessage[<-]{$(\bar\theta_1, B_2)$}{1.5}{-4.1}{-2.5}
    \acmessage[<-]{$(open, D_2)$}{1}{-4.1}{-2.5}
    
    \acellipsis{-4.6}{0.5}
    
    \acmessage[->]{$(T, C)$}{-0.25}{-6.5}{-2.5}
    \acmessage[->]{$(\{ \theta_i \}_i, E)$}{-0.75}{-6.5}{-2.5}
    \acmessage[<-]{$((I_0, I_1), X)$}{-1.25}{-6.5}{-2.5}
    \acmessage[->]{$(t_0, T_0)$}{-1.75}{-6.5}{-2.5}
    \acmessage[->]{$(t_1, T_1)$}{-2.25}{-6.5}{-2.5}
    \acprotocol{$\Pi^5_B$}{-2.5}{-2.45}{0.75}{6}
    \acrightmessage[<-]{$(b, P_B)$}{0.75}{-1.75}{-1.25}
    \acrightmessage[->]{$(a_b, Q)$}{0.25}{-1.75}{-1.25}

    \indist{0.2}{0.5}
    \acpin{(b)}{0.2}{-1.5}

    \acleftmessage[->]{$(\{ \ket{\psi_i} \}_i, A)$}{3.25}{2}{3}
    \acleftmessage[<-]{$(recv, B'_1)$}{2.75}{2}{3}
    \acleftmessage[<-]{$(\bar{x}_1, D'_1)$}{2.25}{2}{3}
    
    \acleftmessage[<-]{$(recv, B'_2)$}{1.5}{2}{3}
    \acleftmessage[<-]{$(\bar\theta_1, D'_2)$}{1}{2}{3}
    
    \acellipsis{2.5}{0.5}
    
    \acleftmessage[->]{$(T, C)$}{-0.25}{2}{3}
    \acleftmessage[->]{$(\{ \theta_i \}_i, E)$}{-0.75}{2}{3}
    \acleftmessage[<-]{$((I_0, I_1), X)$}{-1.25}{2}{3}
    \acleftmessage[->]{$(t_0, T_0)$}{-1.75}{2}{3}
    \acleftmessage[->]{$(t_1, T_1)$}{-2.25}{2}{3}
    
    \acprotocol{$\sigma^5_A$}{3}{-2.45}{0.75}{6}
    \acmessage[->]{$(a_0, P'_0)$}{0.75}{3.75}{4.75}
    \acmessage[->]{$(a_1, P'_1)$}{0.25}{3.75}{4.75}
    \acresource{$OT_A$}{4.75}{0}{1}{1}
    \acrightmessage[<-]{$(b, P_B)$}{0.75}{5.75}{6.25}
    \acrightmessage[->]{$(a_b, Q)$}{0.25}{5.75}{6.25}
    
    \accausality{$T_i \prec P'_i \prec Q$ and $P_B \prec Q$}{3.5}{-3}

    \acreframe{0}{-15}

    \acleftmessage[->]{$(a_0, P_0)$}{0.75}{-7.75}{-7.25}
    \acleftmessage[->]{$(a_1, P_1)$}{0.25}{-7.75}{-7.25}
    \acprotocol{$\Pi^5_A$}{-7.25}{-2.45}{0.75}{6}
    \acmessage[->]{$(\{ \ket{\psi_i} \}_i, A)$}{3.25}{-6.5}{-2.5}
    \acmessage[<-]{$(recv, B'_1)$}{2.75}{-6.5}{-5.1}
    \acmessage[<-]{$(\bar{x}_1, D'_1)$}{2.25}{-6.5}{-5.1}
    \acresource{$BC^1_B$}{-5.1}{2}{1}{1}
    \acmessage[<-]{$(\bar{x}_1, B_1)$}{2.75}{-4.1}{-2.5}
    \acmessage[<-]{$(open, D_1)$}{2.25}{-4.1}{-2.5}
    
    \acmessage[<-]{$(recv, B'_2)$}{1.5}{-6.5}{-5.1}
    \acmessage[<-]{$(\bar\theta_1, D'_2)$}{1}{-6.5}{-5.1}
    \acresource{$BC^2_B$}{-5.1}{0.75}{1}{1}
    \acmessage[<-]{$(\bar\theta_1, B_2)$}{1.5}{-4.1}{-2.5}
    \acmessage[<-]{$(open, D_2)$}{1}{-4.1}{-2.5}
    
    \acellipsis{-4.6}{0.5}
    
    \acmessage[->]{$(T, C)$}{-0.25}{-6.5}{-2.5}
    \acmessage[->]{$(\{ \theta_i \}_i, E)$}{-0.75}{-6.5}{-2.5}
    \acmessage[<-]{$((I_0, I_1), X)$}{-1.25}{-6.5}{-2.5}
    \acmessage[->]{$(t_0, T_0)$}{-1.75}{-6.5}{-2.5}
    \acmessage[->]{$(t_1, T_1)$}{-2.25}{-6.5}{-2.5}

    \indist[\delta]{0.2}{0.5}
    \acpin{(c)}{0.2}{-1.5}

    \acleftmessage[->]{$(a_0, P_0)$}{0.75}{1.5}{2}
    \acleftmessage[->]{$(a_1, P_1)$}{0.25}{1.5}{2}
    \acresource{$OT_B$}{2}{0}{1}{1}
    \acmessage[<-]{$(b, P'_B)$}{0.75}{3}{4}
    \acmessage[->]{$(a_b, Q')$}{0.25}{3}{4}
    
    \acprotocol{$\sigma^5_B$}{4}{-2.45}{0.75}{6}
    \acrightmessage[->]{$(\{ \ket{\psi_i} \}_i, A)$}{3.25}{4.75}{5.75}
    \acrightmessage[<-]{$(\bar{x}_1, B_1)$}{2.75}{4.75}{5.75}
    \acrightmessage[<-]{$(open, D_1)$}{2.25}{4.75}{5.75}
    
    \acrightmessage[<-]{$(\bar\theta_1, B_2)$}{1.5}{4.75}{5.75}
    \acrightmessage[<-]{$(open, D_2)$}{1}{4.75}{5.75}
    
    \acellipsis{5.25}{0.6}
    
    \acrightmessage[->]{$(T, C)$}{-0.25}{4.75}{5.75}
    \acrightmessage[->]{$(\{ \theta_i \}_i, E)$}{-0.75}{4.75}{5.75}
    \acrightmessage[<-]{$((I_0, I_1), X)$}{-1.25}{4.75}{5.75}
    \acrightmessage[->]{$(t_0, T_0)$}{-1.75}{4.75}{5.75}
    \acrightmessage[->]{$(t_1, T_1)$}{-2.25}{4.75}{5.75}
    
    \accausality{$P_0, P_1 \prec Q' \prec T_0, T_1$ and $X \prec P'_B \prec Q'$}{3.5}{-3}

    \acresetframe
\end{tikzpicture}
        \caption{Construction of a $\binom{2}{1}$-OT from $2n$ instances of bit commitment (Theorem~\ref{thm:bc-ot-construction}).}
        \label{fig:bc-ot-construction}
    \end{figure}
\end{completeproof}

\begin{restatable}[Construction $\mathcal{OT} \rightarrow \mathcal{BC}$]{lemma}{otbcconstruction}
    \label{thm:ot-bc-construction}
    $\mathcal{BC}$ can be $2^{-k}$-constructed from $k$ instances of $\mathcal{OT}$. The constructing protocol $\Pi^6$ (Definition~\ref{def:protocol-6}) and the simulators are classical, run in time $\bigO(k)$ and use $\bigO(k)$ space. 
\end{restatable}
\begin{completeproof}
    \begin{figure}[!ht]
        \centering
        \begin{tikzpicture}[scale=1]
    \scriptsize
    
    \acleftmessage[->]{$(x, P)$}{1.75}{-8}{-7.25}
    \acleftmessage[->]{$(open, Q)$}{1.25}{-8}{-7.25}
    \acprotocol{$\Pi^6_A$}{-7.25}{-0.45}{0.75}{4}
    \acmessage[->]{$(s^1_0, S^1_0)$}{3.25}{-6.5}{-5.1}
    \acmessage[->]{$(s^1_1, S^1_1)$}{2.75}{-6.5}{-5.1}
    \acresource{$OT^1$}{-5.1}{2.5}{1}{1}
    \acmessage[<-]{$(b^1, S^1_B)$}{3.25}{-4.1}{-2.5}
    \acmessage[->]{$(s^1_{b^1}, T^1)$}{2.75}{-4.1}{-2.5}
    
    \acmessage[->]{$(s^k_0, S^k_0)$}{1.25}{-6.5}{-5.1}
    \acmessage[->]{$(s^k_1, S^k_1)$}{0.75}{-6.5}{-5.1}
    \acresource{$OT^k$}{-5.1}{0.5}{1}{1}
    \acmessage[<-]{$(b^k, S^k_B)$}{1.25}{-4.1}{-2.5}
    \acmessage[->]{$(s^k_{b^k}, T^k)$}{0.75}{-4.1}{-2.5}

    \acellipsis{-4.6}{2.1}
    
    \acmessage[->]{$(\{ \bar{s}^i_0, \bar{s}^i_1 \}_i = \{ s^i_0, s^i_1 \}_i, X)$}{-0.25}{-6.5}{-2.5}
    \acprotocol{$\Pi^6_B$}{-2.5}{-0.45}{0.75}{4}
    \acrightmessage[->]{$(recv, P')$}{1.75}{-1.75}{-1}
    \acrightmessage[->]{$(x, Q')$}{1.25}{-1.75}{-1}
    
    \accausality{$P \prec S^i_j \prec T^i \prec P'$ and $Q \prec X \prec Q'$}{-4}{-1}

    \indist{0.2}{1.5}
    \acpin{(a)}{0.2}{-0.75}

    \acleftmessage[->]{$(x, P)$}{1.75}{1.25}{2}
    \acleftmessage[->]{$(open, Q)$}{1.25}{1.25}{2}
    \acresource{$BC$}{2}{1}{1}{1}
    \acrightmessage[->]{$(recv, P')$}{1.75}{3}{3.75}
    \acrightmessage[->]{$(x, Q')$}{1.25}{3}{3.75}

    \acreframe{0}{-5.5}

    \acmessage[->]{$(s^1_0, S^1_0)$}{3.25}{-6.5}{-5.1}
    \acmessage[->]{$(s^1_1, S^1_1)$}{2.75}{-6.5}{-5.1}
    \acresource{$OT^1_A$}{-5.1}{2.5}{1}{1}
    \acmessage[<-]{$(b^1, S^1_B)$}{3.25}{-4.1}{-2.5}
    \acmessage[->]{$(s^1_{b^1}, T^1)$}{2.75}{-4.1}{-2.5}
    
    \acmessage[->]{$(s^k_0, S^k_0)$}{1.25}{-6.5}{-5.1}
    \acmessage[->]{$(s^k_1, S^k_1)$}{0.75}{-6.5}{-5.1}
    \acresource{$OT^k_A$}{-5.1}{0.5}{1}{1}
    \acmessage[<-]{$(b^k, S^k_B)$}{1.25}{-4.1}{-2.5}
    \acmessage[->]{$(s^k_{b^k}, T^k)$}{0.75}{-4.1}{-2.5}

    \acellipsis{-4.6}{2.1}
    
    \acmessage[->]{$(\{ \bar{s}^i_0, \bar{s}^i_1 \}_i, X)$}{-0.25}{-6.5}{-2.5}
    \acprotocol{$\Pi^6_B$}{-2.5}{-0.45}{0.75}{4}
    \acrightmessage[->]{$(recv, P')$}{1.75}{-1.75}{-1}
    \acrightmessage[->]{$(x, Q')$}{1.25}{-1.75}{-1}
    
    \accausality{$S^i_j \prec P'' \prec P'$ and $X \prec Q'' \prec Q$}{4}{-1}

    \indist[2^{-k}]{0.2}{1.5}
    \acpin{(b)}{0.2}{-0.75}

    \acleftmessage[->]{$(s^1_0, S^1_0)$}{3.25}{1.5}{2}
    \acleftmessage[->]{$(s^1_1, S^1_1)$}{2.75}{1.5}{2}
    
    \acleftmessage[->]{$(s^k_0, S^k_0)$}{1.25}{1.5}{2}
    \acleftmessage[->]{$(s^k_1, S^k_1)$}{0.75}{1.5}{2}
    
    \acellipsis{1.5}{2.1}
    
    \acleftmessage[->]{$(\{ \bar{s}^i_0, \bar{s}^i_1 \}_i, X)$}{-0.25}{1}{2}
    
    \acprotocol{$\sigma^6_A$}{2}{-0.45}{0.75}{4}
    \acmessage[->]{$(x, P'')$}{1.75}{2.75}{4}
    \acmessage[->]{$(open, Q'')$}{1.25}{2.75}{4}
    \acresource{$BC_A$}{4}{1}{1}{1}
    \acrightmessage[->]{$(recv, P')$}{1.75}{5}{5.75}
    \acrightmessage[->]{$(x, Q')$}{1.25}{5}{5.75}

    \acreframe{0}{-11}

    \acleftmessage[->]{$(x, P)$}{1.75}{-8}{-7.25}
    \acleftmessage[->]{$(open, Q)$}{1.25}{-8}{-7.25}
    \acprotocol{$\Pi^6_A$}{-7.25}{-0.45}{0.75}{4}
    \acmessage[->]{$(s^1_0, S^1_0)$}{3.25}{-6.5}{-5.1}
    \acmessage[->]{$(s^1_1, S^1_1)$}{2.75}{-6.5}{-5.1}
    \acresource{$OT^1_B$}{-5.1}{2.5}{1}{1}
    \acmessage[<-]{$(b^1, S^1_B)$}{3.25}{-4.1}{-2.5}
    \acmessage[->]{$(s^1_{b^1}, T^1)$}{2.75}{-4.1}{-2.5}
    
    \acmessage[->]{$(s^k_0, S^k_0)$}{1.25}{-6.5}{-5.1}
    \acmessage[->]{$(s^k_1, S^k_1)$}{0.75}{-6.5}{-5.1}
    \acresource{$OT^k_B$}{-5.1}{0.5}{1}{1}
    \acmessage[<-]{$(b^k, S^k_B)$}{1.25}{-4.1}{-2.5}
    \acmessage[->]{$(s^k_{b^k}, T^k)$}{0.75}{-4.1}{-2.5}

    \acellipsis{-4.6}{2.1}
    
    \acmessage[->]{$(\{ \bar{s}^i_0, \bar{s}^i_1 \}_i = \{ s^i_0, s^i_1 \}_i, X)$}{-0.25}{-6.5}{-2.5}
    
    \accausality{$P \prec P'' \prec T^i$ and $Q \prec Q'' \prec X$}{4}{-1}

    \indist{0.2}{1.5}
    \acpin{(c)}{0.2}{-0.75}

    \acleftmessage[->]{$(x, P)$}{1.75}{1.25}{2}
    \acleftmessage[->]{$(open, Q)$}{1.25}{1.25}{2}
    \acresource{$BC_B$}{2}{1}{1}{1}
    \acmessage[->]{$(recv, P'')$}{1.75}{3}{4.25}
    \acmessage[->]{$(x, Q'')$}{1.25}{3}{4.25}
    
    \acprotocol{$\sigma^6_B$}{4.25}{-0.45}{0.75}{4}
    
    \acrightmessage[<-]{$(b^1, S^1_B)$}{3.25}{5}{5.75}
    \acrightmessage[->]{$(s^1_{b^1}, T^1)$}{2.75}{5}{5.75}
    
    \acrightmessage[<-]{$(b^k, S^k_B)$}{1.25}{5}{5.75}
    \acrightmessage[->]{$(s^k_{b^k}, T^k)$}{0.75}{5}{5.75}
    
    \acellipsis{5.25}{2.1}
    
    \acrightmessage[->]{$(\{ \bar{s}^i_0, \bar{s}^i_1 \}_i, X)$}{-0.25}{5}{6}
    
    \acresetframe
\end{tikzpicture}
        \caption{Construction of a bit commitment from $k$ instances of $\binom{2}{1}$-OT (Theorem~\ref{thm:ot-bc-construction}).}
        \label{fig:ot-bc-construction}
    \end{figure}
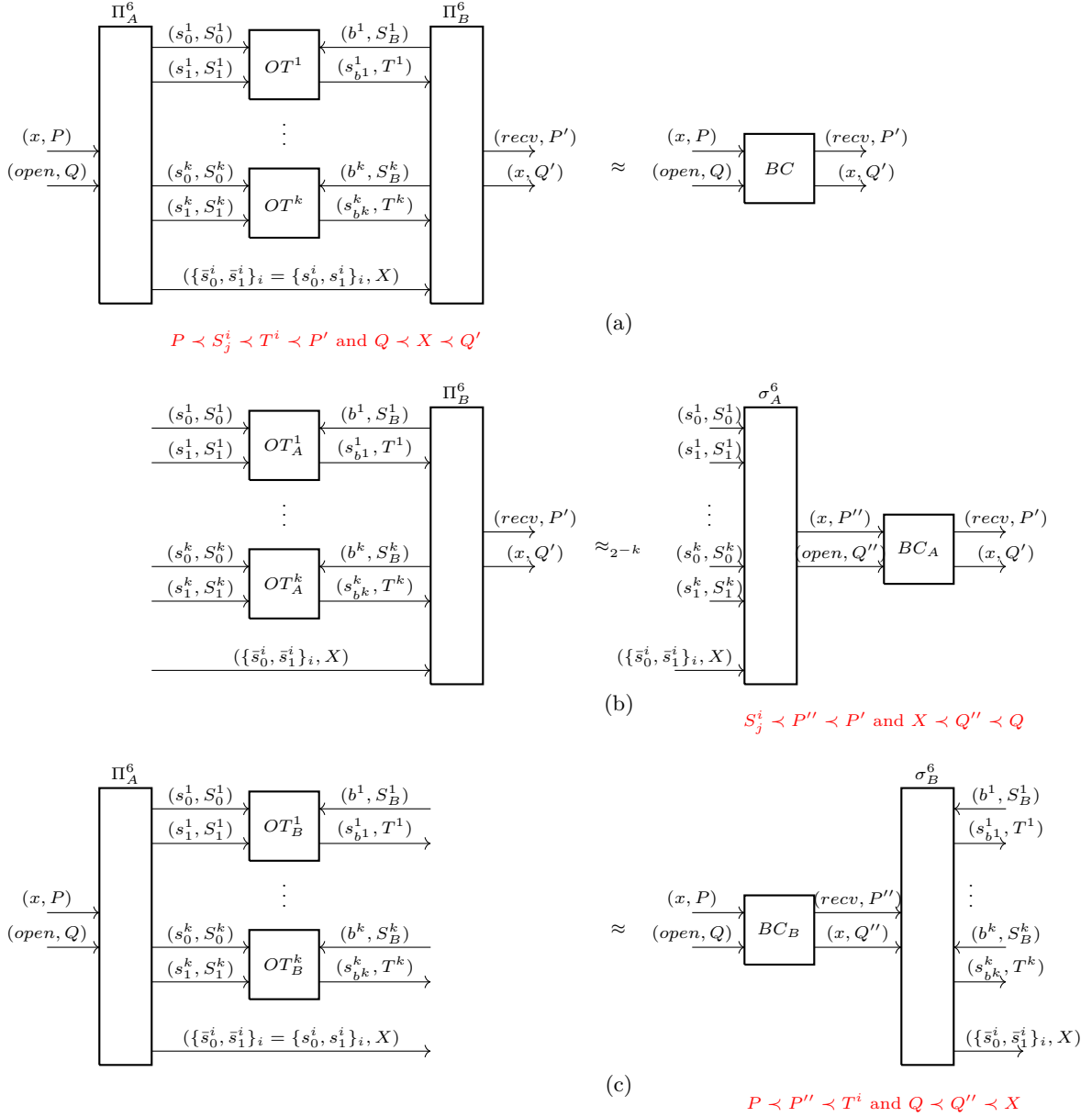
    In the construction given below, $x$ is the input to the bit commitment and $OT_* = OT^1 || \cdots || OT^k$ is the parallel composition of the $k$ instances of oblivious transfer.
    
    \constructionproof{Honest protocol}{$\Pi^6_A OT_* \Pi^6_B \approx BC$}{\ref{fig:ot-bc-construction}(a)}
    One can see that the honest protocol perfectly constructs a bit commitment resource (under honest assumption, the test will never fail).
    
    \constructionproof{Simulation against dishonest Alice}{$OT_* \Pi^6_B \approx_{2^{-k}} \sigma^6_A BC_A$}{\ref{fig:ot-bc-construction}(b)} $\sigma^6_A$ receives all $s_0^i, s_1^i$ immediately, as it simulates the oblivious transfers internally.
    \begin{enumerate}[label=(\arabic*)]
        \item $\sigma^6_A$ commits to $x = s^1_0 \oplus s^1_1$ on its right interface.
        
        \item As soon as $\{ \bar{s}_0^i, \bar{s}_1^i \}_i$ arrive for the opening phase, $\sigma^6_A$ simulates the test done by $\Pi^6_B$ by checking only one of the two bits uniformly at random, for each pair, and also $\bar{s}_0^i \oplus \bar{s}_1^i = x$ for every $i$. If this test succeeds, then $\sigma^6_A$ opens the commitment on its right interface. Otherwise, it aborts just like $\Pi^6_B$.
    \end{enumerate}
    Notice that, when the test succeeds, $OT_* \Pi^6_B$ computes the committed bit $\bar{s}_0^1 \oplus \bar{s}_1^1$ from the bits $\{ \bar{s}_0^i, \bar{s}_1^i \}_i$ delivered during the opening phase, while $\sigma^6_B BC_A$ takes $s_0^1 \oplus s_1^1$ from the bits delivered during the commitment phase.
    Thus, the two systems are perfectly indistinguishable unless the test succeeds with $s^i_0 \oplus s^i_1 \neq \bar{s}^i_0 \oplus \bar{s}^i_1$, which means the bits $\{ \bar{s}_0^i, \bar{s}_1^i \}_i$ sent during the opening phase have one bit flipped for each pair, and these $k$ flipped bits all avoid the test carried out by $\sigma^6_A$. In this case, the bits outputted on the right interface are different in the two systems, but this happens only with probability at most $2^{-k}$. Hence, the difference lemma concludes the construction.
    
    \constructionproof{Simulation against dishonest Bob}{$\Pi^6_A OT_* \approx BC_B \sigma^6_B$}{\ref{fig:ot-bc-construction}(c)}
    $\sigma^6_B$ receives the bits $b^1, \ldots, b^k$ on its right interface, as it also simulates the oblivious transfers internally.
    \begin{enumerate}[label=(\arabic*)]
        \item When the \textit{received} signal arrives from the left interface, it  returns $s^i_{b^i}$ chosen uniformly at random.
        
        \item As soon as the commitment on the left interface is open and $\sigma^6_B$ receives $x$, it can simply construct $\{ \bar{s}^i_0, \bar{s}^i_1 \}$ accordingly, namely $\bar{s}^i_{b^i} \leftarrow s^i_{b^i}, \bar{s}^i_{1 - b^i} \leftarrow x \oplus s^i_{b^i}$.
    \end{enumerate}
    This last construction is perfect.
\end{completeproof}

\end{protibox}

\begin{definition}[Protocol for $\mathcal{BC} \rightarrow \mathcal{OT}$ ~\cite{Unruh10}]
    \label{def:protocol-5}
    Given fixed security parameters $n, k, h$ with $n = k + h$ and $k, h = \Theta(n)$ (e.g.\ $k = h = \frac{n}{2}$), the protocol $\Pi^5 = (\Pi^5_A, \Pi^5_B)$ uses $2n$ instances of bit commitment to implement a $\binom{2}{1}$-oblivious transfer. It works as follows:
    \begin{enumerate}[label=(\arabic*)]
        \item Alice chooses $n$ bits $x_i \in \{ 0, 1 \}$ and $n$ measurement bases $\theta_i \in \{ X, Z \}$ uniformly at random. Then, according to the random choices, she creates $n$ BB84 states  $$\ket{\psi_i} = \ket{x_i}_{\theta_i} \in \{ \ket{0}, \ket{1}, \ket{+}, \ket{-} \}.$$ 
        
        \item These states are sent to Bob, which chooses $\bar\theta_i \in \{ X, Z \}$ uniformly at random, and measures $\ket{\psi_i}$ using $\bar\theta_i$ for every $i$. Let $\bar{x}_i$ be the results of the measurements.
        
        \item Bob uses the $2n$ instances of $\mathcal{BC}$ to commit to $\bar\theta_i$ and $\bar{x}_i$.
        
        \item Alice chooses a test set $T \subseteq [n]$, $|T| = h$ uniformly at random, and sends it to Bob which, in turn, opens the commitments of $\bar{x}_i, \bar\theta_i$ for each $i \in T$ (and aborts the others).
        
        \item Denoting $S \subseteq [n]$ as the subset of states with $\theta_i = \bar\theta_i$, Alice checks $x_i = \bar{x}_i$ for every $i \in S \cap T$. If the test fails, Alice aborts. Otherwise, she continues with the protocol by sending $\theta_i$ to Bob, for every $i \in R := [n] \setminus T$.
        
        \item Bob now has $|R| = k$ bits that were not used in the test, and each of them was correctly measured with probability $\frac{1}{2}$. We use these $k$ bits to construct two subsets $I_0, I_1$ with $|I_0| = |I_1| = \frac{k}{3}$ as we did in Lemma~\ref{thm:rabin-ot-construction}. Bob will abort if $X := |R \cap S| < \frac{k}{3}$ (since he cannot construct a completely known subset).
    \end{enumerate}
    The test carried out in steps 3--5 is needed as a proof that Bob measured $\{ \ket{\psi_i} \}_i$ before committing to the outcomes of their measurements.
\end{definition}


\begin{definition}[Protocol for $\mathcal{OT} \rightarrow \mathcal{BC}$]
    \label{def:protocol-6}
    Given fixed security parameter $k$, the protocol $\Pi^6 = (\Pi^6_A, \Pi^6_B)$ uses $k$ instances of $\binom{2}{1}$-oblivious transfer to construct a bit commitment, and works as follows:
    \begin{enumerate}[label=(\arabic*)]
        \item In the commit phase, Alice chooses bits $s_0^1, \ldots, s_0^k$ uniformly and independently at random, and $s_1^1, \ldots, s_1^k$ such that $s_1^i = s_0^i \oplus x$. Thus, $x = s_0^i \oplus s_1^i$ for every $i \in [k]$.
        
        \item Bob chooses bits $b^1, \ldots, b^k$ uniformly and independently at random, and uses them to choose one of $s_0^i$ and $s_1^i$ through an instance of the $\binom{2}{1}$-OT for every $i$. At this point, for every $i$, Alice has no information about which of $s_0^i, s_1^i$ is known to Bob.
        
        \item During the opening phase, Alice sends all $\{ \bar{s}_0^i, \bar{s}_1^i \}_i = \{ s_0^i, s_1^i \}_i$ to Bob, which will check that they are consistent with what he received from the OT primitives in the commit phase. Moreover, he checks that $\bar{s}_0^i \oplus \bar{s}_1^i$ are all equal for every $i$. If this test fails, Bob aborts. Otherwise, Bob outputs $x = \bar{s}_0^1 \oplus \bar{s}_1^1$ on its right interface.
    \end{enumerate}
\end{definition}

\section{Discussion}
\label{sec:discussion}
\paragraph{Summary of results.} We proved impossibility of composable oblivious transfer and multi-party computation in relativistic and quantum settings, and  provided mutual constructions between different versions of oblivious transfer and bit commitment. We did so in the abstract cryptography framework~\cite{MauRen11}, with cryptographic resources instantiated as causal boxes in Minkowski space~\cite{CausalBoxesFW,Vilasini19}. 

\paragraph{Minimal resources for oblivious transfer.} This works dashes hopes to rely on relativistic constraints to construct composably secure  oblivious transfer, without further resources and assumptions on the behaviour of agents. A next step would be to investigate precisely whether there are weaker resources from which these primitives can be built. For example in~\cite{prokop20} Prokop introduces an `asymmetric quantum beamer' (which sends Bob a series of BB84 qubits, and Alice a limited classical description of the qubits produced) and shows that it can be used to build bit commitment; our results imply that it can also be used to build oblivious transfer. It would be interesting to investigate to which extent this resource is strictly weaker than bit commitment.
 

\paragraph{Cryptography under general relativity.} The only aspect of relativity explored so far is the limited speed of light in special relativity ~\cite{Kent99,Kent12, Vilasini19}.
It would be interesting to extend the theoretical framework to cover general relativity scenarios, like quantum superpositions of large masses, which may cause  true superpositions of causal orders. 
After all, learning which information-processing tasks are allowed by nature is a way to better explore both physics and cryptography.

\newpage
\appendix
\section*{Appendix}
\vspace{0.5cm}

\section{Abstract cryptography: formal definitions}
\label{appendix:AC}
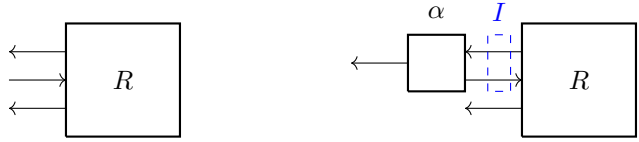
\begin{figure}
        \centering
        \begin{tikzpicture}[scale=1.5]
    \acleftmessage[<-]{}{0.25}{-0.5}{0}
    \acleftmessage[->]{}{0.5}{-0.5}{0}
    \acleftmessage[<-]{}{0.75}{-0.5}{0}
    \acresource{$R$}{0}{0}{1}{1}
    

    \acmessage[<-]{}{0.65}{2.5}{3}
    \acprotocol{$\alpha$}{3}{0.4}{0.5}{0.5}
    \acmessage[<-]{}{0.25}{3.5}{4}
    \acmessage[->]{}{0.5}{3.5}{4}
    \acmessage[<-]{}{0.75}{3.5}{4}
    \acresource{$R$}{4}{0}{1}{1}
    
    \draw[dashed,blue]
        (3.7, 0.4) -- 
        (3.7, 0.9) -- 
        (3.9, 0.9) -- 
        (3.9, 0.4) -- 
        (3.7, 0.4);
    \node[blue] at (3.8,1.1) {$I$};
    
\end{tikzpicture}
        \caption{Example of a resource. The arrows represent inputs (pointing towards the box) and outputs (pointing away from the box). These are the graphical representations of $R$ (left) and $\alpha^I R$ (right). In this case, $\alpha$ is a converter with two inner interfaces (connected to the interfaces of $R$ in the set $I$), and one outer interface.}
        \label{fig:resource-definition}
\end{figure}

In this Appendix we present a more formal review of the Abstract Cryptography framework~\cite{MauRen11}. In a \emph{resource theory} of cryptography, the resources are cryptographic primitives, like bit commitment or key distribution. These can be composed along with others constructions (e.g.\ protocols, trusted third parties) to construct new resources. 
\begin{definition}[Component space \cite{MauRen11}]
A component space is a triple $(\Omega, ||, \delta)$, where
    \begin{itemize}
        \item $\Omega$ is a set of resources;
        \item $||$ is an operation on $\Omega$ called parallel composition;
        \item $\delta : \Omega^2 \rightarrow \R^+_0$ is a pseudo-metric on $\Omega$ such that $(\Omega, \delta)$ is a pseudo-metric space\footnote{A pseudo-metric space is similar to a metric space, with the only difference that the identity of indiscernibles does not hold in general: two elements with $\delta(a, b) = 0$ may not be equal.}.
    \end{itemize}
\end{definition}
For our purposes, a resource $R \in \Omega$ is an abstract system: a closed box with a number of interfaces where inputs are read and outputs are delivered (Figure~\ref{fig:resource-definition}). From a cryptographic point of view, one can imagine a resource as a trusted device, where different parties connect to different interfaces. First we need a space of resources and a measure of closeness between resources that tells us how similar they are. This measure will be given an operational meaning later.

In the rest of the work, we use the notation $R \approx_\varepsilon S$ to denote that $\delta(R, S) \le \varepsilon$ and, if $\varepsilon = 0$, we may also remove the subscript. The pseudo-metric satisfies the triangle inequality by definition, and one can infer that:
\begin{align*}
    \begin{cases}
        R \approx_\varepsilon S \\
        S \approx_{\varepsilon'} T
    \end{cases}
    \Longrightarrow
    R \approx_{\varepsilon + \varepsilon'} T.
\end{align*}
Note that in this work, the pseudo-metric used is the statistical advantage with respect to a class of distinguishers, $d^{\mathbb{D}}$.

\begin{definition}[Constructor space \cite{MauRen11}]
    A constructor space is a triple $(\Gamma, \circ, |)$, where
    \begin{itemize}
        \item $\Gamma$ is a set of converters;
        \item $\circ$ is an operation on $\Gamma$ called serial composition;
        \item $|$ is an operation on $\Gamma$ called parallel composition;
    \end{itemize}
\end{definition}
Here we define $\alpha \in \Gamma$ as a \textit{converter}, which is an abstract system like the resources defined above, with two sets of interfaces, one internal and one external. We can attach a converter to a subset of interfaces of a resource. The notation $\alpha^I R$ denotes a resource obtained by attaching $\alpha$ to the set of interfaces $I$ of $R$ (Figure~\ref{fig:resource-definition}). Throughout the rest of this work we will only consider two-party settings, therefore we will denote as $\alpha R \beta$ a resource obtained by attaching $\alpha$ to the interfaces of the first party (Alice), and $\beta$ to the interfaces of the second party (Bob). 
These structures satisfy properties of \textit{general composability}~\cite{MauRen11}, which we will not discuss here.

\begin{figure}
        \centering
        \begin{tikzpicture}[scale=1.5]
    \acmessage[<-]{}{0.25}{-0.5}{0}
    \acmessage[<-]{}{0.5}{-0.5}{0}
    \acmessage[->]{}{0.75}{-0.5}{0}
    \acresource{$R$}{0}{0}{1}{1}
    \acmessage[->]{}{0.35}{1}{1.5}
    \acmessage[<-]{}{0.65}{1}{1.5}
    
    \acprotocol{$\alpha$}{-1}{0}{0.5}{1}
    \acprotocol{$\beta$}{1.5}{0}{0.5}{1}
    
    \acleftmessage[->]{}{0.65}{-1.5}{-1}
    \acleftmessage[<-]{}{0.35}{-1.5}{-1}
    
    \acrightmessage[<-]{}{0.5}{2}{2.5}
\end{tikzpicture}
        \caption{Graphical representation of the construction $\alpha R \beta$. The subset of interfaces the converters $\alpha, \beta$ connect to will be usually omitted in the notation, as it will be clear from the context. For example, $\alpha, \beta$ may represent operations and protocols implemented by Alice and Bob, respectively, while $R$ could be a shared cryptographic primitive.}
        \label{fig:two-party-resource}
\end{figure}
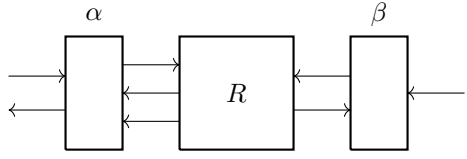

\begin{figure}
        \centering
        \begin{tikzpicture}[scale=1.5]
    \acmessage[<-]{}{0.25}{-0.5}{0}
    \acmessage[<-]{}{0.5}{-0.5}{0}
    \acmessage[->]{}{0.75}{-0.5}{0}
    \acresource{$R$}{0}{0}{1}{1}
    \acmessage[->]{}{0.35}{1}{1.5}
    \acmessage[<-]{}{0.65}{1}{1.5}
    
    \draw[thick]
        (-0.5, 1) -- 
        (-0.5, -0.3) --
        (1.5, -0.3) --
        (1.5, 1) --
        (2, 1) --
        (2, -0.8) --
        (-1, -0.8) --
        (-1, 1) --
        (-0.5, 1);
        
    \node[align=center] at (0.5, -0.56) {$D$};
    
    \draw[->] (0.5,-0.8) -- (0.5,-1.1);
    \node[align=left] at (1.35, -1.1) {$D[R] \in \{ 0, 1 \}$};
\end{tikzpicture}
        \caption{Graphical representation of a distinguisher $D$ observing a resource $R$. $D$ covers all the interfaces of $R$ and returns a bit on its outer interface. These usually represent attackers to our constructions, distinguishing our protocols from ideal resources.}
        \label{fig:distinguisher-definition}
\end{figure}
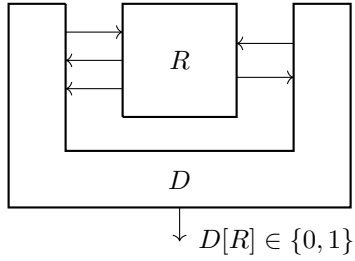

There is a third type of component, besides resources and converters, called  \textit{distinguishers} (Figure~\ref{fig:distinguisher-definition}). These are special types of converters whose internal interfaces attach to all the interfaces of a resource, while the external interface only outputs one bit. Given a resource $R$ and a distinguisher $D$, we have that $D[R]$ is an indicator random variable, and this is important to keep in mind when we define security. The definition of primitive and security with respect to a class of distinguishers can be found in the main text.

\paragraph{Why is the construction of Definition~\ref{def:crypto-security} sufficient for security against dishonest behaviour?}
Consider the following result:
\begin{lemma}
    \label{thm:converter-application}
    Let $D \in \mathbb{D}$ be a distinguisher, and $\alpha \in \Gamma$ be a converter. Moreover, we denote with $D\alpha$ a distinguisher such that, for every $R \in \Omega$:
    \begin{align*}
        D\alpha [R] \equiv D [\alpha R]
    \end{align*}
    If $D \alpha \in \mathbb{D}$ for every $D \in \mathbb{D}$, the following holds:
    \begin{align*}
        R \approx_\varepsilon S \Longrightarrow \alpha R \approx_\varepsilon \alpha S
    \end{align*}
\end{lemma}
\begin{proof}
    We have that $d^\mathbb{D}(R, S) \le \varepsilon$. Therefore:
    \begin{align*}
        d^\mathbb{D}(\alpha R, \alpha S) & = \sup_{D \in \mathbb{D}} \left|\Prob{D\alpha [R] = 1} - \Prob{D\alpha [S] = 1}\right| \\
        & \le \sup_{D \in \mathbb{D}} \left|\Prob{D [R] = 1} - \Prob{D [S] = 1}\right| \\
        & = d^\mathbb{D}(R, S) \le \varepsilon
    \end{align*}
    The inequality follows from the fact that, since $D\alpha \in \mathbb{D}$, it is already considered in the $\sup$ of the right-hand side.
\end{proof}
The original paper~\cite{MauRen11} makes this property more precise by defining an algebra over $(\mathbb{D}, \mathbb{S})$: for example, if we consider $\mathbb{D}$ as the set of polynomial-time distinguishers, the hypotheses of Lemma~\ref{thm:converter-application} would be satisfied by any polynomial-time converter $\alpha$.
Let us consider the case of honest Bob and dishonest Alice (the other case is analogous): therefore Bob will run the honest protocol $\Pi_B$, while Alice may use a dishonest procedure (let us call it $\bar{\Pi}_A$). By applying Lemma~\ref{thm:converter-application}:
\begin{align*}
    R_A \Pi_B \approx_\varepsilon \sigma_A S_A \Longrightarrow \bar{\Pi}_A R_A \Pi_B \approx_\varepsilon \bar{\Pi}_A \sigma_A S_A
\end{align*}
which means that Bob sees a behaviour that is statistically close to the one given by the ideal resource on its interface.

\paragraph{Example.} 
A very common attack on composable security is the man-in-the-middle attack, in which the distinguisher runs another copy  $R'$ of $R$ in parallel and forwards messages between the two resources, creating correlations between the outputs of the $R$ and $R'$ that one would not obtain from two copies of $S$, and therefore reaching  $d^D(R, S) >0$. This is at the heart of impossibility proofs for coin-flipping constructions \cite{Vilasini19}, and also for the oblivious transfer primitives treated in the present work.

\section{Overview of causal boxes}
\label{appendix:CB}
In this section we review the Causal Box framework~\cite{CausalBoxesFW}: causal boxes are powerful because they can model information processing systems in great generality, for example allowing a superposition of the order of messages, or order of messages that are defined during protocol runtimes. Moreover, Portmann et al.~\cite{CausalBoxesFW} showed that Causal Boxes are closed under composition, a feature that is crucial in order to guarantee the general composability properties required by the Abstract Cryptography framework. Another important point is that the formalism allows us to instantiate Causal Boxes in a Minkowski space-time, thus easily taking into account special relativity constraints for the exchange of messages.

\subsection*{Message space and wires}
We model an arbitrary message as a pair $(v, t) \in \mathcal{V} \times \mathcal{T}$, where $\mathcal{V}$ is a message space and $\mathcal{T}$ is a partially ordered set defining the order of the messages. If a message is encoded as a quantum state, the Hilbert space of a single message is $\mathcal{H} = \C^{|\mathcal{V}|} \otimes l^2(\mathcal{T})$, where $l^2(\mathcal{T}) = \text{span}\{ \ket{t} \}_{t \in \mathcal{T}}$. Therefore, $\mathcal{H}$ is spanned by the basis $\{ \ket{v, t} \}_{v \in \mathcal{V}, t \in \mathcal{T}}$.

A causal box receives its inputs and outputs through \textit{wires}, which can carry any number of messages of fixed dimension, or even a superposition of them. The dimension of such messages defines the dimension of the wire: for example, a two-dimensional wire ($|\mathcal{V}| = 2$) can send any number of qubits one after the other (or, again, in superposition) but it cannot carry a qutrit. Hence, we can model a $d$-dimensional wire as a (bosonic) Fock space:
\begin{align*}
    \mathcal{F}(\C^d \otimes l^2(\mathcal{T})) := \text{span}\{ \ket{\Omega} \} \oplus \bigoplus_{n = 1}^\infty \vee^n (\C^d \otimes l^2(\mathcal{T}))
\end{align*}
where $\vee^n \mathcal{H}$ denotes the symmetric subspace of $\mathcal{H}^{\otimes n}$ and $\ket{\Omega}$ is the vacuum state, which represents that no messages are sent through the wire. We take the symmetric subspaces because we want $\ket{(m_1, t_1), (m_2, t_2)} \equiv \ket{(m_2, t_2), (m_1, t_1)}$, i.e.\ the order of messages is already induced by the elements of $\mathcal{T}$. Also wires are proven to be composable~\cite{CausalBoxesFW}: if we have two wires with Fock spaces $\mathcal{F}_A, \mathcal{F}_B$, of dimensions $d_A, d_B$, these will be equivalent to a single wire with Fock space $\mathcal{F}_A \otimes \mathcal{F}_B$. Moreover, since
\begin{align*}
    \mathcal{F}(\mathcal{H}_A) \otimes \mathcal{F}(\mathcal{H}_B) \simeq \mathcal{F}(\mathcal{H}_A \oplus \mathcal{H}_B)
\end{align*}
with $\mathcal{H}_A = \C^{d_A} \otimes l^2(\mathcal{T}), \mathcal{H}_B = \C^{d_B} \otimes l^2(\mathcal{T})$, this new wire has dimension $d = d_A + d_B$. Notice that the above isomorphism also allows us to conclude the opposite: any wire of dimension $d$ can be split into two wires of dimension $d_A + d_B = d$.

\subsection*{Cuts and causality}
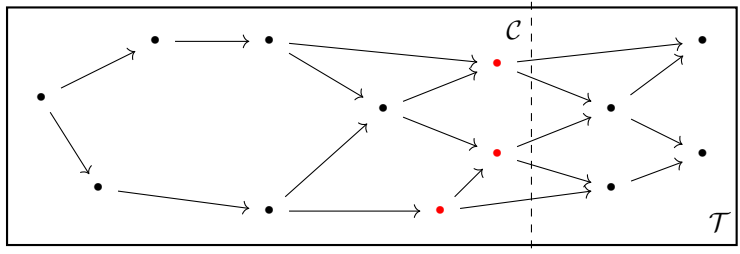
\begin{figure}
        \centering
        \begin{tikzpicture}[scale=1.5]

    \fill [superlightgray] (-4.3, 1.3) rectangle (0.3, -0.8);
    \draw [thick] (-4.3, 1.3) rectangle (2.1, -0.8);
    
    \draw[dashed] (0.3, 1.35) -- (0.3, -0.85);
    
    \node at (0.15, 1.1) {$\mathcal{C}$};
    \node at (1.95, -0.6) {$\mathcal{T}$};

    \node (A) at (-4,0.5) {\textbullet};
    \node (B) at (-3.5,-0.3) {\textbullet};
    \node (C) at (-3,1) {\textbullet};
    \node (D) at (-2,1) {\textbullet};
    \node (E) at (-2,-0.5) {\textbullet};
    \node (F) at (-1,0.4) {\textbullet};
    \node[red] (G) at (-0.5,-0.5) {\textbullet};
    \node[red] (H) at (0,0.8) {\textbullet};
    \node[red] (I) at (0,0) {\textbullet};
    \node (J) at (1,-0.3) {\textbullet};
    \node (K) at (1,0.4) {\textbullet};
    \node (L) at (1.8,0) {\textbullet};
    \node (M) at (1.8,1) {\textbullet};
    
    \draw[->] (A) -- (B);
    \draw[->] (A) -- (C);
    \draw[->] (C) -- (D);
    \draw[->] (B) -- (E);
    \draw[->] (D) -- (F);
    \draw[->] (E) -- (F);
    \draw[->] (E) -- (G);
    \draw[->] (D) -- (H);
    \draw[->] (F) -- (H);
    \draw[->] (F) -- (I);
    \draw[->] (G) -- (I);
    \draw[->] (G) -- (J);
    \draw[->] (I) -- (J);
    \draw[->] (H) -- (K);
    \draw[->] (I) -- (K);
    \draw[->] (K) -- (L);
    \draw[->] (J) -- (L);
    \draw[->] (K) -- (M);
    \draw[->] (H) -- (M);
\end{tikzpicture}
        \caption{Example of a cut $\mathcal{C}$ in a finite set $\mathcal{T}$ with a partial order. Here $\mathcal C$ is the set of all points on the left of the dashed line.
        The highlighted points form the minimal frontier of the cut: all points in the cut are in their causal past.}
        \label{fig:cut-example}
\end{figure}

Now we would like to formalize a reasonable notion of causality which will be satisfied by causal boxes. First we need to define a cut: we can think of it as a partition through space-time where we only consider the points that came before the `cut' (Figure~\ref{fig:cut-example}). 
 
\begin{definition}[Cut~\cite{CausalBoxesFW}]
    Given a partially ordered set $\mathcal{T}$, a cut is a subset $\mathcal{C} \subseteq \mathcal{T}$ such that, for some set of points $\mathcal{P}$:
    \begin{align*}
        \mathcal{C} & = \bigcup_{t \in \mathcal{P}} \mathcal{T}^{\le t}
    \end{align*}
    where $\mathcal{T}^{\le t} := \{ p \in \mathcal{T} \,|\, p \le t \}$. A cut is said to be \textit{bounded} if there is $t \in \mathcal{T}$ such that $\mathcal{C} \subseteq \mathcal{T}^{\le t}$. Moreover, we denote the set of all cuts in $\mathcal{T}$ with $\mathfrak{C}(\mathcal{T})$, and the set of all bounded cuts with $\bar{\mathfrak{C}}(\mathcal{T})$.
\end{definition}

We can call the set of points $\mathcal{P}$ the \emph{frontier} of the cut $\mathcal{C}$. Notice that it is not unique for a cut: indeed any $\mathcal{P}'$ with $\mathcal{P} \subseteq \mathcal{P}' \subseteq \mathcal{C}$ is a valid frontier of $\mathcal{C}$. Indeed, one can see that, for finite $\mathcal{T}$, the minimal frontier of the cut is a vertex cut of the digraph defined by the order (Figure~\ref{fig:cut-example}). A more compact (but less straightforward) way to characterize cuts is:
\begin{align*}
    \mathcal{C} = \bigcup_{t \in \mathcal{C}} \mathcal{T}^{\le t}
\end{align*}

Now we introduce a causality function, which defines a sort of gap between the positions of inputs and outputs of information processing tasks: an output at positions in $\mathcal{C} \in \mathfrak{C}(\mathcal{T})$ must necessarily be caused only by inputs in its causal past, that is at positions in $\chi(\mathcal{C}) \subsetneq \mathcal{C}$. For example, if we take $\mathcal{T} = \Q$ (equipped with its natural order), a gap of $\delta > 0$, namely $\chi((-\infty, x]) = (-\infty, x - \delta]$ will satisfy Definition~\ref{def:causality-function}, and it imposes a time gap of at least $\delta$ between inputs and correlated outputs.

\begin{definition}[Causality function~\cite{CausalBoxesFW}]
    \label{def:causality-function}
    A function $\chi : \mathfrak{C}(\mathcal{T}) \rightarrow \mathfrak{C}(\mathcal{T})$ is said to be a causality function if the following conditions hold:
    \begin{enumerate}[label=(\arabic*)]
        \item For any two cuts $\mathcal{C}, \mathcal{D} \in \mathfrak{C}(\mathcal{T})$, $\chi(\mathcal{C} \cup \mathcal{D}) = \chi(\mathcal{C}) \cup \chi(\mathcal{D})$;
        \item For any two cuts  $\mathcal{C}, \mathcal{D} \in \mathfrak{C}(\mathcal{T})$, $\mathcal{C} \subseteq \mathcal{D} \Longrightarrow \chi(\mathcal{C}) \subseteq \chi(\mathcal{D})$;
        \item For any cut $\mathcal{C} \in \mathfrak{C}(\mathcal{T}) \setminus \{ \emptyset \}$, $\chi(\mathcal{C}) \subsetneq \mathcal{C}$;
        \item For any cut $\mathcal{C} \in \mathfrak{C}(\mathcal{T})$ and $t \in \mathcal{T}$, $\exists n \in \N$ such that $t \not\in \chi^n(\mathcal{C})$.
    \end{enumerate}
    where $\chi^n = \chi \circ \cdots \circ \chi$ applies the function $n$ times.
\end{definition}

Note that Conditions (1) and (2) guarantee consistency: if an output on $\mathcal{C}$ can be computed only from inputs on $\chi(\mathcal{C})$, and an output on $\mathcal{D}$ can be computed only from inputs on $\chi(\mathcal{D})$, then outputs on $\mathcal{C} \cup \mathcal{D}$ can be computed only from inputs in $\chi(\mathcal{C}) \cup \chi(\mathcal{D})$. Moreover, if inputs from $\chi(\mathcal{C})$ are available to compute an output on $\mathcal{C}$, then they must also be available to compute any output on $\mathcal{D} \supseteq \mathcal{C}$.
    
Condition (4) is necessary to avoid ill-defined systems: consider $\mathcal{T} = \mathbb{Q}^+$ and a system outputting in position $1 - t/2$ from an input in position $1 - t$ for every $0 < t \le 1$. This corresponds to a causality function $\chi([0, 1 - t/2]) = 1 - t$ which satisfies conditions (1)-(3). However, setting the initial position to $0$ and looping back the output to the input produces an infinite amount of messages before position $1$ is reached, namely $1 - 1/2^n$. The problem here is that the gap between input and output tends to $0$, and condition (4) imposes that a position $t$ goes out of the possible input positions after a finite number of steps.

\subsection*{Definition of causal box}
We are now ready to define a causal box:
\begin{definition}[Causal box~\cite{CausalBoxesFW}]
    A $(d_X, d_Y)$-causal box $\Phi$ with input wire dimension $d_X$ and output wire dimension $d_Y$ is a set of completely positive trace-preserving maps
    \begin{align*}
        \Phi = \left\{ \Phi^{\mathcal{C}} : \mathcal{S}(\mathcal{F}_X^{\chi(\mathcal{C})}) \rightarrow \mathcal{S}(\mathcal{F}_Y^{\mathcal{C}}) \right\}_{\mathcal{C} \in \bar{\mathfrak{C}}(\mathcal{T})}
    \end{align*}
    for some causality function $\chi(\mathcal{C})$, satisfying \textit{mutual consistency}, i.e.
    \begin{align*}
        \Phi^{\mathcal{C}} & = \Phi^{\mathcal{C}} \circ \tr_{\mathcal{T} \setminus \chi(\mathcal{C})} \\
        \Phi^{\mathcal{C}} & = \tr_{\mathcal{D} \setminus \mathcal{C}} \circ \Phi^{\mathcal{D}}
    \end{align*}
    for any two cuts $\mathcal{C}, \mathcal{D} \in \mathfrak{C}(\mathcal{T})$ with $\mathcal{C} \subseteq \mathcal{D}$. Here, $\mathcal{S}(\mathcal{H})$ denotes the density operator space over $\mathcal{H}$ and $\tr_{\mathcal{H}}$ denotes the partial trace operator over $\mathcal{H}$.
\end{definition}
\noindent Mutual consistency imposes that the map $\Phi^{\mathcal{C}}$ acting on the cut $\mathcal{C}$ should ignore any information that is not in $\chi(\mathcal{C})$. Moreover, for any two cuts $\mathcal{C} \subseteq \mathcal{D}$, tracing out any output from $\Phi^\mathcal{D}$ arriving at positions in the gap $\mathcal{D \setminus C}$ should give the same result as $\Phi^{\mathcal{C}}$. The original paper by Portmann~et~al.~\cite{CausalBoxesFW} also extends Choi-Jamio\l{}kowski and Stinespring representations to Causal Boxes.

\subsection*{Abstract Cryptography in the relativistic quantum setting}
We can now concretely define the abstract systems behind the resources, converters, and distinguishers presented in Appendix~\ref{appendix:AC} above as Causal Boxes. In the rest of their work, Portmann et al.~\cite{CausalBoxesFW} also provide general composition operations between Causal Boxes, proving the properties required by Maurer and Renner~\cite{MauRen11} in order to guarantee the desired general composability properties of the framework. Moreover, a pseudo-metric $\delta$ defining statistical distance between Causal Boxes is provided in the same work. The last element to instantiate is the partially ordered set $\mathcal{T}$ used to define causality. In order to model relativistic effects, we use the following:
\begin{figure}
        \centering
        \begin{tikzpicture}

    \draw[->,thick] (-0.2, 0) -- (6, 0);
    \draw[->,thick] (0, -0.2) -- (0, 6);
    
    \node[align=left] at (6.3, 0) {$x$};
    \node[align=left] at (0.3, 6) {$t$};
    
    \node (P) at (4, 1) {};
    \node (Q) at (2, 1.5) {};
    \node (R) at (3, 4.7) {};

    \node at (P) {\textbullet};
    \node at (Q) {\textbullet};
    \node at (R) {\textbullet};
    
    \draw[dashed] (P) -- (5.8, 5.5);
    \draw[dashed] (P) -- (2.2, 5.5);
    
    \draw[dashed] (Q) -- (3.6, 5.5);
    \draw[dashed] (Q) -- (0.4, 5.5);
    
    \node[align=center, below] at (P) {$P$};
    \node[align=center, below] at (Q) {$Q$};
    \node[align=center, above] at (R) {$R$};
    
    \fill [opacity=0.1,blue]
        (2, 1.5) -- (3.6, 5.5) -- (0.4, 5.5) -- cycle;
    \fill [opacity=0.1,yellow]
        (4, 1) -- (5.8, 5.5) -- (2.2, 5.5) -- cycle;
\end{tikzpicture}
        \caption{\textbf{Graphical representation of (1D) Minkowski space-time}. The horizontal axis $x$ is one dimension of space, and the vertical axis $t$ is time. Each point in the space generates a causal cone (its \emph{future light cone}), whose slope is exactly the speed of light $c$. In the figure, $R$ is in the intersection of the future light cones of $P$ and $Q$, thus it is in the causal future of both points.}
        \label{fig:minkowski-space-time}
\end{figure}
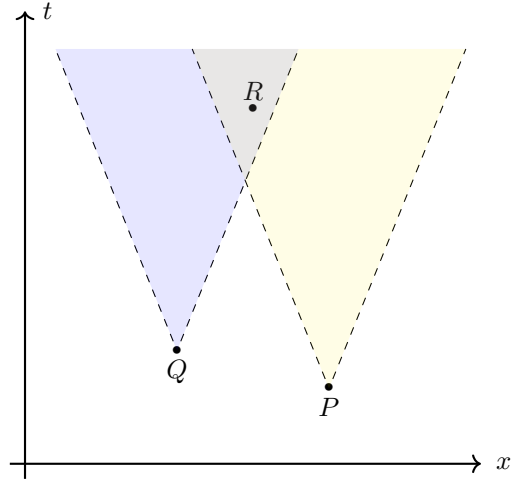

\begin{definition}[Minkowski space-time]
    A Minkowski space-time $\mathcal{M} \simeq \R^4$ is a four-dimensional space where a point $P \in \mathcal{M}$ is a tuple $P = (\vec{x}, t)$ of a position $\vec{x}$ in space and a timestamp $t$.
    
    We define a causal order $\prec$ on the space-time: we say that $P = (\vec{x}_P, t_P)$ is in the \textit{causal past} of $Q = (\vec{x}_Q, t_Q)$ (i.e.\ $P \prec Q$) if and only if the following holds:
\begin{align*}
    ||\vec{x}_Q - \vec{x}_P||_2 \le c \cdot (t_Q - t_P),
\end{align*}
where $c$ is the speed of light.
\end{definition}
 In other words, light can reach $\vec{x}_Q$ from $\vec{x}_P$ in time $t_Q - t_P$. One can see that this order can be used to define a causality function satisfying Definition~\ref{def:causality-function}. Different parties in a relativistic protocol should negotiate a common reference frame in order to not have ambiguities in representing the space-time points. However, it is important to notice that security does not depend on the chosen frame, as it will only depend on the causal order $\prec$ defined above, which is invariant under Lorentz transformations. From now on, along with messages, we will also specify the timestamp of the position in the space-time in which they are sent.
This completes our framework, since Causal Boxes instantiated with a Minkowski space-time not only model quantum effects\footnote{Indeed, Portmann et al.~\cite{CausalBoxesFW} remark that Causal Boxes can model any non-signalling process with either quantum and classical inputs.}, but also relativistic ones.

\ifexportproofs
    \section{Proofs of all results}
    \label{appendix:proofs}
    \subsection{Probability theory results used in proofs}
    \label{appendix:prob}
    
\subsubsection{Concentration bounds}

These results are used in the proofs of Lemma~\ref{thm:rabin-ot-construction} and Theorem~\ref{thm:bc-ot-construction}.

\begin{theorem}[simplified Chernoff bound, relative error~\cite{tulsiani2013probability}]
    \label{thm:chernoff-bound}
    Let $X_1, \ldots, X_n$ be independent indicator random variables defined under a probability space $(\Omega, \mathcal{F}, \mathbb{P})$. Defining $X = \sum_i X_i$ and $\mu = \E{X}$, the following bounds hold:
    \begin{align*}
        \Prob{X \ge (1 + \delta) \mu } \le e^{-\delta^2 \mu / 3} \\
        \Prob{X \le (1 - \delta) \mu } \le e^{-\delta^2 \mu / 2}
    \end{align*}
\end{theorem}

\begin{theorem}[Hoeffding's inequality for hypergeometric distributions~\cite{hoeffding63}]
    \label{thm:hoeffding-hyperg}
    Let $H$ be an hypergeometric random variable representing $h$ extractions without replacement with $x$ initial success objects over $n$ total objects defined under a probability space $(\Omega, \mathcal{F}, \mathbb{P})$. The following bounds hold for $0 < t < \frac{x}{n}$:
    \begin{align*}
        \Prob{ H \le \E{H} - t h } \le e^{-2 t^2 h} \\
        \Prob{ H \ge \E{H} + t h } \le e^{-2 t^2 h}
    \end{align*}
\end{theorem}

\subsubsection{Bounds for distinguishing advantages}
These two lemmas are extensively used across all the proofs in order to bound distinguishing advantages. The first lemma tells us that, if two systems are perfectly indistinguishable unless an observable event $Z$ happens, the statistical distance of any distinguisher is bounded by the probability of such event.

\begin{lemma}[Difference lemma]
    \label{thm:diff-lemma-1}
    Let $(\Omega, \mathcal{F}, \mathbb{P})$ be a probability space, and consider three events $X, Y, Z \in \mathcal{F}$ such that $\Prob{X \cap \neg Z} = \Prob{Y \cap \neg Z}$. The following bound holds:
    \begin{align*}
        \left| \Prob{X} - \Prob{Y} \right| \le \Prob{Z}
    \end{align*}
\end{lemma}
\begin{proof}
    \begin{align*}
        \left| \Prob{X} - \Prob{Y} \right| & = \left| \Prob{X \cap Z} + \Prob{X \cap \neg Z} - \Prob{Y \cap Z} - \Prob{Y \cap \neg Z} \right| \\
        & = \left| \Prob{X \cap Z} - \Prob{Y \cap Z} \right| \\
        & = \Prob{Z} \left| \Prob{X \given Z} - \Prob{Y \given Z} \right| \le \Prob{Z}
    \end{align*}
\end{proof}

The second lemma is used when we want to tell apart two systems: if an event $Z$ can happen only in one of the two systems, then the distinguishing advantage of a distinguisher which outputs $1$ if and only if $Z$ occurs will be at least the probability of $Z$.
\begin{lemma}[Statistical separation lemma]
    \label{thm:diff-lemma-2}
    Let $(\Omega, \mathcal{F}, \mathbb{P})$ be a probability space, and consider $X, Y, Z \in \mathcal{F}$ such that $Z \subseteq X$ and $X \cap Y = \emptyset$. The following bound holds:
    \begin{align*}
        \left| \Prob{Z \given X} - \Prob{Z \given Y} \right| \ge \Prob{Z}
    \end{align*}
\end{lemma}
\begin{proof}
    \begin{align*}
        \left| \Prob{Z \given X} - \Prob{Z \given Y} \right| & = \left| \Prob{Z \given X} - 0 \right| = \frac{\Prob{Z \cap X}}{\Prob{X}} \ge \Prob{Z \cap X} = \Prob{Z}
    \end{align*}
\end{proof}
    
\def\completeproofrequired{1}

\newpage
\subsection{Equivalence of Oblivious Transfer primitives}
\label{apx:equivalence-proofs}

\rototconstruction*

\newpage
\otrotconstruction*

\newpage
\otrabinconstruction*

\newpage
\rabinotconstruction*

\newpage \newpage
\subsection{Impossibility results}
\label{apx:impossibility-proofs}

\rotimpossibility*

This proof technique looks very `classical', in the sense that no quantum information seems to be involved. However, it is worth noticing that the initial assumption on the protocol $\Pi$ is very general, as includes any kind of protocol, also quantum, relativistic and non-signalling ones.

\rabinimpossibility*

\subsection{Oblivious String Transfer}
\label{apx:impossibility-string-proofs}

\rotstringimpossibility*
\begin{completeproof}
    The proof is identical to the one given for Theorem~\ref{thm:rot-impossibility}, with the only difference that $s_{b'}, s'_{b'}$ are strings of $s$ bits. Therefore, if $\sigma_{BA}$ does not receive information about $s_{b'}$, the probability that the output on the right is different is:
    \begin{align*}
        \Prob{s_{b'} \neq s'_{b'} \given b \neq b'} = 1 - \frac{1}{2^s}
    \end{align*}
\end{completeproof}

\rabinstringimpossibility*
\begin{completeproof}
    We follow the same reasoning as for Theorem~\ref{thm:rabin-impossibility}, except for the fact that, when the event $E$ occurs, $\sigma_{BA}$ needs to match a string of $s$ bits instead of a single bit, which translates to:
    \begin{align*}
        \Prob{x_A \neq x_B \given E} = 1 - \frac{1}{2^s}
    \end{align*}
\end{completeproof}

\subsection{Multi-party computation}
\label{apx:impossibility-mpc-proofs}

\mpcimpossibility*
\begin{completeproof}
    Suppose for a contradiction there is a two-party protocol $\Pi = (\Pi_A, \Pi_B)$, ran by Alice and Bob respectively, such that
    \begin{align*}
        \Pi_A \Pi_B & \approx_\varepsilon C^{and} \\
        \Pi_B & \approx_\varepsilon \sigma_A C^{and}_A \\
        \Pi_A & \approx_\varepsilon C^{and}_B \sigma_B
    \end{align*}
    for $\varepsilon < \frac{1}{12}$. We apply the triangle inequality once again and obtain:
    \begin{align*}
        C^{and}_B \sigma_{BA} C^{and}_A \approx_{3\varepsilon} C^{and}
    \end{align*}
    
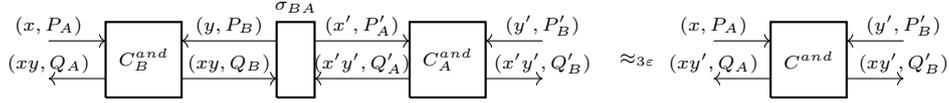
\begin{figure}[H]
        \centering
        \begin{tikzpicture}[scale=1]
    \scriptsize

    \acleftmessage[->]{$(x, P_A)$}{0.25}{-4.25}{-3.5}
    \acleftmessage[<-]{$(xy, Q_A)$}{-0.25}{-4.25}{-3.5}
    \acresource{$C^{and}_B$}{-3.5}{-0.5}{1}{1}
    
    \acmessage[<-]{$(y, P_B)$}{0.25}{-2.5}{-1.25}
    \acmessage[->]{$(xy, Q_B)$}{-0.25}{-2.5}{-1.25}
    \acprotocol{$\sigma_{BA}$}{-1.25}{-0.5}{0.5}{1}
    \acmessage[->]{$(x', P'_A)$}{0.25}{-0.75}{0.5}
    \acmessage[<-]{$(x'y', Q'_A)$}{-0.25}{-0.75}{0.5}
    
    \acresource{$C^{and}_A$}{0.5}{-0.5}{1}{1}
    \acrightmessage[<-]{$(y', P'_B)$}{0.25}{1.5}{2.25}
    \acrightmessage[->]{$(x'y', Q'_B)$}{-0.25}{1.5}{2.25}
    
    \indist[3\varepsilon]{3.5}{0}
    
    \acleftmessage[->]{$(x, P_A)$}{0.25}{4.5}{5.25}
    \acleftmessage[<-]{$(x y', Q_A)$}{-0.25}{4.5}{5.25}
    \acresource{$C^{and}$}{5.25}{-0.5}{1}{1}
    \acrightmessage[<-]{$(y', P'_B)$}{0.25}{6.25}{7}
    \acrightmessage[->]{$(x y', Q'_B)$}{-0.25}{6.25}{7}
    
    \acresetframe
\end{tikzpicture}
        \label{fig:mpc-no-go-construction}
    \end{figure}
    We consider a distinguisher $D$ which inputs $x, y'$ chosen uniformly at random and guesses the constructed resource if and only if $xy \neq xy'$ or $x'y' \neq xy'$. By the usual application of the statistical separation lemma:
    \begin{align*}
        3\varepsilon \ge d^{D}(C^{and}_B \sigma_{BA} C^{and}_A, C^{and}) \ge \Prob{xy \neq xy' \vee x'y' \neq xy'}.
    \end{align*}
    Hence, we conclude the proof by finding a lower bound for this probability:
    \begin{align*}
        &\Prob{xy \neq xy' \vee x'y' \neq xy'} \\
         =& \quad
            \frac{1}{4} \Prob{xy \neq xy' \vee x'y' \neq xy' \given x = 0, y' = 0} 
             + \frac{1}{4} \Prob{xy \neq xy' \vee x'y' \neq xy' \given x = 0, y' = 1} \\
            & + \frac{1}{4} \Prob{xy \neq xy' \vee x'y' \neq xy' \given x = 1, y' = 0} 
             + \frac{1}{4} \Prob{xy \neq xy' \vee x'y' \neq xy' \given x = 1, y' = 1} \\
        =& \quad \frac{1}{4} \Prob{0 \neq 0 \vee 0 \neq 0 \given x = 0, y' = 0}  
             + \frac{1}{4} \Prob{0 \neq 0 \vee x' \neq x \given x = 0, y' = 1}  \\
            & + \frac{1}{4} \Prob{y \neq y' \vee 0 \neq 0 \given x = 1, y' = 0}  
             + \frac{1}{4} \Prob{y \neq y' \vee x' \neq x \given x = 1, y' = 1} \\
        =& \frac{1}{4} \Prob{x' \neq x \given x = 0, y' = 1}  
             + \frac{1}{4} \Prob{y \neq y' \given x = 1, y' = 0}  
            + \frac{1}{4} \Prob{y \neq y' \vee x' \neq x \given x = 1, y' = 1}.
    \end{align*}
    Now, note that we can have $Q'_A \prec P_B$ or $Q_B \prec P'_A$, but not both. Assume without loss of generality $Q'_A \not\prec P_B$ (the other case is analogous). Therefore, $y$ cannot depend on $xy'$ (and, in particular, is independent from $y'$) and the lower bound will become:
    \begin{align*}
        \Prob{xy \neq xy' \vee x'y' \neq xy'} & \ge \frac{1}{4} \Prob{y \neq y' \given x = 1, y' = 0} + \frac{1}{4} \Prob{y \neq y' \given x = 1, y' = 1} \\
        & = \frac{1}{2} \Prob{y \neq y' \given x = 1} = \frac{1}{4}.
    \end{align*}
    which gives us $\varepsilon \ge \frac{1}{12}$ for any possible choice of the causal order used by $\sigma_{BA}$. This leads to a contradiction.
\end{completeproof}

\subsection{Oblivious Transfer and Bit Commitment}
\label{apx:equivalence-ot-bc-proofs}

\unruhconstruction*

\newpage
\otbcconstruction*

\fi

\vspace{2cm} 


\end{document}